\documentclass[11pt,a4paper]{article}

\usepackage[pagebackref=false,colorlinks=true,citecolor=blue,filecolor=blue,%
linkcolor=blue,anchorcolor=blue,urlcolor=blue]{hyperref}

\usepackage[english]{babel} 

\usepackage{microtype} 

\usepackage{amsmath,amsfonts,amsthm} 

\usepackage[svgnames]{xcolor} 

\usepackage[small, labelfont=bf, up, textfont=it]{caption} 

\usepackage{booktabs} 

\usepackage{lastpage} 

\usepackage{graphicx} 

\usepackage{enumitem} 
\setlist{noitemsep} 

\usepackage{sectsty} 
\allsectionsfont{\usefont{OT1}{phv}{b}{n}} 

\usepackage{multirow}
\usepackage{amssymb}
\usepackage{hepunits}
\usepackage{hepnames}
\usepackage{tensor}
\usepackage{braket}

\usepackage{xpatch}
\makeatletter
\xpatchcmd\@HepConStyle
 {\edef\@upcode{\updefault}}
 {\ifdefined\shapedefault\edef\@upcode{\shapedefault}\else\edef\@upcode{\updefault}\fi}
 {}{}
\makeatother

\usepackage{geometry} 

\usepackage{setspace}
\usepackage[T1]{fontenc} 
\usepackage[utf8]{inputenc} 

\usepackage{XCharter} 

\usepackage{fancyhdr} 
\fancypagestyle{firstpage}{ 
	\fancyhf{}
}

\newcommand{\authorstyle}[1]{{\large\usefont{OT1}{phv}{b}{n}\color{Black}#1}} 

\newcommand{\institution}[1]{{\footnotesize\usefont{OT1}{phv}{m}{sl}\color{Black}#1}} 

\usepackage{titling} 

\newcommand{\HorRule}{\noindent\color{DarkGoldenrod}\rule{\linewidth}{1pt}} 

\pretitle{
	\vspace{-30pt} 
	\HorRule\vspace{0pt} 
	\fontsize{32}{36}\usefont{OT1}{phv}{b}{n}\selectfont 
	\color{DarkRed} 
}

\posttitle{\par\vskip 25pt} 

\preauthor{} 

\postauthor{ 
	\vspace{0pt} 
	\par\HorRule 
	\vspace{0pt} 
}

\usepackage{lettrine} 
\usepackage{fix-cm}	

\newcommand{\initial}[1]{ 
	\lettrine[lines=3,findent=4pt,nindent=0pt]{
		\color{DarkGoldenrod}
		{#1}
	}{}%
}

\usepackage{xstring} 

\newcommand{\lettrineabstract}[1]{
	\StrLeft{#1}{1}[\firstletter] 
	\initial{\firstletter}\textbf{\StrGobbleLeft{#1}{1}} 
}

\newcommand{\req}[1]{(\ref{#1})}
\newcommand{\nn}{\nonumber}

\def\otimesy{\stackrel{y}{\otimes}}
\def\otimesz{\stackrel{z}{\otimes}}

\def\alphaS{\alpha_s}

\def\GPD{\text{GPD}}
\def\tH{\widetilde{H}}
\def\tE{\widetilde{E}}

\def\GPD{H}
\def\tGPD{\widetilde{H}}
\def\CFF{\mathcal{H}}
\def\tCFF{\widetilde{\mathcal{H}}}

\def\GPDe{E}
\def\tGPDe{\widetilde{E}}
\def\CFFe{\mathcal{E}}
\def\tCFFe{\widetilde{\mathcal{E}}}

\def\caseI{(\text{I})}
\def\caseII{(\text{II})}
\def\caseIII{(\text{III})}
\def\caseIV{(\text{IV})}
\def\Mev{\relax\ifmmode{\text{MeV}}\else{MeV{ }}\fi}

\def\veps{\varepsilon}

\def\expal{\delta}
\def\zcor{w}

\def\muR{\mu_R}
\def\muF{\mu_F}

\def\ss{s_0}

\def\by{\overline{y} \,}
\def\bz{\overline{z} \,}

\def\bv{\overline{v} \,}

\def\balpha{\bar{\alpha} \, }

\def\MM{{\rm M}}

\def\NN{{\rm N}}
\def\PP{{\rm P}}
\def\SS{{\rm S}}

\def\MA{{\rm PP}}
\def\MV{{\rm SP}}
\def\MAO{{\rm PP(0)}}
\def\MAf{{\rm PP(1)}}
\def\MVf{{\rm SP(1)}}
\def\MVs{{\rm SP(2)}}

\def\PV{\mathcal{P}}

\def\Li{\relax\ifmmode{\text{Li}_{2}}\else{Li$_2${ }}\fi}

\def\Ieta{i \epsilon}

\def\a{i}
\def\b{j}
\def\ei{e_\a^2}
\def\ej{e_\b^2}
\def\eij{e_{\a} e_{\b}}

\def\Mamp{\mathcal{T}}

\def\MampPP{\mathcal{T}_{\PP \PP}}
\def\MampSP{\mathcal{T}_{\SS \PP}}
\def\MAi{\Mamp_{A}^{(\a)}}
\def\MAj{\Mamp_{A}^{(\b)}}
\def\MAij{\Mamp_{A}^{(\a\b)}}
\def\MBi{\Mamp_{B}^{(\a)}}
\def\MBj{\Mamp_{B}^{(\b)}}
\def\MBij{\Mamp_{B}^{(\a\b)}}
\def\MAiFive{\Mamp_{A5}^{(\a)}}
\def\MAjFive{\Mamp_{A5}^{(\b)}}
\def\MAijFive{\Mamp_{A5}^{(\a\b)}}
\def\MBiFive{\Mamp_{B5}^{(\a)}}
\def\MBjFive{\Mamp_{B5}^{(\b)}}
\def\MBijFive{\Mamp_{B5}^{(\a\b)}}
\def\MBijFiveDelta{\Mamp_{\Delta B5}^{(\a\b)}}

\def\qsum{(+)}
\def\qdiff{(-)}
\def\qprod{(0)}
\def\ggres{(g)}

\newcommand{\mixAzy}{\bz y - \alpha \, z \,\by}

\def\Mp{m}
\def\pT{\kappa}

\def\ssM{s\raisebox{-0.4ex}{$'$}}
\def\tsM{t\raisebox{-0.2ex}{$'$}}
\def\usM{u\raisebox{-0.4ex}{$'$}}

\def\smin{\text{min}}
\def\smax{\text{max}}
\def\sMin{\text{Min}}
\def\sMax{\text{Max}}
\def\sTrans{\text{Trans}}

\graphicspath{{figs/}}

\numberwithin{equation}{section}

\title{
\Huge
\begin{center}
Hard exclusive photoproduction of photon-meson pairs: 
pseudoscalar channels $\pi$, $\eta$ and $\eta'$
\end{center}
}

\author{
\begin{center}
    \authorstyle{Nikola Crnkovi\'{c}\textsuperscript{1}},
    \authorstyle{Goran Duplan\v{c}i\'{c}\textsuperscript{1}},
    \authorstyle{Saad Nabeebaccus\textsuperscript{2}},
    \authorstyle{Kornelija Passek-K.\textsuperscript{1}} 
    \authorstyle{Bernard Pire\textsuperscript{3}},
    \authorstyle{Lech Szymanowski\textsuperscript{4}},
    \authorstyle{Samuel Wallon\textsuperscript{5}},
	\newline\newline 
	\textsuperscript{1}\institution{Division of Theoretical Physics, Ruđer Bošković Institute,
    HR-10002 Zagreb, Croatia}\\ 
    \textsuperscript{2}\institution{Department of Physics and Astronomy, University of Manchester, Manchester M13 9PL, United Kingdom}\\ 
    \textsuperscript{3}\institution{Centre de Physique Théorique, CNRS, École Polytechnique, I.P. Paris, 91128 Palaiseau, France}\\ 
    \textsuperscript{4}\institution{National Centre for Nuclear Research, NCBJ, 02-093 Warsaw, Poland}\\ 
    \textsuperscript{5}\institution{Université Paris-Saclay, CNRS, IJCLab, 91405 Orsay, France}\\ 
\end{center}
}

\begin{document}


\maketitle 

\thispagestyle{firstpage} 

\lettrineabstract{%
We investigate the hard exclusive photoproduction of 
photon-meson pairs
at leading-twist and leading-order in perturbative QCD,
and focus on pseudoscalar
mesons $\text{M} \in \{\pi^\pm, \pi^0, \eta, \eta'\}$.
Compact analytical expressions are obtained for the
amplitudes involving quark generalized parton distributions,
with the two-gluon components of the $\eta$ and $\eta'$
distribution amplitudes included.
The numerical analysis is performed in the moderate-$\xi$ region,
where valence-quark GPDs are expected to be important.
In this region, we find a strong impact of the pion-pole term 
in $\gamma\pi^\pm$ production, 
and a non-negligible effect for neutral mesons.  
We also observe a marked dependence of
$\gamma\eta'$ photoproduction on two-gluon contributions.
This process offers enhanced sensitivity to the shape of the GPDs
already at leading-order,
while the tested dependence on the meson distribution amplitude
and the renormalization scale introduces
further theoretical uncertainties, the latter emphasizing the need 
for next-to-leading-order corrections.
Our results provide a concise analytical framework and
a numerical baseline for future studies.
}

\clearpage


\tableofcontents


\section{Introduction}
\label{sec:intro}

In contrast to parton distribution functions (PDFs),
generalized parton distributions (GPDs) probed in hard exclusive
processes encode the multidimensional structure of the nucleon
through three variables:
the parton’s average longitudinal momentum fraction $x$,
the skewness $\xi$, and the momentum transfer $t$.
They enter the process amplitudes through convolutions
with the corresponding hard subprocess amplitudes.
Consequently, extracting detailed information about GPDs,
in particular their full dependence on the longitudinal
momentum fraction $x$, remains a difficult task that
requires combining input from several reactions.
Most of the present knowledge comes from
deeply virtual Compton scattering (DVCS)
and deeply virtual meson production (DVMP),
for which factorization has been rigorously proven 
\cite{Ji:1998xh,Collins:1998be,Collins:1996fb}.
These reactions are supported by abundant experimental data,
and their analyses have been carried out up to
next-to-leading order (NLO) and beyond;
for the phenomenological status see
\cite{dHose:2016mda,Kumericki:2016ehc,Favart:2015umi,Cuic:2023mki}
and references therein.
These processes primarily access the chiral-even GPDs:
parity-even (vector) $H$ and $E$, and parity-odd (axial)
$\tH$ and $\tE$, while they lack sensitivity to the quark
transversity GPDs 
at leading twist.

Photon-meson photoproduction,
$\gamma \NN \to \gamma\MM\, \NN'$,
has been proposed as a promising alternative to DVCS and DVMP,
offering complementary access to GPDs
\cite{Boussarie:2016qop,Duplancic:2018bum}.
Similar to DVMP, this reaction probes quark flavors and includes
gluon contributions already at leading order.
An additional advantage is its sensitivity to transversity GPDs,
which can be accessed at leading twist in the production of
transversely polarized vector mesons.

Unlike DVCS ($\gamma^* \NN \to \gamma \NN$) and
DVMP ($\gamma^* \NN \to \MM \NN'$), which are $2 \to 2$ processes
involving essentially a single large scale, namely the incoming photon virtuality,
photon-meson photoproduction is a $2 \to 3$ reaction that 
is characterized by two large scales and features more complex kinematics and a more
elaborate leading-order (LO) hard-scattering amplitude.
While the former processes provide mainly moment-type constraints,
effectively probing GPDs at $x = \pm \xi$ at LO,
the additional kinematic degrees of freedom in
photon-meson photoproduction lead to a more detailed sensitivity
to the $x$-dependence of GPDs \cite{Qiu:2023mrm}.
A similar mechanism appears in double deeply virtual Compton
scattering (DDVCS), where the presence of an additional scale
also makes the process sensitive to $x\ne \pm \xi$ region already at LO \cite{Deja:2023ahc}.

Several other $2 \to 3$ processes have also been proposed.
These include meson-meson photoproduction
\cite{Ivanov:2002jj,Enberg:2006he,ElBeiyad:2010pji,Siddikov:2022bku},
which provides access to transversity GPDs;
two-photon photoproduction
\cite{Pedrak:2017cpp,Pedrak:2020mfm,Grocholski:2021man,Grocholski:2022rqj},
which has been studied at next-to-leading order (NLO);
and the pion-nucleon to photon-photon process
\cite{Qiu:2022bpq,Qiu:2024mny},
which represents the crossed counterpart of
photon-meson photoproduction.

Photon-meson photoproduction has recently attracted
considerable attention.
The vector-meson channels, $\gamma \rho_L$ and $\gamma \rho_T$
\cite{Boussarie:2016qop},
and the pseudoscalar channels, such as $\gamma \pi$
\cite{Duplancic:2018bum,Qiu:2023mrm},
have been investigated at leading order.
In particular, the sensitivity of $\gamma \pi$ photoproduction
to the GPD structure has been explored in detail in
\cite{Qiu:2023mrm}, together with its crossed counterpart,
$\pi \NN \to \gamma \gamma \NN'$
\cite{Qiu:2024mny}.
The feasibility of measuring these reactions at JLab, COMPASS, and EIC,
as well as in ultraperipheral collisions 
at the LHC,
has also been extensively studied
\cite{Qiu:2023mrm,Duplancic:2022ffo,Duplancic:2023kwe}.
Photoproduction of photon-heavy-meson pairs has likewise been analyzed
\cite{Siddikov:2024blb},
in which gluon contributions are leading 
and, due to the large meson mass, remain theoretically well behaved.

The factorization of photon-meson photoproduction has recently been established \cite{Qiu:2022pla}. However, it was pointed out in \cite{Nabeebaccus:2023rzr,Nabeebaccus:2024mia} that there are Glauber pinch singularities, contributing at the leading power, which spoil the factorisation in cases where two-gluon exchanges with the nucleon sector are allowed. 
There are potential remedies to save the phenomenology, 
e.g. in a modified factorisation framework \cite{Botts:1989kf,Fucilla:2025wow}.
However, it is not the task of the present paper to address this. 
In this paper, we restrict ourselves to the contributions of quark GPDs at leading order, 
where there are no divergences appearing from the previously-mentioned 
factorisation breaking effects.

One of the motivations for the present work was the analysis
of contributions involving the two-gluon DA. 
We show that, in contrast to the two-gluon exchange
contributions associated with gluon GPDs, the two-gluon components
of the $\eta$ and $\eta'$ distribution amplitudes do not lead to
end-point or integration singularities. This is expected to be the case, since the Glauber pinch singularities that were identified in \cite{Nabeebaccus:2023rzr,Nabeebaccus:2024mia} only affect cases where the Glauber exchange occurs between partons that connect incoming and outgoing asymptotic states.
Therefore, the gluonic content of the $\eta$ and $\eta'$ mesons
can be straightforwardly incorporated within the factorized framework
for $\gamma \NN \to \gamma \eta (\eta') \NN'$,
unlike the case of gluon GPDs in the nucleon sector discussed above.

In this work, we revisit the photoproduction of a photon-meson pair,
$\gamma \NN \to \gamma \MM \, \NN'$,
to systematize the results for pseudoscalar mesons
$\MM \in \{ \pi^\pm, \pi^0, \eta, \eta' \}$ and to include the two-gluon
contributions to the $\eta$ and $\eta'$ channels.
We obtain compact analytical expressions for the quark GPD
contributions, which clarify the structure of the results and
allow efficient numerical implementation.

Our phenomenological analysis is performed in the moderate-$\xi$
kinematic region ($\xi > 0.1$).
Only valence quarks were taken into account, and
the effects of sea quarks and, in particular, gluons
from the nucleon -- relevant only in the neutral meson channels --
are not considered.
As discussed above, the latter would require a modified framework,
and we leave the inclusion of these contributions for
future work.
Since valence quarks are significant in this $\xi$ region, we
believe that the valence-quark approximation in the neutral meson
channel provides a consistent picture and useful insight into the
contributions.
In the case of charged pions, the sea-quark contributions cancel in
the model used and gluons do not contribute.
This kinematic region is of interest for
the JLab and COMPASS/AMBER experiments in the near future.
In this regime, the arguments for the dominance of
the $H$ and $\tH$ GPDs based on the smallness of $\xi$
\cite{Duplancic:2018bum,Duplancic:2022ffo}
are not applicable, and we therefore include the
$E$ and $\tE$ GPDs in our analysis as well.
Moreover, for moderate $\xi$, the pion-pole contributions
can be significant, and we perform a dedicated study of the
pion-pole contributions in all meson channels.

Furthermore, we discuss the meson structure in terms of
DAs and its impact on
$\gamma \MM$ photoproduction. 
The shape of pion DA, in addition to determining the internal structure
of the produced pion, also affects the pion-pole contribution.
For $\eta$ and $\eta'$ mesons, we analyze the role of
the two-gluon components.
The evolution of DAs and the dependence on the factorization scale
are consistently taken into account, while the impact of the
renormalization scale is examined and its implications for NLO
calculations discussed.
These aspects are investigated within the GPD model proposed in
\cite{Goloskokov:2005sd,Goloskokov:2007nt,Goloskokov:2009ia}.

The aim of this work is to extend and systematize the analytical
results, to investigate the individual contributing parts and the
associated theoretical uncertainties within the kinematical range
of interest, and to provide a reference numerical baseline for
future experimental measurements, whose data would in turn enable
a more precise determination of GPDs.

The paper is organized as follows.
Section~\ref{sec:theory} presents the theoretical framework
for evaluating photon-meson photoproduction amplitudes.
We describe the collinear kinematics relevant for factorization
and specify the perturbative and nonperturbative components,
in particular the DAs and GPDs.
Subsection~\ref{sec:ares} contains the main analytical results of this
work, namely the compact analytical expressions for the
subprocess amplitudes, including the two-gluon contributions to
$\gamma \eta$ and $\gamma \eta'$ photoproduction.
Section~\ref{sec:obs} summarizes the elements needed for the
evaluation of observables used in the numerical analysis.
Section~\ref{sec:numres} then presents a detailed discussion of
the numerical results and their phenomenological implications.
Section~\ref{sec:sum} concludes the paper with final remarks.

Additional material is provided in the Appendices.
Appendices~\ref{app:kin} and \ref{app:DAev}
give further details on kinematics and meson DAs and their evolution.
Appendices~\ref{app:THq}, \ref{app:PV} and \ref{app:nres}
present selected technical and numerical results.
In particular, Appendix~\ref{app:THq} contains the general unsymmetrized
form of the quark subprocess amplitudes,
Appendix~\ref{app:PV} explains the integration of the subprocess
building blocks, 
and Appendix~\ref{app:nres} gathers, for convenience,
the figures illustrating the main numerical results.

\section{Perturbative QCD framework for photon-meson photoproduction}
\label{sec:theory}

\newsavebox{\leftfig}
\newsavebox{\rightfig}

\sbox{\leftfig}{\includegraphics[width=0.3\textwidth]{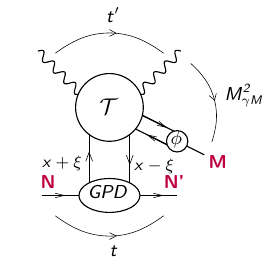}}
\sbox{\rightfig}{\includegraphics[width=0.3\textwidth]{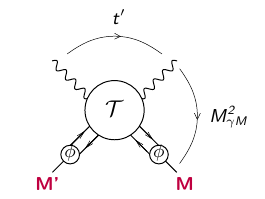}}

\newlength{\raisefig}
\setlength{\raisefig}{\dimexpr .35\ht\leftfig - .8\ht\rightfig\relax}

\newlength{\arrowraise}
\setlength{\arrowraise}{.27\ht\leftfig}

\begin{figure}[t]
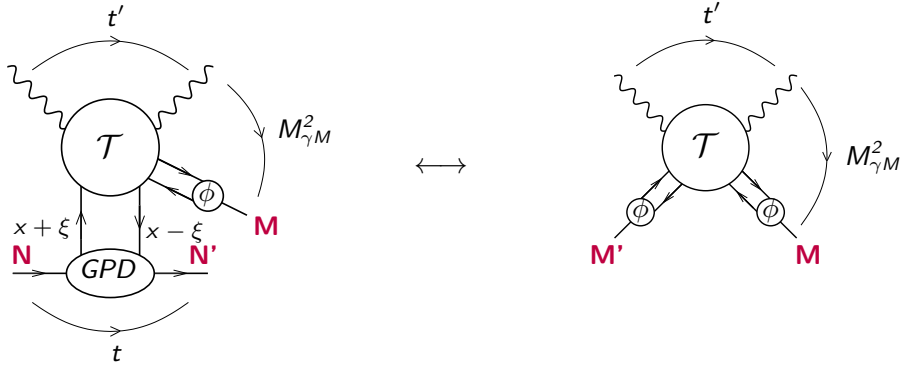

\centering
\begin{tabular}{ccc}
  \raisebox{\raisefig}{\usebox{\leftfig}} &
  \hspace{0.55cm}\raisebox{\arrowraise}{$\longleftrightarrow$}\hspace{0.5cm} &
  \usebox{\rightfig}
\end{tabular}
\caption{
Left: 
Factorization of $\gamma \NN \to \gamma \MM \, \NN'$ process.
Right: 
Brodsky--Lepage factorization for $\gamma \MM' \to \gamma \MM$ process.
}
\label{fig:fact-gMgM}
\end{figure}
%
%
%

In this work, we study the photoproduction of a photon-meson pair,
\begin{equation}
\gamma(q) + \NN(p_1) 
\to 
\gamma(k) + \MM(p_{\MM}) 
+ \NN'(p_2)
\, ,
\label{eq:proc}
\end{equation}
where $\MM$ denotes a pseudoscalar meson,
$\MM \in \{\pi^\pm, \pi^0, \eta, \eta'\}$.
The relevant scalar Lorentz invariants
for this process include
\begin{equation}
\begin{array}{lll}
S_{\gamma \NN} = (q + p_1)^2
\, ,
\qquad
&
t = (p_2 - p_1)^2
\, ,
\qquad
&
M_{\MM \NN'}^2 = (p_{\MM} + p_2)^2
\, ,
\\[0.3cm]
\ssM = M_{\gamma \MM}^2 = (k + p_{\MM})^2
\, ,
\qquad
&
\tsM = (q - k)^2
\, ,
\qquad
&
\usM = (q - p_{\MM})^2
\, .
\end{array}
\label{eq:scalars}
\end{equation}
Here, $S_{\gamma \NN}$ denotes the invariant mass of the incoming photon-nucleon pair,
$t = (p_2 - p_1)^2$ is the squared momentum transfer between nucleons,
$M_{\MM \NN'}^2$ represents the invariant mass of the outgoing meson-nucleon system,
and $\ssM = M_{\gamma \MM}^2$ corresponds to the invariant mass of the outgoing $\gamma \MM$ pair.
The variables $\ssM$, $\tsM$, and $\usM$ also serve as
the Mandelstam variables of the partonic subprocesses
$\gamma \, ( q_1 \bar{q}_2 ) \to \gamma ( q_1 \bar{q}_2 )$
and
$\gamma \, ( q \bar{q} ) \to \gamma ( g g )$,
which are discussed in more detail in the next subsection. 
One easily verifies that
\begin{equation}
\ssM + \tsM + \usM = t + m_{\MM}^2
\, .
\label{eq:stutm}
\end{equation}
The photoproduction of the $(\gamma \MM)$ pair, Eq.~\eqref{eq:proc},
is fully specified by the meson and nucleon masses,
$m_{\MM}$ and $\Mp_{\NN}$,
together with five independent Lorentz-invariant quantities%
\footnote{For fixed particle masses, a process with $N$ external
particles ($N\geq 4$) is characterized by $3N - 10$ independent Lorentz-invariant
scalar products.}.
These can be taken either from Eq.~\eqref{eq:scalars}
or expressed in terms of the dimensionless variables
introduced in App.~\ref{app:kin}.
The factorization constraints further reduce the number
of independent scalar quantities required.
As explained below, we use four invariants:
$\ssM = M_{\gamma \MM}^2$, $\usM$, $t$,
and $S_{\gamma \NN}$,
or equivalent combinations of dimensionless variables.

The factorization of photon-meson photoproduction
has been proved in Ref.~\cite{Qiu:2022pla}, up to the subtleties regarding two-gluon exchanges with the nucleon sector mentioned in Sec.~\ref{sec:intro}.
Figure~\ref{fig:fact-gMgM} illustrates
the key ingredients of this factorization.
When the momentum transfer to the outgoing nucleon, $t$,
is small compared to the large invariant mass
of the photon-meson pair, $M_{\gamma \MM}^2$,
a kinematic configuration arises that closely resembles
timelike Compton scattering (TCS)~\cite{Berger:2001xd}, where the final-state virtual photon with $Q^2 > 0$
is replaced by the $\gamma \MM$ system.
The factorization conditions require a large momentum transfer $\tsM$,
corresponding to large scattering angles,
i.e., a large transverse momentum of the outgoing photon.
It follows that the factorization of the $\gamma \NN \to \gamma \MM \NN'$
process in terms of nucleon GPDs
is analogous to the large-angle Brodsky-Lepage factorization
of the $\gamma \MM' \to \gamma \MM$ process,
with the GPD corresponding to the meson $\MM'$ distribution amplitude%
\footnote{This analogy parallels
DVCS ($\gamma^* \NN \to \gamma \NN$)
and the meson transition form factor ($\gamma^* \gamma \to \MM'$),
as well as
DVMP ($\gamma^* \NN \to \MM \NN'$)
and the meson electromagnetic or
meson-to-meson transition form factor ($\gamma^* \MM' \to \MM$).}.

We refer to Refs.~\cite{Boussarie:2016qop,Duplancic:2018bum,Duplancic:2022ffo}
and App.~\ref{app:kin} for a detailed account of the general kinematics
of the process
$\gamma \, \NN \to \gamma \, \MM \, \NN'$.
As usual, the momentum transfer between nucleons is defined as
\begin{equation}
\Delta^{\mu} = p_2^{\mu} - p_1^{\mu}
\, ,
\label{eq:Delta}
\end{equation}
$t = \Delta^2$,
and the longitudinal momentum transfer 
(i.e., the skewness), is defined in terms of light-cone variables as
\begin{equation}
\xi = -\frac{\Delta^+}{P^+}
\, ,
\qquad
\text{where}
\qquad
P^{\mu} = p_1^{\mu} + p_2^{\mu}
\, .
\label{eq:xidef}
\end{equation}
In what follows, we summarize the key ingredients required for
the calculation of the subprocess amplitudes
within the collinear approximation.
For this purpose both the transverse momentum between nucleons, $\Delta_{\perp}$,
and the hadron masses are neglected,
except for the nucleon mass retained in the skewness parameter
\begin{equation}
\xi = \frac{s'}{2 \left( S_{\gamma \NN} - \Mp_{\NN}^2 \right) - s'}
\, ,
\label{eq:xi}
\end{equation}
derived in Eq.~(\ref{eq:xigen}).

\subsection{Collinear kinematics of the hard-scattering subprocess}
\label{sec:kinematics}

When the factorization conditions are satisfied,
the amplitude for the process
$\gamma \NN \to \gamma \MM \, \NN'$
can be represented as a convolution of
the GPD, the meson DA,
and the hard subprocess amplitude
$
\gamma \, q_1 
\to \gamma q_2 \left( q_1 \bar{q}_2 \right)
$
or, equivalently,
\begin{equation}
\gamma \, \left( q_1 \bar{q}_2 \right)
\to \gamma \left( q_1 \bar{q}_2 \right)
\, .
\label{eq:sub-qqqq}
\end{equation}
Effectively, we are considering the
$\gamma (\MM') \to \gamma \MM$ process,
where $(\MM')$ denotes a partonic state carrying meson quantum numbers,
but in our case described through GPDs.
For neutral $C = +1$ states, the quark $(q \bar{q})$ Fock components
can be replaced by gluon Fock states, $gg$.
The contributions of gluon GPDs
to the photoproduction of light neutral pseudoscalar mesons
pose challenges due to Glauber pinch singularities,
and the distinct cross-over behavior of gluon GPDs at 
$|x|=\xi$ compared 
to the end-point behavior of DAs
\cite{Nabeebaccus:2023rzr,Nabeebaccus:2024mia}.
In this work, we calculate the quark GPD contributions,
while the $gg$ configurations are included solely in the
outgoing meson.
The corresponding subprocess amplitude then reads
$
\gamma \, q 
\to \gamma q \left( gg \right)
$,
i.e.,
\begin{equation}
\gamma \, \left( q \bar{q} \right)
\to \gamma \left( g g \right)
\, .
\label{eq:sub-qqgg}
\end{equation}

The subprocess momenta are given by
\begin{eqnarray}
q^\mu + (-\Delta)^\mu &=& k^\mu + p_{\MM}^\mu
\, ,
\end{eqnarray}
with
\begin{eqnarray}
(-\Delta)^\mu &=& y\,(-\Delta)^\mu + \by\,(-\Delta)^\mu\,,
\nonumber \\[0.2cm]
p_{\MM}^\mu &=& z\,p_{\MM}^\mu + \bz\,p_{\MM}^\mu
\, ,
\end{eqnarray}
decomposed in the collinear approximation
in terms of the longitudinal
momentum fractions $z$ and $y$ carried by the partons.
Here we adopt the usual shorthand $\bv = 1 - v$.
While the meson momentum fractions satisfy $0 < z < 1$,
the variable $y$,
in the case of photoproduction of $\gamma \MM$ pair,
serves as a convenient parametrization defined by
\begin{equation}
y = \frac{\xi + x}{2\xi} 
\, , 
\quad
\by = \frac{\xi - x}{2\xi}
\, .
\label{eq:yxix}
\end{equation}
Here $x$ denotes the usual “average” parton
momentum fraction, $-1 < x < 1$,
and consequently $(\xi - 1)/(2\xi) < y < (\xi + 1)/(2\xi)$.

In the collinear limit,
where $t = \Delta^2 = 0$ and $p_{\MM}^2 = 0$,
the subprocess Mandelstam variables introduced in \req{eq:scalars},
$\ssM = (q + (-\Delta))^2$,
$\tsM = (p_{\MM} - (-\Delta))^2$,
and $u' = (k - (-\Delta))^2$,
satisfy
\begin{equation}
\ssM + \tsM + \usM = 0
\, .
\end{equation}
The dimensionless variable $\alpha$,
introduced in App.~\ref{app:kin},
takes the form
\begin{equation}
\alpha = \frac{-\usM}{\ssM} 
\, , 
\qquad
\balpha = \frac{-\tsM}{\ssM}
\, ,
\label{eq:alpha}
\end{equation}
and $0<\alpha<1$. 
The scattering angle $\theta$
of the outgoing photon
in the center-of-mass frame of the subprocess in \eqref{eq:sub-qqqq} or \eqref{eq:sub-qqgg} (i.e. defined wrt to $q-\Delta$ system)
can then be expressed through $\alpha$ as
\begin{equation}
\cos \theta = 2\alpha - 1
\, .
\label{eq:theta-alpha}
\end{equation}
Following \req{eq:fvectors},
the subprocess momenta 
are given by
\begin{equation}
\begin{alignedat}{2}
q^{\mu}   &= n^{\mu} 
\, , 
&\qquad
k^{\mu}   &= 
\alpha\, n^{\mu}
+ \balpha\,(2\xi)\, p^{\mu}
+ \pT_{\perp}^{\mu}
\, , \\[0.3em]
(-\Delta)^{\mu} &= 2\xi\, p^{\mu}
\, , 
&\qquad
p_{\MM}^{\mu} &= 
\balpha\, n^{\mu}
+ \alpha\,(2\xi)\, p^{\mu}
- \pT_{\perp}^{\mu}
\, ,
\end{alignedat}
\label{eq:fvcollkin}
\end{equation}
with the corresponding
Sudakov decomposition
in terms of the light-cone vectors $p$ and $n$
introduced in App.~\ref{app:kin}.
The momentum $\pT_{\perp}$
represents the relative transverse
momentum between the outgoing photon and the meson,
as defined in \req{eq:fvectors}.
According to \req{eq:stucol},
the invariant mass of the photon-meson pair
and the relative transverse momentum are related by
\begin{equation}
-\pT_{\perp}^2 
= \alpha \, \bar{\alpha} \, \ssM 
= \frac{\sin^2 \theta}{4}\, \ssM 
\label{eq:pT}
\, ,
\end{equation}
and both $s'$ and the angle $\theta$,
or equivalently the transverse momentum $\kappa_\perp$,
must be sufficiently large to ensure that the process
lies in the hard–scattering regime.

Following Refs.~\cite{Boussarie:2016qop,Duplancic:2018bum,Duplancic:2022ffo,Duplancic:2023kwe},
in this work
we adopt the axial gauge for photon polarizations,
\begin{equation}
\varepsilon_q \cdot p =
\varepsilon_k \cdot p = 0
\, .
\label{eq:axgauge}
\end{equation}
The polarization vectors can then be written as
\begin{align}
\varepsilon_q^{\mu} &= 
\varepsilon_{q_{\perp}}^{\mu}
\, , 
\nonumber \\[0.3em]
\varepsilon_k^{\mu} &=
-
\frac{\varepsilon_{k_{\perp}}\!\cdot k_{\perp}}{p\!\cdot k}\, p^{\mu}
+
\varepsilon_{k_{\perp}}^{\mu}
\, .
\label{eq:polvectors}
\end{align}

The $2 \to 2$ subprocess
$\gamma (\MM') \to \gamma \MM$
is fully characterized by two scalar
invariants, which we choose as $\ssM$ and $\alpha$.
After accounting for the limit of small transverse momentum
$\Delta_{\perp}$,
the full process $\gamma \NN \to \gamma \MM \, \NN'$
is described by four invariants:
$\ssM$, $\alpha$, $t$, and $S_{\gamma \NN}$.
The skewness $\xi$ is defined as in \req{eq:xi}.

\subsection{Factorization and amplitude decomposition}
\label{sec:ampl1}

The factorization of the
$\gamma \NN \to \gamma \MM \, \NN'$
process proceeds similarly to DVCS and DVMP,
with the amplitude being decomposed using
Dirac spinor bilinears associated to the corresponding GPDs.
In this work, we focus on the photoproduction
of pseudoscalar mesons $\PP$,
and considering allowed quantum numbers,
two possible contributing configurations arise
\begin{alignat}{3}
\gamma(\PP) \to \gamma\PP
&:\qquad
& \tilde{h}^+ &= \bar{u}(p_2)\gamma^+\gamma^5 u(p_1)
\: \to \tH,
&\qquad
\tilde{e}^+ &= \bar{u}(p_2)\frac{\Delta^+\gamma^5}{2 \Mp_\NN}\,u(p_1)
\: \to \tE
\, ,
\nn \\[0.4cm]
\gamma(\SS) \to \gamma\PP
&:\qquad
& h^+ &= \bar{u}(p_2)\gamma^+ u(p_1)
\: \to H,
&\qquad
e^+ &= \bar{u}(p_2)\frac{i\sigma^{+\Delta}}{2 \Mp_\NN}\,u(p_1)
\: \to
 E
\, ,
\label{eq:spin_bil_aa}
\end{alignat}
where $(\PP)$ and $(\SS)$ denote states with pseudoscalar
and scalar quantum numbers, respectively%
\footnote{Besides providing compact notation,
this convention clarifies the relation between the subprocess amplitudes
and those obtained in the crossed
process $\gamma \gamma \to \MM' \MM$
\cite{Baier:1982vlv,Baier:1985pd,Baier:1985wv,
Brodsky:1981rp,Nizic:1987sw,Benayoun:1989ng}.
}.
Specifically,
the parity is $P=-1$ for $\PP$ and $P=+1$ for $\SS$,
while for neutral states the charge parity is $C=+1$.
The total amplitude $\mathcal{M}$ of the process
can therefore be written as a sum
\begin{equation}
\mathcal{M}= 
\mathcal{M}^\MA + \mathcal{M}^\MV = 
\left(\tCFF \frac{\tilde{h}^+}{P^+}
+\tCFFe \frac{\tilde{e}^+}{P^+} \right)
+
\left( \CFF \frac{h^+}{P^+}
+\CFFe \frac{e^+}{P^+} \right)
\, ,
\label{eq:MMAMV}
\end{equation}
where, by analogy with DVCS,
we define the Compton-meson form factors (CMFFs)
$\tCFF$, $\tCFFe$, $\CFF$, and $\CFFe$,
named after the corresponding GPDs.
These form factors factorize into a convolution of
the relevant GPD, the meson DA,
and the subprocess amplitude
$\Mamp$ associated with subprocesses
\req{eq:sub-qqqq} and
\req{eq:sub-qqgg}:
\begin{align}
\left(
\begin{array}{c}
\tCFF_{\MM} \\
\tCFFe_{\MM}
\end{array}
\right)
&\sim
\left(
\begin{array}{c}
\tGPD(x,\xi)\\
\tGPDe(x,\xi) \\
\end{array}
\right)
\stackrel{x}{\otimes}
\MampPP(x,\xi,z)
\stackrel{z}{\otimes}
\phi_{\MM}(z)
=
\left(
\begin{array}{c}
\tGPD(\xi (2 y -1),\xi)
\\
\tGPDe(\xi (2 y -1),\xi)
\end{array}
\right)
\stackrel{y}{\otimes}
\MampPP (y,z)
\stackrel{z}{\otimes}
\phi_{\MM}(z)
\, ,
\\[0.5cm]
\left(
\begin{array}{c}
\CFF_{\MM}
\\
\CFFe_{\MM}
\end{array}
\right)
&\sim
\left(
\begin{array}{c}
\GPD(x,\xi)
\\
\GPDe(x,\xi)
\end{array}
\right)
\stackrel{x}{\otimes}
\MampSP(x,\xi,z)
\stackrel{z}{\otimes}
\phi_{\MM}(z) =
\left(
\begin{array}{c}
\GPD(\xi (2 y -1),\xi)
\\
\GPDe(\xi (2 y -1),\xi)
\end{array}
\right)
\stackrel{y}{\otimes}
\MampSP (y,z)
\stackrel{z}{\otimes}
\phi_{\MM}(z)
\, .
\end{align}
Here we introduce the notation
\begin{equation}
\stackrel{x}{\otimes} \equiv 
\int_{-1}^1 \frac{d x}{2 \xi} 
=\int_{\frac{1}{2}\left(1- \frac{1}{\xi}\right)}^{\frac{1}{2}\left(1+ \frac{1}{\xi}\right)}
dy 
\equiv 
\stackrel{y}{\otimes} 
\, ,
\qquad
\stackrel{z}{\otimes} \equiv 
\int_{0}^1 d z\,,
\label{eq:otimes}
\end{equation}
where \req{eq:yxix} has been taken into account.

We define the tensor amplitude $\mathcal{M}^{\mu \nu}$ through
\begin{equation}
\mathcal{M}(\lambda_q,\lambda_k)= 
\mathcal{M}_{\mu \nu} 
\: 
\varepsilon_{q}^{\mu}(\lambda_q) 
\varepsilon_{k}^{*\nu}(\lambda_k)
\, .  
\label{eq:Mmunu}
\end{equation}
It is then convenient to introduce its general Lorentz decomposition,
which, when applied to the subprocess amplitudes $\Mamp$
and with gauge invariance properly accounted for,
facilitates the translation between different photon gauges
\cite{Duplancic:2018bum}
and enables a direct comparison with established results in the literature,
particularly for the crossed process
$\gamma \gamma \to \MM \MM'$
\cite{Baier:1982vlv,Baier:1985pd,Baier:1985wv,
Brodsky:1981rp,Nizic:1987sw,Benayoun:1989ng}.
In this work, for the photon polarizations,
we adopt the axial gauge
\req{eq:axgauge},
for which generally\footnote{The notation for the numbers in the round brackets are inspired from \cite{Duplancic:2018bum}.}
\begin{align}
\mathcal{M}_{\rm ax}^\MA &=
\left(
\varepsilon_{q} \!\cdot\!
\varepsilon_{k}^*
\right)
\:
\mathcal{M}^\MAO
+
(\varepsilon_{q} \!\cdot\! k)
\:
(q\! \cdot \! \varepsilon_{k}^*)
\:
\mathcal{M}^\MAf ,
\nonumber \\[0.5cm]
\mathcal{M}_{\rm ax}^\MV &=
(q\! \cdot \! \varepsilon_{k}^*)
\:
\varepsilon^{\varepsilon_{q} q (-\Delta) k}\,
\:
\mathcal{M}^\MVf
+
(\varepsilon_{q}\!\cdot\! k)
\:
\varepsilon^{\varepsilon_{k}^* q (-\Delta) k}\,
\:
\mathcal{M}^\MVs
\, .
\end{align}
Following
Refs.~\cite{Boussarie:2016qop,Duplancic:2018bum,Duplancic:2022ffo,Duplancic:2023kwe}
and using \req{eq:polvectors},
we express the Compton-meson form factors through the tensor decomposition
\begin{equation}
\begin{alignedat}{2}
\widetilde{\mathcal{H}} &= \widetilde{\mathcal{H}}_A \, T_A + \widetilde{\mathcal{H}}_B \, T_B,
&\qquad
\widetilde{\mathcal{E}} &= \widetilde{\mathcal{E}}_A \, T_A + \widetilde{\mathcal{E}}_B \, T_B, \\[0.4em]
\mathcal{H} &= \mathcal{H}_{A5} \, T_{A5} + \mathcal{H}_{B5} \, T_{B5},
&\qquad
\mathcal{E} &= \mathcal{E}_{A5} \, T_{A5} + \mathcal{E}_{B5} \, T_{B5}
\, ,
\end{alignedat}
\label{eq:rast_full}
\end{equation}
which employs the modified tensor structures%
\footnote{
Their relations to the four-vectors of the process are given by
$
T_A = \varepsilon_{q} \!\cdot\!  \varepsilon_{k}^*
$,
$
T_B =
- \alpha\, (\varepsilon_{q} \!\cdot\! k)
(\varepsilon_{k}^* \!\cdot\! q)
$, $
T_{A_5}
=
\tfrac{-\alpha}{2\xi}\,
\varepsilon^{\varepsilon_{q} q (-\Delta) k}\,
(\varepsilon_{k}^*\!\cdot\! q)
$,
$
T_{B_5}
=
\tfrac{1}{2\xi}\,
\varepsilon^{\varepsilon_{k}^* q (-\Delta) k}\,
(\varepsilon_{q}\!\cdot\! k)
$.
}
\begin{equation}
\begin{aligned}
T_A &\equiv
\varepsilon_{q_{\perp}} \!\cdot\!
\varepsilon_{k_{\perp}}^* 
\, ,
&
T_B &\equiv
(\varepsilon_{q_{\perp}} \!\cdot\! k_{\perp})
\:
(k_{\perp} \!\cdot\! \varepsilon_{k_{\perp}}^*)
\, ,
\\[0.3cm]
T_{A_5} &\equiv
(k_{\perp} \!\cdot\! \varepsilon_{k_{\perp}}^*)
\: \varepsilon^{n p \varepsilon_{q_{\perp}} k_{\perp}}
\, ,
&
\qquad T_{B_5} &\equiv
(k_{\perp} \!\cdot\! \varepsilon_{q_{\perp}})
\: \varepsilon^{n p \varepsilon_{k_{\perp}}^* k_{\perp}}
\, .
\end{aligned}
\label{eq:TATB5}
\end{equation}

\subsection{Nonperturbative inputs: DAs and GPDs}
\label{sec:GPDDAs}

\subsubsection{Meson distribution amplitudes}

At leading twist, pseudoscalar mesons
can be represented by their $q_i \bar{q}_j$ Fock components,
with the corresponding distribution amplitudes 
defined through hadronic matrix elements as 
\begin{align}
\langle
\MM(p_\MM)|\bar{q}_i(-\zcor)
\, \mathcal{C}_\MM^{ij} \,
\gamma^+ \gamma_5\, q_j(\zcor)|0\rangle
& =
i\, p_\MM^+ f_\MM
\int_0^1 d z\, e^{-i (z-\bz) p_\MM^+ \zcor^-}
\phi_\MM(z)
\, ,
\label{eq:defphiqFT}
\end{align}
for a generic meson momentum $p_\MM$ in the $+$ light-cone direction.
The matrix $\mathcal{C}_\MM^{ij}$ encodes the flavour structure of the meson:
\begin{equation}
\begin{array}{llcl @{\qquad} llcl}
\pi^{\pm} : & \displaystyle
u \bar{d} ( d \bar{u} )
&\rightarrow& \;\displaystyle
{\cal C}_{\pi^{\pm}} = \frac{1}{2}\, \left( \lambda_1 \mp \lambda_2 \right)\,,
&
\eta_{8} : &\displaystyle
\frac{1}{\sqrt{6}} (u \bar{u} + d \bar{d} - 2 s \bar{s})
&\rightarrow& \;\displaystyle
{\cal C}_{\eta_8} = \frac{1}{\sqrt{2}}\, \lambda_8 \,,
\\[0.4cm]
\pi^0 : & \displaystyle
\frac{1}{\sqrt{2}} (u \bar{u} - d \bar{d})
&\rightarrow& \;\displaystyle
{\cal C}_{\pi^0} = \frac{1}{\sqrt{2}}\, \lambda_3 \,,
&
\eta_{1} : & \displaystyle
\frac{1}{\sqrt{3}} (u \bar{u} + d \bar{d} + s \bar{s})
&\rightarrow& \;\displaystyle
{\cal C}_{\eta_1} = \frac{1}{\sqrt{n_f}}\, \mathbf{1}_f \,,
\end{array}
\label{eq:flavour}
\end{equation}
with $\lambda_i$ being the usual SU(3) Gell-Mann matrices, and $n_f=3$.
Additionally, the two-gluon ($gg$) Fock component contributes to the
flavor-singlet pseudoscalar meson $\eta_1$,
whose gluon distribution amplitude is defined as
\begin{align}
\langle \eta_1(p_\MM) |
G^{+\nu}(-\zcor)\,
\widetilde{G}_{\nu}^{\;+}(\zcor)
| 0 \rangle
&=
\frac{(p_\MM^+)^2}{2\sqrt{n_f}}\,
f_{1}
\int_0^1 d z \;
e^{-i (z-\bz) p_\MM^+ \zcor^-}\,
\phi_{\eta_1 g}(z)
\, ,
\label{eq:defphigFT}
\end{align}
where
$\widetilde{G}^{\mu\nu} = \tfrac{1}{2}\, \epsilon^{\mu\nu\gamma\delta} G_{\gamma\delta}$, with the convention that $\epsilon^{0123}=+1$.
The quark and gluon distribution amplitudes,
$\phi_{\eta_{1}q}\equiv\phi_{\eta_{1}}$ and $\phi_{\eta_{1}g}$,
mix under evolution.

Due to pseudoscalar nature and isospin invariance,
the pion%
\footnote{As usual pion DA is taken the same 
for $\pi^+$, $\pi^-$ and $\pi^0$.} 
and quark eta DAs  are symmetric in $(z \to \bz)$,
while two-gluon DAs are antisymmetric:
\begin{equation}
\phi_{\pi (\eta_8, \eta_{1q})}(z)
=
\phi_{\pi (\eta_8, \eta_{1q})}(\bz)
\, ,
\qquad
\phi_{\eta_{1g}}(z)
=
 - \phi_{\eta_{1g}}(\bz)
\, .
\label{eq:DAsym}
\end{equation}
The quark distribution amplitudes
$\phi_{\pi (\eta_8, \eta_{1q})}$
are normalized to unity,
while the flavor-singlet gluon DA,
$\phi_{\eta_{1g}}$,
is, through QCD evolution,
closely linked to its quark counterpart
$\phi_{\eta_{1q}}$.
The convention for the gluon DA adopted here
follows Refs.~\cite{Kroll:2002nt,Kroll:2012gsh,Kroll:2019nby,Kroll:2021zss}.

\begin{figure}[t]
\centering
\begin{tabular}{cc}
\includegraphics[width=0.47\textwidth]{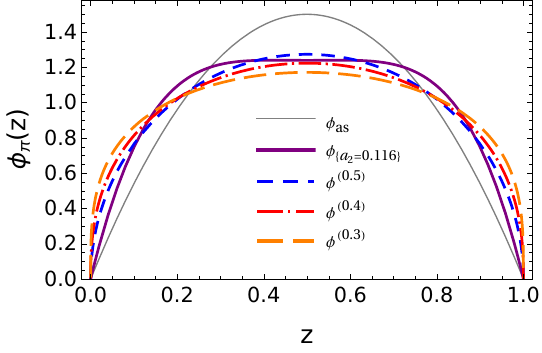}
&
\includegraphics[width=0.47\textwidth]{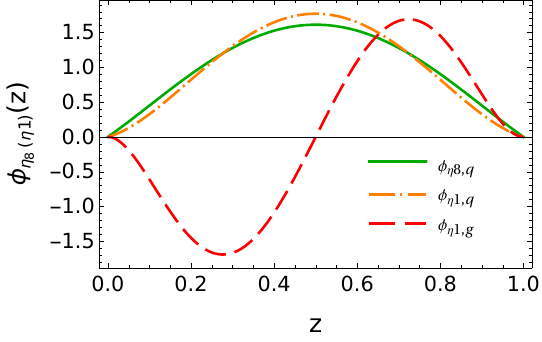}
\end{tabular}
\caption{
Left: Pion DAs.
Shown are the asymptotic DA ($a_2 = 0$),
the lattice-motivated DA with $a_2^\pi(\mu_0 = 2~\text{GeV}) = 0.116$~\cite{RQCD:2019osh}
in the Gegenbauer representation~\req{eq:phiGegen-Q},
and the beta-function ansatz~\req{eq:phi_expal}
for $\eta = 0.5$, $0.4$, and $0.3$.
Right: $\eta_8$ and $\eta_1$ DAs obtained from
the Gegenbauer expansions~\req{eq:phiGegen-Q} and~\req{eq:phiGegen-G},
with coefficients evaluated at $\mu_0 = 2~\text{GeV}$:
$a_2^8 = -0.039$, $a_2^1 = -0.057$, and $a_2^g = 0.38$~\cite{Kroll:2021zss}.
}
\label{fig:DAs}
\end{figure}

It is often convenient to express the meson distribution amplitude 
as an expansion in the eigenfunctions of the LO
evolution kernel-namely, the Gegenbauer polynomials
$C_n^{3/2}(2z-1)$ and $C_n^{5/2}(2z-1)$
for the quark~\req{eq:defphiqFT} and gluon~\req{eq:defphigFT} cases, respectively.
The evolution of DA in this representation is available in a closed analytical form
at both LO and NLO accuracy.
Experimentally and on the lattice, however,
only the first few expansion coefficients are accessible,
with $a_2$ being the most reliably determined.
Consequently, in this work we use the truncated expansions
\begin{align}
\phi_{\MM}(z;\muF) &= 6z(1-z)
\left[
1 + a_2^{\MM}(\muF)\, C_2^{3/2}(2z-1)
\right],
\label{eq:phiGegen-Q}
\\[0.3cm]
\phi_{\eta_{1g}}(z;\muF) &= 30z^2(1-z)^2
\, a_2^{g}(\muF)\, C_{1}^{5/2}(2z-1).
\label{eq:phiGegen-G}
\end{align}
The leading-order evolution  
is described in App.~\ref{app:DAev}.

For $M = \{\pi^+, \pi^-, \pi^0\}$ we compare two forms,
\begin{equation}
a_2^{\pi,\text{as}} = 0,
\qquad
a_2^\pi(\mu_0 = 2~\text{GeV}) = 0.116\substack{+16 \\ -17}
\, ,
\label{eq:a2-pion}
\end{equation}
corresponding respectively to the asymptotic DA
and the lattice moment-based result ~\cite{RQCD:2019osh}.
Recent studies suggest that the pion DA may in fact be broader,
as indicated by momentum-fraction dependent lattice approaches
that can, in principle, reconstruct the full DA shape
~\cite{LatticeParton:2022zqc,Gao:2022vyh}
and by various phenomenological analyses~\cite{Li:2022qul}.
To explore this possibility,
we also consider a symmetric beta-function ansatz,
\begin{equation}
\phi^{(\expal)}_\pi(z) =
\frac{\Gamma(2+2\expal)}{\Gamma^2(1+\expal)}
\, z^{\expal}(1-z)^{\expal},
\label{eq:phi_expal}
\end{equation}
where the parameter $\expal$ controls the end-point behavior:
smaller values correspond to broader DAs.
We test several values-$\expal = 1/2$ (as in~\cite{Brodsky:2006uqa}),
and the broader $\expal = 0.4$ and $\expal = 0.3$
suggested by the lattice results.
The corresponding pion DAs are illustrated in Fig.~\ref{fig:DAs}.

For $\eta$ and $\eta'$ mesons, the octet state $\MM = \eta_8$
is described at leading twist by its $q\bar{q}$ component
and quark DA~\req{eq:phiGegen-Q},
while the singlet $\MM = \eta_1$
also includes a gluon Fock state
described by $\phi_{\eta_{1g}}$~\req{eq:phiGegen-G}.
The two $\eta_1$  DAs, $\phi_{\eta_1}$ and $\phi_{\eta_{1g}}$,
mix under QCD evolution.
The definition of the gluon DA and the associated
flavor-singlet evolution is convention dependent.
In this work, we follow the convention of
Ref.~\cite{Kroll:2019nby}, outlined in App.~\ref{app:DAev}.
The Gegenbauer coefficients are taken from the analysis
of $\eta$ and $\eta'$ transition form factors
in Ref.~\cite{Kroll:2012gsh}
\begin{equation}
a_2^8 = -0.05 \pm 0.02 \, , \qquad
a_2^1 = -0.12 \pm 0.01 \, , \qquad
a_2^g = 0.63 \pm 0.17 \, 
\qquad
\mbox{at}
\; \:
\mu_0 = 1 \,\mbox{GeV}
\, .
\label{eq:a2-eta}
\end{equation}
The corresponding DAs, evolved to the scale 
$\mu = 2~\text{GeV}$,
are depicted in Fig.~\ref{fig:DAs}.

The physical $\eta$ and $\eta'$ mesons are described
in the octet-singlet basis as mixtures of $\eta_8$ and $\eta_1$.
This basis provides a natural framework for incorporating
the two-gluon flavour-singlet components.
Moreover, given the quality of the available data,
the particle dependence and the mixing behaviour
are incorporated solely through the decay constants.
Within the two-angle mixing scheme,
the decay constants are given by 
\begin{align}
f_\eta^8 &= f_8 \cos{\theta_8} \, , \qquad
f_\eta^1 = -f_1 \sin{\theta_1} \, ,
\nonumber\\
f_{\eta'}^8 &= f_8 \sin{\theta_8} \, , \qquad
f_{\eta'}^1 = f_1 \cos{\theta_1} 
\, .
\label{eq:etamix}
\end{align}
The phenomenological values of these parameters,
taken from Ref.~\cite{Feldmann:1998vh}, are
\begin{equation}
\begin{alignedat}{3}
f_\pi &= 131\,\mathrm{MeV},
&\qquad
f_8 &= (1.26 \pm 0.06)\, f_\pi,
&\qquad
\theta_8 &= -(21.2 \pm 1.4)^\circ,
\\[0.3em]
&&
f_1 &= (1.17 \pm 0.04)\, f_\pi,
&\qquad
\theta_1 &= -(9.2 \pm 1.4)^\circ.
\end{alignedat}
\label{eq:pheno}
\end{equation}
The corresponding helicity amplitudes for $\eta'$ and $\eta$ production read
\begin{align}
{\cal M}_{\eta}  &
= \cos{\theta_8}\, {\cal M}_{\eta_8} - \sin{\theta_1}\, {\cal M}_{\eta_1}
\, , \nonumber\\[0.3cm]
{\cal M}_{\eta'} &
= \sin{\theta_8}\, {\cal M}_{\eta_8} + \cos{\theta_1}\, {\cal M}_{\eta_1} 
\, ,
\quad
\mbox{with} 
\;
{\cal M}_{\eta_1}
=
\left( {\cal M}_{\eta_{1q}} + {\cal M}_{\eta_{1g}}\right)
\, .
\label{eq:Metaetap}
\end{align}

\subsubsection{Generalized parton distributions}

Similarly to the meson distribution amplitudes
discussed in the preceding section,
the leading-twist
generalized parton distributions 
are conventionally defined through hadronic matrix elements
in terms of Dirac spinor bilinears \req{eq:spin_bil_aa}:
\begin{align}
\langle
p(p_2)|\bar{q}(-\zcor) \, \gamma^+ \gamma_5 \; q(\zcor)|p(p_1)\rangle
& =
\int_{-1}^{1} dx \: e^{-i x P^+ \zcor^-}
\left(
\tilde{h}^+ \tH^q(x, \xi, t)
        + \tilde{e}^+ \tE^q(x, \xi, t)
\right)
\, ,
\label{eq:deftHEqFT} \\[0.3cm]
\langle p(p_2)|\bar{q}(-\zcor) \, \gamma^+ \, q(\zcor)|p(p_1)\rangle
& =
\int_{-1}^{1} dx \: e^{-i x P^+ \zcor^-}
\left(
h^+ H^q(x, \xi, t) + e^+ E^q(x, \xi, t)
\right)
\, ,
\label{eq:defHEqFT}
\end{align}
which correspond to parity-odd (axial) and parity-even (vector) GPDs, respectively.
The proton GPDs of quark flavour $q$
are denoted as $\tH^q\equiv \tH^q_p$, with analogous notation for the remaining GPDs.
Isospin symmetry relates proton and neutron distributions according to
$\tH^u_p=\tH^d_n$, $\tH^d_p=\tH^u_n$, and $\tH^s_p=\tH^s_n$.
In the case of $\gamma \pi^\pm$ photoproduction, transition GPDs contribute,
and isospin relations lead to \cite{Diehl:2003ny}
$
\tH^{du}_{pn}=
\tH^{ud}_{np}=
\tH^{u}_{p}-\tH^{d}_{p}
$.
Quark GPDs do not, in general, exhibit definite symmetry under
the $(x \to -x)$ transformation.
We therefore introduce the standard symmetric and antisymmetric
combinations, which possess the required symmetry properties
\cite{Diehl:2003ny}
\begin{equation}
\begin{alignedat}{2}
\tGPD^{q(\pm)}(x,\xi)
&\equiv
\tGPD^q(x,\xi)\pm\tGPD^q(-x,\xi)
\, ,
&\qquad
\GPD^{q(\pm)}(x,\xi)
&\equiv
H^q(x,\xi)\mp H^q(-x,\xi)
\, ,
\end{alignedat}
\label{eq:GPDsym}
\end{equation}
and analogously for $\tE$ and $E$ GPDs%
\footnote{
Note that the $(\pm)$ superscript convention
corresponds to the $t$-channel $C$-parity
of the two partons.
Multiplying it by the intrinsic parity,
$(-1)$ for parity-odd and $(+1)$ for parity-even GPDs,
defines the signature $\sigma$,
for which the quark GPDs satisfy
$\text{GPD}(x) = -\sigma\, \text{GPD}(-x)$.
}. 

In this work, we employ the GPDs proposed in
\cite{Goloskokov:2005sd,Goloskokov:2007nt,Goloskokov:2009ia}
and summarized in \cite{Kroll:2012sm},
often referred to as the GK model.
In this framework, the GPDs are constructed from their zero-skewness forms,
whose products with suitable weight functions are interpreted
as double distributions generating the skewness dependence
of the GPDs \cite{Musatov:1999xp}.
For the adopted parameterization of the zero-skewness GPDs,
together with the chosen weight function,
the corresponding double-distribution integral
can be evaluated analytically, 
and the Regge-inspired $t$-dependence added \cite{Goloskokov:2007nt}.
The parameterization of GPDs used in this work corresponds to
initial scale $\mu_0=2$ GeV.

Additionally, we include the pion-pole contribution
to $\tE$, following \cite{Kroll:2012sm},
\begin{equation}
\tE_{\text{pole}}^u(x,\xi,t) =
-\tE_{\text{pole}}^d(x,\xi,t) =
\Theta(|x| \le \xi)
\,
\frac{F_\PP(t)}{4\xi}
\,
\phi_\pi\!\left(\frac{x+\xi}{2\xi}\right)
\, ,
\label{eq:pipole}
\end{equation}
where $F_\PP(t)$ represents the pseudoscalar form factor of the nucleon. 
Chiral symmetry constraints and low-energy theorems imply
that the pole part of this form factor can be written as 
\begin{equation}
F_\PP(t) =
\Mp_\NN f_\pi
\frac{2\sqrt{2}\, g_{\pi \NN \NN}\, F_{\pi \NN \NN}(t)}
{\Mp_\pi^2 - t}
\, .
\end{equation}
Here, $g_{\pi \NN \NN} = 13.1$ denotes the pion-nucleon coupling constant,
and $F_{\pi \NN \NN}(t)$ is the pion-nucleon vertex form factor,
parameterized as
\begin{equation}
F_{\pi \NN \NN}(t) =
\frac{\Lambda_\NN^2 - \Mp_\pi^2}
     {\Lambda_\NN^2 - (t - t_0)}
\, ,
\end{equation}
with $\Lambda_\NN = 0.44~\text{GeV}$ and $t_0$ defined in~\req{eq:t0}.
When the pion-pole term is included,
the replacement $\tE \to \tE + \tE_{\text{pole}}$
(or equivalently,
$
\widetilde{\mathcal{E}} 
\to \widetilde{\mathcal{E}}+\widetilde{\mathcal{E}}_{\text{pole}}
$)
is applied in our calculations.

The pion-pole contribution to $\tE$
can be viewed as a specific example of a resonance-exchange
contribution to GPDs.
In particular, the pion exchange contributing to the region 
$-\xi \le x \le \xi$ of $\tE$
may become significant at small $t$ 
because of the proximity of the pion pole at $t=\Mp_\pi^2$.
This behavior is expected from chiral dynamics, where
the pion is the lightest $t$-channel state and couples strongly
through PCAC and the $\pi NN$ interaction.
The relative importance of the pion pole increases with $\xi$
because it contributes only in the ERBL region $|x| \le \xi$. As
$\xi$ grows, this region widens while the DGLAP domains shrink, so a
larger fraction of the GPD is sensitive to $t$-channel exchange.

In the case of $\gamma \MM$ photoproduction, it corresponds
to the right-hand diagram in Fig.~\ref{fig:fact-gMgM} with $\MM'=\pi$,
that is, to the amplitude of the subprocess $\gamma \pi \to \gamma \MM$.
In DVCS, the pion pole contributes via the transition form factor
$\gamma^* \gamma \to \pi^0$,
and may become important for some observables at moderate $\xi$ 
\cite{Belitsky:2001ns}.
In DVMP, the pion-pole term contributes for charged pions
but vanishes for $\pi^0$,
and it has been found to be significant for $\xi > 0.1$
\cite{Mankiewicz:1998kg,Vanderhaeghen:1999xj,Goloskokov:2009ia}.
In that case, it corresponds to the 
pion electromagnetic form factor
$\gamma^* \pi^\pm \to \pi^\pm$, 
and since perturbative QCD predictions
lie below the experimental measurements,
the phenomenological analyses for DVMP used the parametrization
of the experimental data for the pion form factor
\cite{Goloskokov:2009ia}.
In the present study, such an approach is not possible,
as there are no data available for Compton scattering on a pion
in the kinematical region of interest.
The crossed process 
$\gamma \gamma \to \MM' \MM$ has also been measured,
and there as well, the theoretical predictions lie below the data.
Hence, our estimate of the pion-pole contribution based on the pion DA
may underestimate its actual impact.

\subsection{Analytical expressions for subprocess amplitudes}
\label{sec:ares}

\begin{figure}[t]
\centering
\begin{tabular}{cccc}
\includegraphics[width=0.15\textwidth]{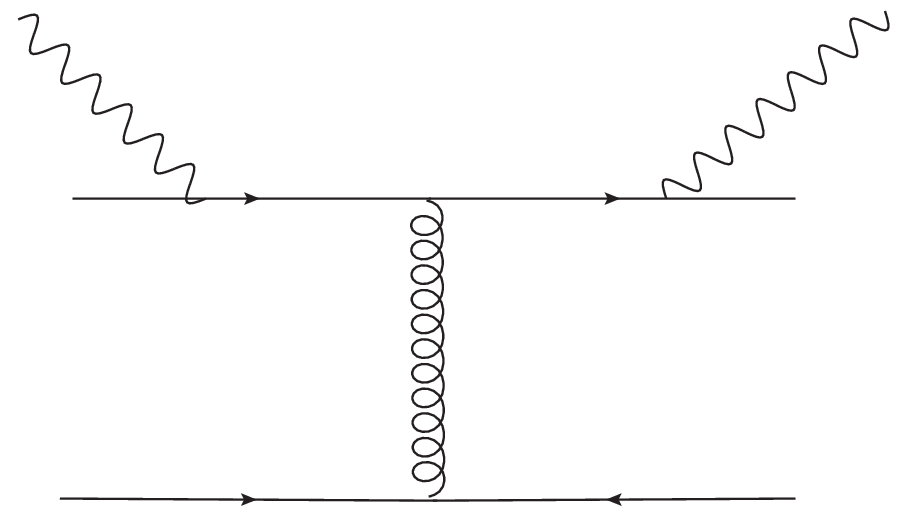}
&
\includegraphics[width=0.15\textwidth]{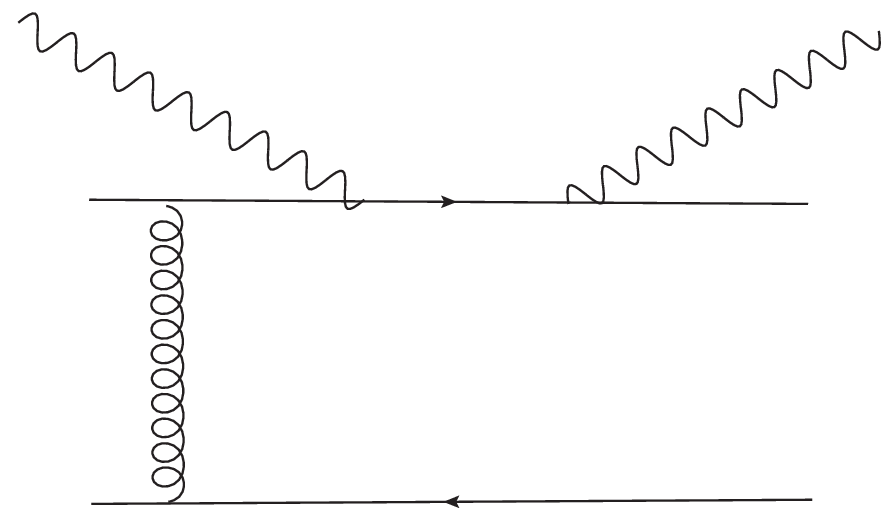}
&
\includegraphics[width=0.15\textwidth]{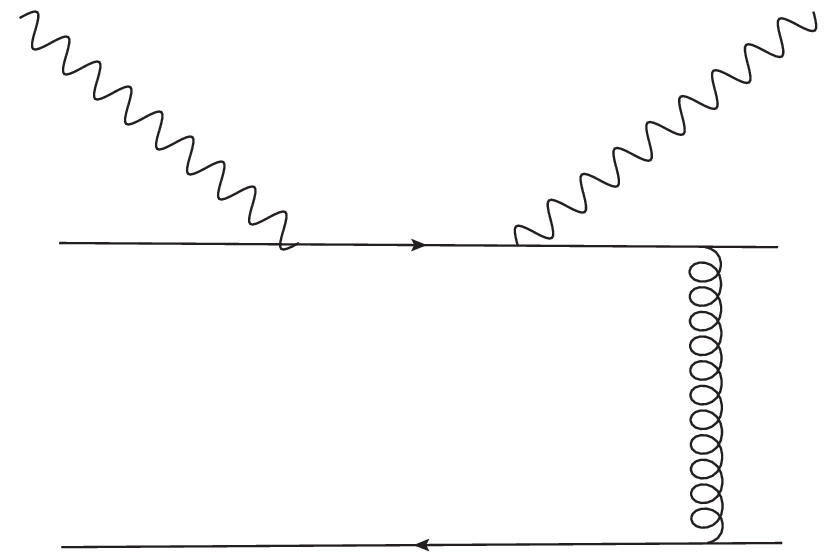}
&
\\
$A_{1}$ & 
$B_{1}$ & 
$B_{3}$ &
\\ 
\includegraphics[width=0.15\textwidth]{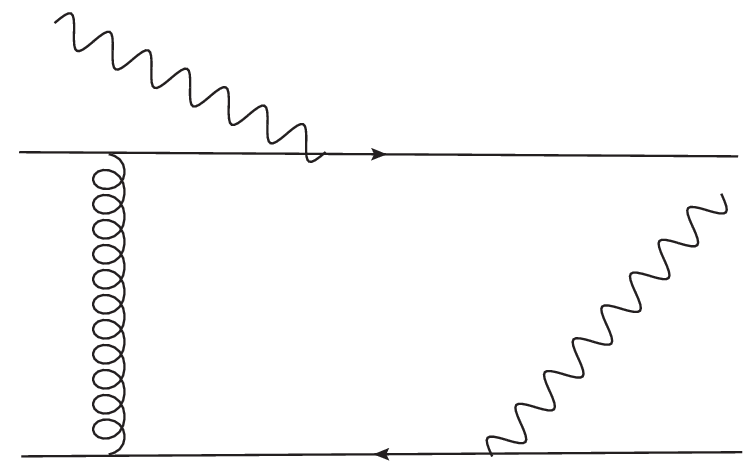}
&
\includegraphics[width=0.15\textwidth]{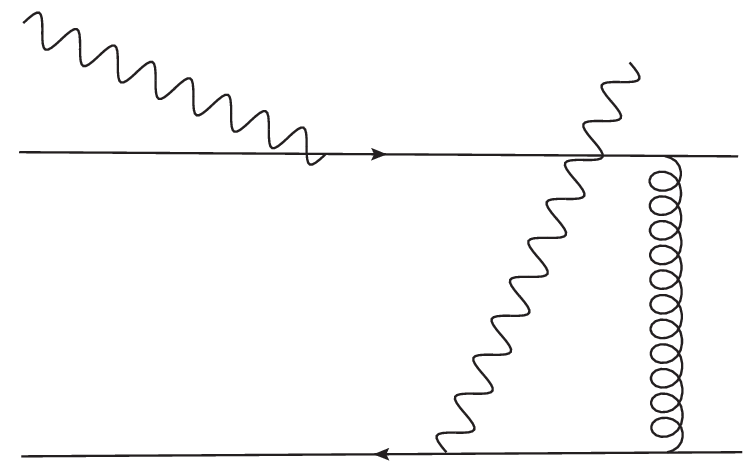}
&
\includegraphics[width=0.15\textwidth]{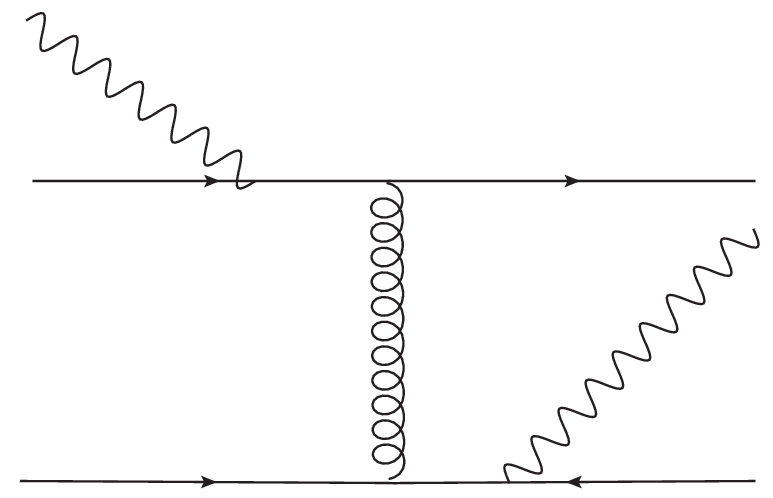}
&
\includegraphics[width=0.15\textwidth]{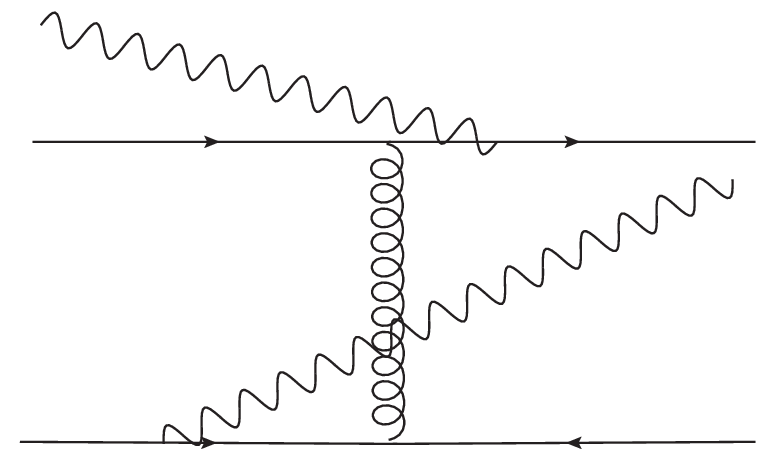}
\\
$C_{1}$ & 
$C_{2}$ & 
$D_{1}$ &
$D_{4}$ 
\end{tabular}
\caption{Representative Feynman diagrams contributing 
to 
$\gamma 
( q_1 \bar{q}_2 )
\to \gamma ( q_1 \bar{q}_2 )
$ \req{eq:sub-qqqq}.}
\label{fig:diagQ}
\end{figure}

\begin{figure}[t]
\centering
\begin{tabular}{ccc}
\includegraphics[width=0.2\textwidth]{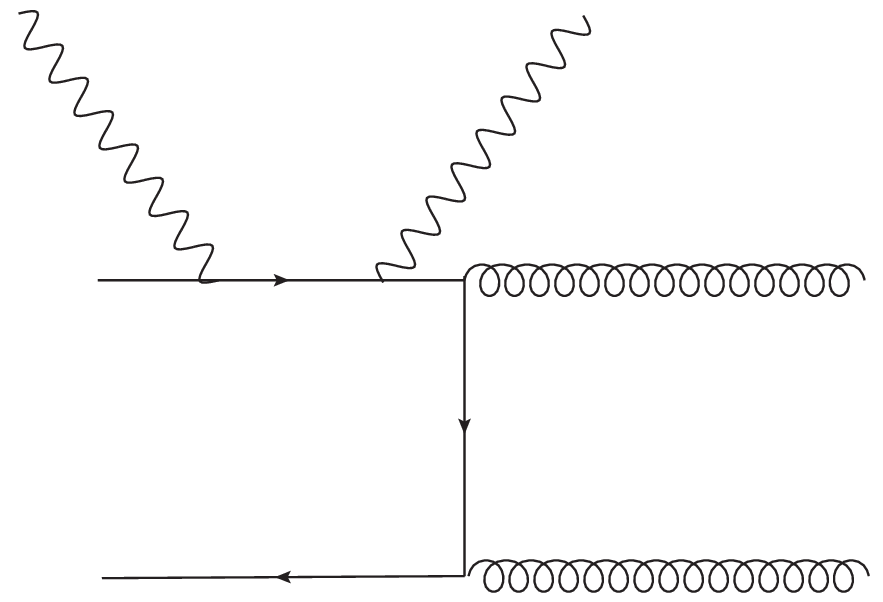}
&
\includegraphics[width=0.2\textwidth]{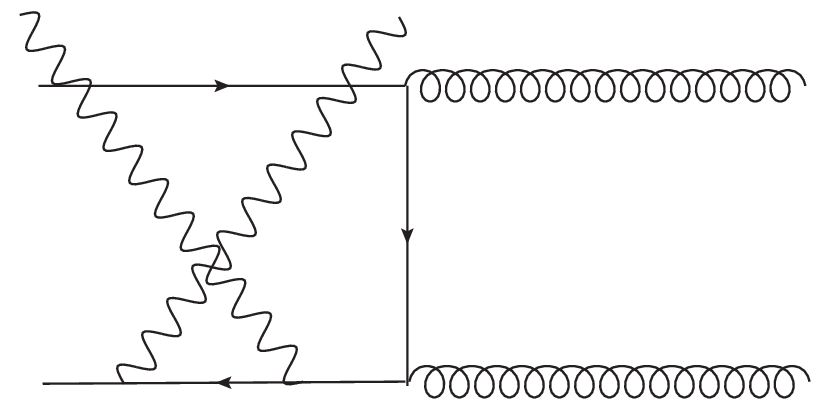}
&
\includegraphics[width=0.2\textwidth]{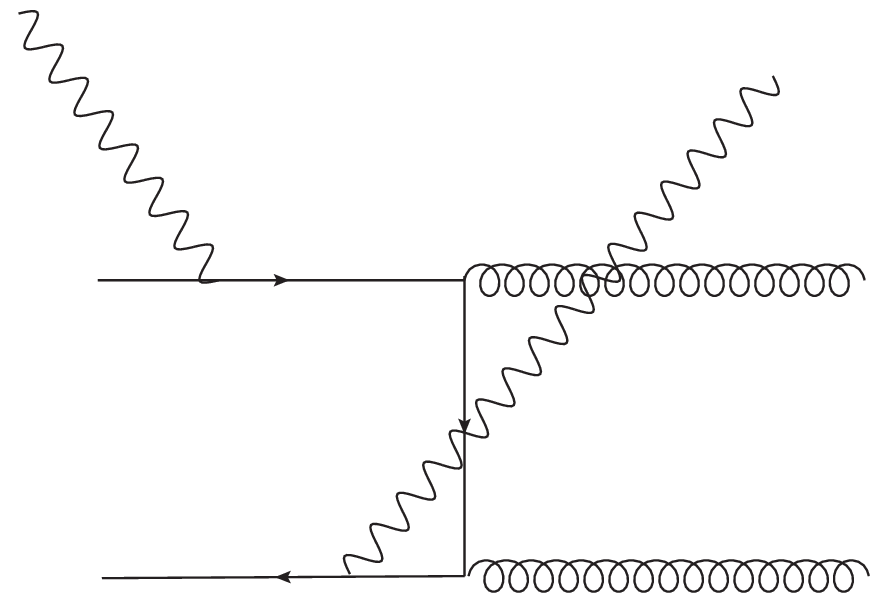}
\\
$A_{1}^g$ & 
$A_{6}^g$ & 
$B_{1}^g$ \\ 
\includegraphics[width=0.2\textwidth]{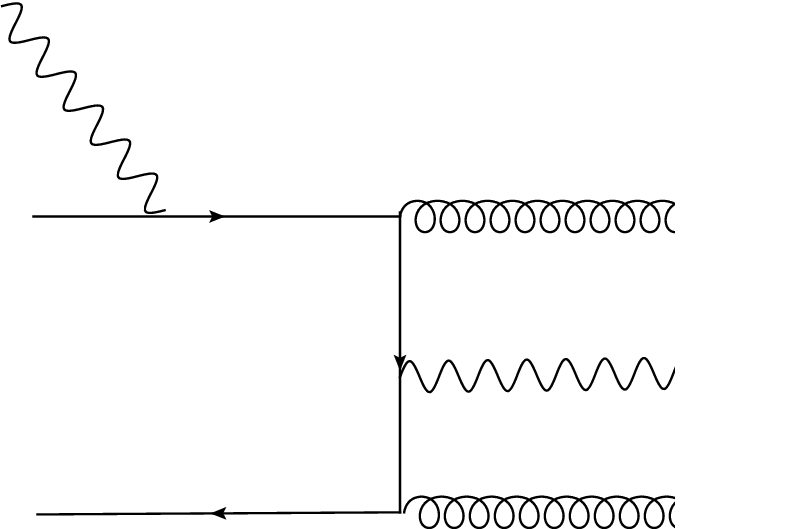}
&
\includegraphics[width=0.2\textwidth]{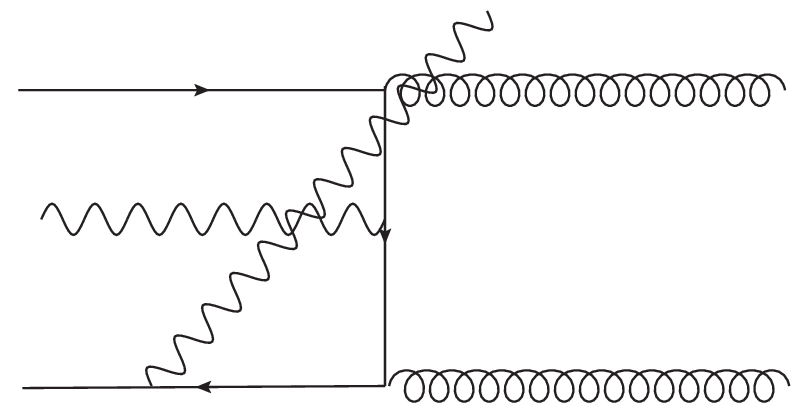}
&
\includegraphics[width=0.2\textwidth]{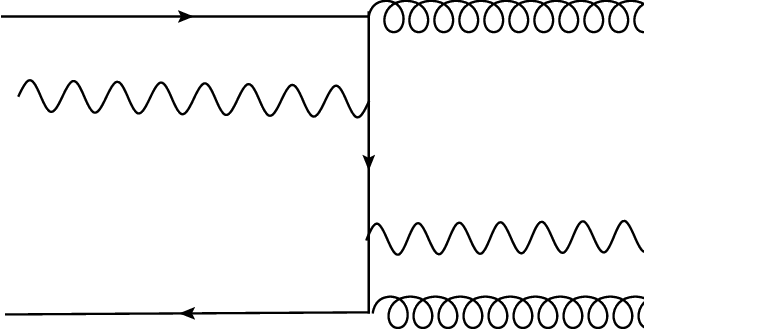}
\\
$C_{1}^g$ & 
$C_{5}^g$ & 
$D_{1}^g$  
\end{tabular}
\caption{Representative Feynman diagrams contributing 
to 
$\gamma 
( q \bar{q} )
\to \gamma ( g g )
$ \req{eq:sub-qqgg}.} 
\label{fig:diagG}
\end{figure}

One of the main results of this work
is the completion and systematization of the expressions for
the subprocess amplitudes
$\gamma ( q_1 \bar{q}_2 ) \to \gamma ( q_1 \bar{q}_2 ) $ \req{eq:sub-qqqq}
and
$\gamma ( q \bar{q} ) \to \gamma ( g g ) $ \req{eq:sub-qqgg}.
The latter are calculated here for the first time
in the context of photon-light-meson photoproduction.
Representative Feynman diagrams are shown in
Figs.~\ref{fig:diagQ} and \ref{fig:diagG}, 
while the contributions are evaluated using the projectors
given in App. \ref{app:DAev}.

The same diagrams also contribute to the $t$-channel subprocesses 
(as opposed to the above $s$-channel case),
$\gamma  \gamma \to ( q_1 \bar{q}_2 ) ( q_2 \bar{q}_1 ) $
and
$\gamma \gamma \to ( q \bar{q} ) ( g g ) $,
and thus to the hadronic processes
$\gamma \gamma \to \MM \MM'$ and
$\MM \NN \to \gamma \gamma \NN'$.
The former has been extensively studied in the past
\cite{Baier:1982vlv,Baier:1985pd,Baier:1985wv,
Brodsky:1981rp,Nizic:1987sw,Benayoun:1989ng}, even at NLO \cite{Duplancic:2006nv},
while the latter has more recently attracted interest
in the framework of GPD-based descriptions
\cite{Qiu:2022bpq,Qiu:2024mny}, being the crossed process
of $\gamma \MM$ photoproduction.
The general $2 \to 2$ subprocess amplitudes
$\gamma ( q_1 \bar{q}_2 ) \to \gamma ( q_1 \bar{q}_2 ) $ 
and
$\gamma ( q \bar{q} ) \to \gamma ( g g ) $ 
are identical across these applications
and can be, as in \req{eq:Mmunu},
expressed in terms of Mandelstam variables and
photon polarization vectors. 
The specific choice of photon gauge, reference frame,
and kinematical setup - in particular,
the range of momentum fractions (whether linked to DAs or GPDs) and
the definitions and signs of Mandelstam variables
(for instance, $\ssM \leftrightarrow \tsM$ between $s$- and $t$-channel subprocesses) -
differentiate these cases.
To illustrate this point,
although the same diagrams contribute,
Compton scattering on a meson $\MM$
($\gamma \MM \to \gamma \MM$)
contains cuts, i.e., imaginary parts, already at LO,
whereas the crossed meson-pair photoproduction
($\gamma \gamma \to \MM \MM$)
is real at LO.
In the literature, the fully general forms of these amplitudes
are not usually presented; however, 
as outlined in \cite{Duplancic:2018bum},
using the general Lorentz decomposition of $\Mamp_{\mu \nu}$ and
imposing gauge invariance constraints,
one can derive relations between the amplitudes in different gauges. 
We applied this method to compare our results
with those for $\gamma \gamma \to \MM \MM'$
from the literature-particularly the gluon contributions
from \cite{Baier:1985wv}-and found agreement up to
$i \epsilon$ terms, which depend on kinematics.

Taking into account the decomposition introduced
in Sec.~\ref{sec:ampl1},
we express the amplitudes corresponding to the subprocess
$\gamma ( q_1 \bar{q}_2 ) \to \gamma ( q_1 \bar{q}_2 )$
in the following form:
\begin{eqnarray}
\MampPP^{[\a \b]} &=&
\left( \ei \, \MAi +
\ej \, \MAj +
\eij \, \MAij \right) T_A 
+
\left( \ei \, \MBi +
\ej \, \MBj +
\eij \, \MBij \right) T_B 
\, ,
\label{eq:TPPij}
\end{eqnarray}
and 
\begin{eqnarray}
\MampSP^{[\a \b]}&=&
\left( \ei \, \MAiFive +
\ej \,\MAjFive  +
\eij \, \MAijFive \right) T_{A5} 
+
\left( \ei \, \MBiFive +
\ej \,\MBjFive  +
\eij \, \MBijFive \right) T_{B5} 
\, ,
\label{eq:TSPij}
\end{eqnarray}
where the definitions in \req{eq:TATB5} apply.
A total of twenty diagrams contribute,
with representative ones shown in Fig. \ref{fig:diagQ}:
the $A$-type (4) and $B$-type (8)
diagrams enter the $\ei$- and $\ej$-proportional terms,
while the $C$-type (4) and $D$-type (4)
diagrams contribute to the $\eij$-proportional terms.
These can be grouped into four gauge-invariant sets
related by symmetry relations:
\begin{equation}
\begin{alignedat}{2}
\Mamp_{X}^{(\a)}(y,z)
&= \Mamp_{X}^{(\b)}(\by,\bz)
\, ,
&\qquad
\Mamp_{X}^{(\a \b)}(y,z)
&= \Mamp_{X}^{(\a \b)}(\by,\bz)
\, ,
\\[0.3cm]
\Mamp_{X5}^{(\a)}(y,z)
&= -\,\Mamp_{X5}^{(\b)}(\by,\bz)
\, ,
&\qquad
\Mamp_{X5}^{(\a \b)}(y,z)
&= -\,\Mamp_{X5}^{(\a \b)}(\by,\bz)
\, ,
\end{alignedat}
\end{equation}
that is, they are symmetric under $(y,z) \to (\by, \bz)$ for $\PP \PP$ contributions
and antisymmetric for $\SS \PP$ ones.
The complete and systematized results,
organized using the compact $(y,z)$ notation
and applicable to the photoproduction of any
light pseudoscalar meson-including kaons described by non-symmetric DAs-
are provided for completeness in Appendix~\ref{app:THq}.
Our concise analytical forms are consistent with previous work%
\footnote{Note that in this study we adopt the convention for $T_{B5}$ which is opposite both \cite{Boussarie:2016qop} and \cite{Duplancic:2018bum}.}
 \cite{Duplancic:2018bum}.

In this work we focus on pseudoscalar mesons whose quark distribution
amplitudes are symmetric under $(z \to \bz)$ \req{eq:DAsym}.
Exploiting this property, and introducing the symmetric
($\tH^{q(+)}$, $\tE^{q(+)}$, $H^{q(-)}$, $E^{q(-)}$)
and antisymmetric
($\tH^{q(-)}$, $\tE^{q(-)}$, $H^{q(+)}$, $E^{q(+)}$)
GPD combinations \req{eq:GPDsym},
we can write the results in a compact analytical form.
We recall that, according to the definition of the variable $y$ \req{eq:yxix},
the $(x \to -x)$ symmetry corresponds to $(y \to \by)$.
To this end, we reorganize
the convolution of the $\Mamp^{[ij]}$ amplitudes with the GPDs
(the convolution with the symmetric DA being understood) as
\begin{equation}
\left\{
\begin{array}{c}
\MampPP^{[\a \b]} \otimesy \tGPD^q 
\\[0.3cm]
\MampSP^{[\a \b]} \otimesy \GPD^q 
\end{array}
\right\}
:
(e_\a -e_\b)^2 
\left\{
\begin{array}{c}
\Mamp_{X}^{\qsum} \otimesy \frac{\tGPD^{q(+)}}{2}
\\[0.3cm]
\Mamp_{X5}^{\qsum} \otimesy \frac{\GPD^{q(+)}}{2} 
\end{array}
\right\}
+(e_\a^2 -e_\b^2) 
\left\{
\begin{array}{c}
\Mamp_{X}^{\qdiff} \otimesy \frac{\tGPD^{q(-)}}{2}
\\[0.3cm]
\Mamp_{X5}^{\qdiff} \otimesy \frac{\GPD^{q(-)}}{2} 
\end{array} \right\}
+e_\a e_\b 
\left\{
\begin{array}{c}
\Mamp_{X}^{\qprod} \otimesy \frac{\tGPD^{q(+)}}{2}
\\[0.3cm]
\Mamp_{X5}^{\qprod} \otimesy \frac{\GPD^{q(+)}}{2} 
\end{array}
\right\}
\label{eq:THGPDsym}
\end{equation}
multiplied by corresponding tensor structures $T_{X(5)}$,
where
$X\in\{A, B\}$,
while
$\Mamp_{X(5)}^{\qsum}$,
$\Mamp_{X(5)}^{\qdiff}$,
and
$\Mamp_{X(5)}^{\qprod}$
are defined through
\begin{eqnarray}
\frac{1}{2}
\left( \Mamp_{X(5)}^{(\a)} 
+ \Mamp_{X(5)}^{(\b)} \right)
\to \Mamp_{X(5)}^{\qsum} 
\, ,
\qquad
\frac{1}{2}
\left( \Mamp_{X(5)}^{(\a)} 
- \Mamp_{X(5)}^{(\b)} \right)
\to \Mamp_{X(5)}^{\qdiff} 
\, ,
\qquad
\Mamp_{X(5)}^{(\a)} 
+\Mamp_{X(5)}^{(\b)} 
+\Mamp_{X(5)}^{(\a \b)} 
\to
\Mamp_{X(5)}^{\qprod} 
\, .
\nn \\
\end{eqnarray}
By further exploiting the symmetry properties of the DA and GPDs (hence the notation $\to$),
we simplify
$\Mamp_{X(5)}^{\qsum}$,
$\Mamp_{X(5)}^{\qdiff}$,
and
$\Mamp_{X(5)}^{\qprod}$.
The resulting compact analytical expressions are summarized
in Table~\ref{tab:anres}.

The gluonic subprocess
$\gamma ( q \bar{q} ) \to \gamma ( g g )$
contributes to photon-meson photoproduction via the flavour-singlet channel.
Its amplitude is obtained from 24 Feynman diagrams, six of which are displayed in Fig.~\ref{fig:diagG},
and the remaining ones follow from
$(z \to \bz)$, $(y \to \by)$, and $(y,z) \to (\by,\bz)$
exchanges.
The total amplitude takes the form 
\begin{equation}
\MampPP^{[g g]} =
\Big\{
 e_q^2 \, 
\left( \Mamp_A^{[gg]} \, T_A 
+
 \Mamp_B^{[gg]} \, T_B \right)
\Big\}
+
\big\{
(y \to \by)
\big\}
-
\big\{
(z \to \bz)
\big\}
-
\big\{
(y,z) \to (\by,\bz)
\big\}
\, ,
\end{equation}
and
\begin{equation}
\MampSP^{[g g]} =
\Big\{
 e_q^2 \, 
\left( \Mamp_{A5}^{[gg]} \, T_{A5} 
+
 \Mamp_{B5}^{[gg]} \, T_{B5} \right)
\Big\}
-
\big\{
(y \to \by)
\big\}
-
\big\{
(z \to \bz)
\big\}
+
\big\{
(y,z) \to (\by,\bz)
\big\}
\, ,
\end{equation}
which leads to
\begin{equation}
\MampPP^{[g g]}(y,z) =
 -\MampPP^{[g g]}(\by,\bz) 
\, ,
\qquad
\MampSP^{[g g]}(y,z) =
\MampSP^{[g g]}(\by,\bz) 
\, .
\end{equation}
These symmetry relations,
together with the antisymmetric nature of the gluon distribution amplitude
\req{eq:DAsym},
select in the convolution only those GPDs
with definite symmetry properties \req{eq:GPDsym}:
\begin{equation}
\left\{
\begin{array}{c}
\MampPP^{[g g]} \otimesy \tGPD^q 
\\[0.3cm]
\MampSP^{[g g]} \otimesy \GPD^q 
\end{array}
\right\} \otimesz \phi_{\eta_{1g}}
=
e_q^2 
\left\{
\begin{array}{c}
\Mamp_{A}^{\ggres} \otimesy \frac{\tGPD^{q(+)}}{2}
\, T_A
\\[0.3cm]
\Mamp_{A5}^{\ggres} \otimesy \frac{\GPD^{q(+)}}{2} 
\, T_{A5}
\end{array}
\right\}
\otimesz \phi_{\eta_{1g}}
+
e_q^2 
\left\{
\begin{array}{c}
\Mamp_{B}^{\ggres} \otimesy \frac{\tGPD^{q(+)}}{2}
\, T_B
\\[0.3cm]
\Mamp_{B5}^{\ggres} \otimesy \frac{\GPD^{q(+)}}{2} 
\, T_{B5}
\end{array}
\right\}
\otimesz \phi_{\eta_{1g}}
\, .
\label{eq:THgGPDsym}
\end{equation}
Making further use of the DA and GPD symmetry relations,
we obtain a concise analytical forms for
$\Mamp_{X(5)}^{\ggres}$,
which are presented in Table~\ref{tab:anres}.

Relations analogous to
\req{eq:THGPDsym} and \req{eq:THgGPDsym} apply also to
$\tE^q$ and $E^q$.
While all terms from \req{eq:THGPDsym}
contribute to $\gamma \pi^\pm$ photoproduction,
only the
$\Mamp_{X(5)}^{\qprod}$ terms contribute to the photoproduction
of neutral mesons.
In the flavour-singlet channel, i.e., for $\gamma \eta_1$ photoproduction,
the $\Mamp_{X(5)}^{\ggres}$ terms contribute as well.

\vspace*{0.15cm}
\noindent\textit{Structure of the subprocess amplitudes} -
As seen from Table \ref{tab:anres},
the building blocks of the subprocess amplitudes
for
$\gamma ( q_1 \bar{q}_2 ) \to \gamma ( q_1 \bar{q}_2 )$ and
$\gamma ( q \bar{q} ) \to \gamma ( g g )$
reduce, after using the symmetry properties of the GPD and DA, to
\begin{equation}
\frac{1}{z} 
\, ,
\qquad  
\frac{1}{y\pm \Ieta}
\, ,
\qquad
\frac{f(z)}{\mixAzy + \Ieta}\,.
\label{eq:bb}
\end{equation}
The first two terms convey moment-type information
on DAs and GPDs, which also appear at LO
in the meson transition form factor and in DVCS,
as well as in the meson electromagnetic form factor and in DVMP.
Note that the NLO corrections to these processes
provide additional insight into the behavior of GPDs and DAs
beyond the moment-type approximation.
The $\pm \Ieta$ designations appearing in our results reflect the
presence of timelike and spacelike scales.
The third term in \req{eq:bb}, i.e., the "mixed" contributions
appearing in $\Mamp_{A(5)}^{(0)}$ and $\Mamp_{A(5)}^{(g)}$,
in principle already at LO, reveals further information on DAs
and GPDs, as it introduces an external scale into the propagators.
The same building blocks also appear
in the analytical expressions for amplitudes
evaluated in different photon gauges.

\vspace*{0.15cm}
\noindent\textit{Principal Value (PV) representation} -
Using Eqs.~\req{eq:1y-PV} and \req{eq:mix-PV},
the functions $\Mamp_{X(5)}$ from Table~\ref{tab:anres}
can be expressed in the principal-value and $\delta$-function
representations, as listed in Table~\ref{tab:anresPV}.
This form is convenient for numerical implementation
and subsequent integration,
as detailed in Appendix~\ref{app:PV}.
In this representation, one clearly separates
real and imaginary parts,
allowing for a more transparent analysis
of the relative contribution sizes.
The $\delta(y)$ terms vanish for
the $\gamma \MM' \to \gamma \MM$ case,
i.e.~for pion-pole contributions,
since $\tilde{E}_{\text{pole}}$ is expressed in terms 
of pion DA which vanishes at the end points.
When convoluted with other GPDs, these terms can be sizable
and correspond to the GPD values at $x=\pm \xi$,
thus carrying the same cross-over line information
as obtained in LO DVCS and DVMP.
The ''mixed'' terms are those that make a difference,
although their numerical impact is less pronounced. 


{\renewcommand{\arraystretch}{3}
\begin{table}
\caption{
Functions $\Mamp_{X(5)}$ ($X \in \{A,B\}$) contributing to the subprocess amplitudes
$\gamma ( q_1 \bar{q}_2 ) \to \gamma ( q_1 \bar{q}_2 )$
\req{eq:THGPDsym} and
$\gamma ( q \bar{q} ) \to \gamma ( g g )$
\req{eq:THgGPDsym}.
}
\label{tab:anres}
\centering
\resizebox{\textwidth}{!}{%
\begin{tabular}{cll}
\hline

&
\multicolumn{1}{c}{$A$}
&
\multicolumn{1}{c}{$B$}
\\
\hline
$\Mamp_X^{\qsum}$
&
$ \displaystyle
\frac{4}{\ssM \, \alpha \balpha\,  z}
\left(
\frac{\alpha (2 \alpha -1)}{y+\Ieta}
+\frac{1+(\alpha-3)(2 \alpha -1)}{y-\Ieta}
\right)
$
&
$ \displaystyle
\frac{-8}{\ssM^2 \, \alpha^2 \balpha \, z}
\left(
\frac{\alpha}{y+\Ieta}
+\frac{\alpha-2}{y-\Ieta}
\right)
$
\\
$\Mamp_X^{\qdiff}$
&
$ \displaystyle
\frac{4}{\ssM \, \balpha \, z}
\left(
\frac{2 \alpha -1}{y+\Ieta}
-\frac{2 \alpha -5}{y-\Ieta}
\right)
$
&
$ \displaystyle
\frac{-8}{\ssM^2 \, \alpha \balpha \, z}
\left(
\frac{1}{y+\Ieta}
-\frac{1}{y-\Ieta}
\right)
$
\\
$\Mamp_X^{\qprod}$
&
$ \displaystyle
\frac{8}{\ssM \, \alpha}
\left(
\frac{2 \alpha-1}{\balpha \,z \, (y+\Ieta)}
-\frac{2 \alpha -1 - \balpha^2}{\balpha \, z \, (y-\Ieta)}
- \frac{(\bz+\alpha z)^2 +\alpha}{z \bz (\mixAzy  +\Ieta)}
\right)
$
&
$ \displaystyle
\frac{-16}{\ssM^2 \, \alpha^2 \balpha \, z}
\left(
\frac{1}{y+\Ieta}
-\frac{1}{y-\Ieta}
\right)
$
\\
$\Mamp_X^{\ggres}$
&
$ \displaystyle
\frac{8}{\ssM \, \alpha \balpha}
\left(
\frac{2 \alpha-1}{z^2 \, (y+\Ieta)}
+\frac{1}{z^2 \, (y-\Ieta)}
- \frac{(z+\alpha \bz)(\bz+\alpha z)^2}{z^2 \bz^2 (\mixAzy  +\Ieta)}
\right)
$
&
$ \displaystyle
\frac{-16}{\ssM^2 \, \alpha^2 \balpha \, z^2}
\left(
\frac{1}{y+\Ieta}
-\frac{1}{y-\Ieta}
\right)
$
\\
\hline

&
\multicolumn{1}{c}{$A5$}
&
\multicolumn{1}{c}{$B5$}
\\
\hline
$\Mamp_{X5}^{\qsum}$
& 
$ \displaystyle
\frac{16 \xi i}{\ssM^3 \, \alpha^2 \, \balpha^2\, z}
\left(
\frac{\alpha}{y+\Ieta}
+\frac{3 \alpha -4}{y-\Ieta}
\right)
$
&
$ \displaystyle
\frac{-16 \xi i}{\ssM^3 \, \alpha \, \balpha^2\, z}
\left(
\frac{2 \alpha-1}{y+\Ieta}
+\frac{2 \alpha -3}{y-\Ieta}
\right)
$
\\
$\Mamp_{X5}^{\qdiff}$
& 
$ \displaystyle
\frac{16 \xi i}{\ssM^3 \, \alpha \, \balpha^2\, z}
\left(
\frac{1}{y+\Ieta}
-\frac{1}{y-\Ieta}
\right)
$
&
$ \displaystyle
\frac{-16 \xi i (2 \alpha -1)}{\ssM^3 \, \alpha \, \balpha^2\, z}
\left(
\frac{1}{y+\Ieta}
-\frac{1}{y-\Ieta}
\right)
$
\\[0.5cm]
$\Mamp_{X5}^{\qprod}$
& 
$ \displaystyle
\frac{32 \xi i}{\ssM^3 \, \alpha^2}
\left(
\frac{1}{\balpha^2 \, z \, (y+\Ieta)}
+\frac{\alpha^2-2}{\balpha^2 \, z \, (y-\Ieta)}
-
\frac{\alpha z^2-\bz^2}{ z \bz\, (\mixAzy +\Ieta)}
\right)
$
&
$ \displaystyle
-\Mamp_{A5}^{\qprod}(y,z)
+ \Mamp_{\Delta B5}^{\qprod}(y,z) 
$
\\
& 
\multicolumn{2}{r}{
$ \displaystyle
\Mamp_{\Delta B5}^{\qprod}=
\frac{64 \xi i}{\ssM^3 \, \alpha^2 \balpha\, z}
\left(
\frac{1}{y+\Ieta}
-\frac{1}{y-\Ieta}
\right)
$}
\\[0.5cm]
$\Mamp_{X5}^{\ggres}$
& 
$ \displaystyle
\frac{32 \xi i}{\ssM^3 \, \alpha^2 \balpha^2}
\left(
\frac{1}{z^2 (y+\Ieta)}
-\frac{1}{z^2 (y-\Ieta)}
- \frac{(\bz+\alpha z)(\alpha-\balpha^2 z \bz)}{z^2 \bz^2 (\mixAzy  +\Ieta)}
\right)
$
&
$ \displaystyle
-\Mamp_{A5}^{\ggres}(y,z)
+ \Mamp_{\Delta B5}^{\ggres}(y,z) 
$
\\
& 
\multicolumn{2}{r}{
$ \displaystyle
\Mamp_{\Delta B5}^{\ggres}=
\frac{64 \xi i}{\ssM^3 \, \alpha^2 \balpha \, z^2}
\left(
\frac{1}{y+\Ieta}
-\frac{1}{y-\Ieta}
\right)
$}
\\
\hline
\end{tabular}
}
\end{table}
}
\renewcommand{\arraystretch}{1}

{\renewcommand{\arraystretch}{3}
\begin{table}
\caption{Principal-value and $\delta$-function representations of the 
functions $\Mamp_{X(5)}$ ($X \in \{A,B\}$) listed in Table~\ref{tab:anres}, 
which contribute to the subprocess amplitudes
$\gamma ( q_1 \bar{q}_2 ) \to \gamma ( q_1 \bar{q}_2 )$
\req{eq:THGPDsym} and
$\gamma ( q \bar{q} ) \to \gamma ( g g )$
\req{eq:THgGPDsym}.}
\label{tab:anresPV}
\centering
\resizebox{\textwidth}{!}{%
\begin{tabular}{cll}
\hline

&
\multicolumn{1}{c}{$A$}
&
\multicolumn{1}{c}{$B$}
\\
\hline
$\Mamp_X^{\qsum}$
&
$ \displaystyle
\frac{8}{\ssM \, z}
\left(
\frac{2 \balpha}{\alpha} \, \PV \frac{1}{y}
- i \pi \, \frac{3 \alpha -2}{\alpha \balpha} \, \delta(y)
\right)
$
&
$ \displaystyle
\frac{16}{\ssM^2 \, z}
\left(
\frac{1}{\alpha^2} \, \PV \frac{1}{y}
+ i \pi \, \frac{1}{\alpha^2 \balpha} \, \delta(y)
\right)
$
\\
$\Mamp_X^{\qdiff}$
&
$ \displaystyle
\frac{8}{\ssM \, z}
\left(
\frac{2}{\balpha} \, \PV \frac{1}{y}
- i \pi \, \frac{2 \alpha -3}{\balpha} \, \delta(y)
\right)
$
&
$ \displaystyle
\frac{16 i \pi \, \delta(y)}{\ssM^2 \,\alpha \balpha \, z}
$
\\
$\Mamp_X^{\qprod}$
&
$ \displaystyle
\frac{8}{\ssM}
\left[
\frac{\balpha}{\alpha \, z} \, \PV \frac{1}{y}
+ i \pi \, \frac{\balpha^2-2(2 \alpha-1)}{\alpha \balpha \, z} \, \delta(y)
\right.
$
&
$ \displaystyle
\frac{32 i \pi \, \delta(y)}{\ssM^2 \,\alpha^2 \balpha \, z}
$
\\
&
$ \displaystyle
\left.
 + 
\frac{\alpha+(\bz+\alpha z)^2}{\alpha z \bz (y+\alpha \by)}
\left(
\PV\!\left(
\frac{1}{z-\frac{y}{y+\alpha \by}}
\right)
+ i \pi \, 
\delta\!\left(
z-
\frac{y}{y+\alpha \by}
\right)
\right)
\right]
$
&
\\
$\Mamp_X^{\ggres}$
&
$ \displaystyle
\frac{8}{\ssM}
\left[
\frac{2}{\balpha \, z^2} \, \PV \frac{1}{y}
+ i \pi \, \frac{2}{\alpha \, z^2} \, \delta(y)
\right.
$
&
$ \displaystyle
\frac{32 i \pi \, \delta(y)}{\ssM^2 \,\alpha^2 \balpha \, z^2}
$
\\
&
$ \displaystyle
\left.
 + 
\frac{(z+\alpha \bz)(\bz+\alpha z)^2}{\alpha \balpha \, z^2 \bz^2 (y+\alpha \by)}
\left(
\PV\!\left(
\frac{1}{z-\frac{y}{y+\alpha \by}}
\right)
+ i \pi \, 
\delta\!\left(
z-
\frac{y}{y+\alpha \by}
\right)
\right)
\right]
$
&
\\
\hline

&
\multicolumn{1}{c}{$A5$}
&
\multicolumn{1}{c}{$B5$}
\\
\hline
$\Mamp_{X5}^{\qsum}$
&
$ \displaystyle
\frac{-32 \xi i }{\ssM^3 \, z}
\left(
\frac{2}{\alpha^2 \balpha} \, \PV \frac{1}{y}
+ i \pi \, \frac{2-\alpha}{\alpha^2 \balpha^2} \, \delta(y)
\right)
$
&
$ \displaystyle
\frac{32 \xi i}{\ssM^3 \, z}
\left(
\frac{2}{\alpha \balpha} \, \PV \frac{1}{y}
+ i \pi \, \frac{1}{\alpha \balpha^2} \, \delta(y)
\right)
$
\\
$\Mamp_{X5}^{\qdiff}$
&
$ \displaystyle
\frac{-32 \xi i}{\ssM^3 \alpha \balpha^2 \, z}
i \pi \, \delta(y)
$
&
$ \displaystyle
\frac{32 \xi i (2 \alpha-1)}{\ssM^3 \alpha \balpha^2 \, z}
i \pi \, \delta(y)
$
\\
$\Mamp_{X5}^{\qprod}$
&
$ \displaystyle
\frac{-32 \xi i}{\ssM^3}
\left[
\frac{1+\alpha}{\alpha^2 \balpha \, z} \, \PV \frac{1}{y}
+ i \pi \, \frac{(3- \alpha^2)}{\alpha^2 \balpha^2 \, z} \, \delta(y)
\right.
$
&
$ \displaystyle
-\Mamp_{A5}^{\qprod}(y,z)
+\frac{-128 \xi i }{\ssM^3 \,\alpha^2 \balpha \, z}
i \pi \, \delta(y)
$
\\
&
$ \displaystyle
\left.
 + 
\frac{\bz^2-\alpha z^2}{\alpha^2 z \bz (y+\alpha \by)}
\left(
\PV\!\left(
\frac{1}{z-\frac{y}{y+\alpha \by}}
\right)
+ i \pi \, 
\delta\!\left(
z-
\frac{y}{y+\alpha \by}
\right)
\right)
\right]
$
&
\\
$\Mamp_{X5}^{\ggres}$
&
$ \displaystyle
\frac{-32 \xi i}{\ssM^3}
\left[
 i \pi \, \frac{2}{\alpha^2 \balpha^2 \, z^2} \, \delta(y)
\right.
$
&
$ \displaystyle
-\Mamp_{A5}^{\ggres}(y,z)
+\frac{-128 \xi i }{\ssM^3 \,\alpha^2 \balpha \, z^2}
i \pi \, \delta(y)
$
\\
&
$ \displaystyle
\left.
 + 
\frac{(\bz+\alpha z)(\balpha^2 z \bz-\alpha)}{\alpha^2 \balpha^2 z^2 \bz^2 (y+\alpha \by)}
\left(
\PV\!\left(
\frac{1}{z-\frac{y}{y+\alpha \by}}
\right)
+ i \pi \, 
\delta\!\left(
z-
\frac{y}{y+\alpha \by}
\right)
\right)
\right]
$
&
\\
\hline
\end{tabular}
}
\end{table}
}
\renewcommand{\arraystretch}{1}


\subsection{Compton-meson form factors in selected channels}

In the preceding sections we have introduced
all the necessary elements for determining
the Compton-meson form factors defined in
Sec.~\ref{sec:ampl1},
Eqs.~(\ref{eq:MMAMV}-\ref{eq:otimes}).
The quark contribution
to the Compton-meson form factor
for the process
$\gamma \NN \to \gamma \MM \NN'$
($M \in \{ \pi^\pm,\pi^0,\eta_8,\eta_{1} \}$)
is generally expressed as
\begin{eqnarray}
\lefteqn{
\tCFF_{\MM; \NN \NN'}(\xi,t,\ssM,\alpha; \muR, \muF)}
\nonumber 
\\ & = & 
N_\MM(\muR) \, 
\left(\sum_{\a, \b, k} \left( \mathcal{C}_\MM^{\a\b}
\, f^{\a k}_{\NN \NN'} \right) \,
\tGPD^{k}(x,\xi,t;\muF)
\stackrel{x}{\otimes}
\Mamp^{[\a \b]}_{\PP \PP}(x, \xi , z; \ssM, \alpha)
\stackrel{z}{\otimes}
\phi_\MM(z;\muF) 
\right)
\, ,
\nonumber \\
\end{eqnarray}
with the normalization factor
\begin{equation}
N_\MM(\muR) = -\pi^2 \alpha \, \alphaS(\muR) \frac{C_F f_{\MM}}{N_c}
\, ,
\end{equation}
where $\mathcal{C}_\MM^{\a\b}$ is defined in Eq.~\req{eq:flavour},
and the functions
$f^{\a \b}_{\NN \NN'}$ are process-dependent
and encode isospin symmetry:
\begin{equation}
\begin{alignedat}{2}
f^{u k}_{pn} & = \delta^{ku}-\delta^{kd} \, ,
&\qquad
f^{d k}_{np} & = \delta^{ku}-\delta^{kd} \, ,
\\
f^{i k}_{pp} & = \delta^{ik} \, ,
&\qquad
f^{i k}_{nn} & = 
\delta^{iu}\delta^{kd}
+ \delta^{id}\delta^{ku}
+ \delta^{is}\delta^{ks}
 \, .
\end{alignedat}
\end{equation}
The convolution
$\tH^k \otimes \Mamp^{[ij]}_{\PP \PP}$
is to be replaced by the symmetrized forms defined in
\req{eq:THGPDsym},
employing the building blocks summarized in Table~\ref{tab:anres}.
The gluon contribution
to the Compton-meson form factor
for the process
$\gamma \NN \to \gamma \eta_{1g} \NN$
reads
\begin{eqnarray}
\lefteqn{
\tCFF_{\eta_{1g}; \NN \NN}(\xi,t,\ssM,\alpha; \muR, \muF)}
\nonumber 
\\ & = & 
N_{\eta_1}(\muR) \, 
\left(\sum_{k}
\, f^{\a k}_{\NN \NN}  \,
\frac{\tGPD^{k (+)}(x,\xi,t;\muF)}{2}
\stackrel{x}{\otimes}
\, e_\a^2 \; 
\Mamp^{(g)}_{\PP \PP}(x, \xi , z; \ssM, \alpha)
\stackrel{z}{\otimes} \phi_{\eta_{1g}}(z;\muF) 
\right)
\, .
\end{eqnarray}
The components of
$\Mamp^{(g)}$ corresponding to the decomposition
\req{eq:THgGPDsym}
are given in Table~\ref{tab:anres}.
Analogous expressions hold for $\tCFFe$
as well as for the vector CMFFs, $\CFF$ and $\CFFe$,
evaluated using $\Mamp_{\SS \PP}$.

It is convenient to introduce combinations
of GPDs that naturally arise in the calculation of specific
photoproduction channels of interest, namely
$\gamma p \to \gamma \pi^+ n$,
$\gamma n \to \gamma \pi^- p$,
$\gamma p \to \gamma \pi^0 p$,
$\gamma p \to \gamma \eta p$,
and
$\gamma p \to \gamma \eta' p$:
\begin{equation} 
\begin{alignedat}{2}
\tGPD_{\pi^{\pm}} &=
\tGPD^{u}
- \tGPD^{d}
\, ,
&\qquad
\tGPD_{\pi^0} &=
\frac{1}{\sqrt{2}}
\left(
e_u^2 \, \tGPD^{u}
- e_d^2 \, \tGPD^{d}
\right)
\, ,
\\[0.3em]
\tGPD_{\eta_8} &=
\frac{1}{\sqrt{6}}
\left(
e_u^2 \, \tGPD^{u}
+ e_d^2 \, \tGPD^{d}
- 2 e_s^2 \, \tGPD^{s}
\right)
\, , 
&\qquad
\tGPD_{\eta_1} &=
\frac{1}{\sqrt{3}}
\left(  
e_u^2 \, \tGPD^{u}
+ e_d^2 \, \tGPD^{d}
+ e_s^2 \, \tGPD^{s}
\right)\, , 
\label{eq:GPDcomb}
\end{alignedat}
\end{equation}
and analogously for $\tE$, $H$, and $E$ GPDs.
The CMFFs for $\gamma \pi^+$ photoproduction then read
\begin{equation}
\begin{alignedat}{2}
\tCFF_{\pi^+}
 & =  
N_\pi \,
\tGPD_{\pi^\pm}
\stackrel{x}{\otimes}
\Mamp^{[u d]}_{\PP \PP}
\stackrel{z}{\otimes}
\phi_\pi
 \, ,
&\quad \quad
\CFF_{\pi^+}
 & =  
N_\pi \,
\GPD_{\pi^\pm}
\stackrel{x}{\otimes}
\Mamp^{[u d]}_{\SS \PP}
\stackrel{z}{\otimes}
\phi_\pi
 \, ,
\end{alignedat}
\end{equation}
while for $\gamma \pi^-$ photoproduction, $\Mamp^{[ud]}$ is replaced by $\Mamp^{[du]}$.
In both cases, the symmetrized expressions of \req{eq:THGPDsym} are employed.
For the neutral flavour-octet mesons $\pi^0$ and $\eta_8$,
the CMFFs take the form
\begin{equation}
\begin{alignedat}{2}
\tCFF_{\pi^0}
 & =  
N_\pi
\,
\frac{\tGPD_{\pi^0}^{(+)}}{2}
\stackrel{x}{\otimes}
\Mamp^{\qprod}_{\PP \PP}
\stackrel{z}{\otimes}
\phi_\pi
 \, ,
&\quad \quad
\CFF_{\pi^0}
 & =  
N_\pi
\,
\frac{\GPD_{\pi^0}^{(+)}}{2}
\stackrel{x}{\otimes}
\Mamp^{\qprod}_{\SS \PP}
\stackrel{z}{\otimes}
\phi_\pi
 \, ,
\end{alignedat}
\end{equation}
and
\begin{equation}
\begin{alignedat}{2}
\tCFF_{\eta_8}
 & =  
N_{\eta_8}
\,
\frac{\tGPD_{\eta_8}^{(+)}}{2}
\stackrel{x}{\otimes}
\Mamp^{\qprod}_{\PP \PP}
\stackrel{z}{\otimes}
\phi_{\eta_8}
\, ,
&\qquad
\CFF_{\eta_8}
 & =  
N_{\eta_8}
\,
\frac{\GPD_{\eta_8}^{(+)}}{2}
\stackrel{x}{\otimes}
\Mamp^{\qprod}_{\SS \PP}
\stackrel{z}{\otimes}
\phi_{\eta_8}
 \, .
\end{alignedat}
\end{equation}
The CMFFs for the flavour-singlet $\eta_1$ meson
consist of both quark and gluon contributions:
\begin{equation}
\begin{alignedat}{1}
\tCFF_{\eta_{1}}
 & =  
 \,
N_{\eta_1}
\,
\frac{\tGPD_{\eta_1}^{(+)}}{2}
\stackrel{x}{\otimes}
\left(
\Mamp^{\qprod}_{\PP \PP}
\quad
\Mamp^{\ggres}_{\PP \PP}
\right)
\stackrel{z}{\otimes}
\left(
\begin{array}{c}
\phi_{\eta_1}
\\[0.3cm]
\phi_{\eta_{1g}}
\end{array}
\right)
\, ,
\\
\CFF_{\eta_{1}}
 & =  
 \,
N_{\eta_1}
\,
\frac{\tGPD_{\eta_1}^{(+)}}{2}
\stackrel{x}{\otimes}
\left(
\Mamp^{\qprod}_{\SS \PP}
\quad
\Mamp^{\ggres}_{\SS \PP}
\right)
\stackrel{z}{\otimes}
\left(
\begin{array}{c}
\phi_{\eta_1}
\\
\phi_{\eta_{1g}}
\end{array}
\right)
\, ,
\end{alignedat}
\end{equation}
which mix under DA evolution.
Finally, the CMFFs of the physical states,
i.e., the $\eta$ and $\eta'$ mesons,
are obtained through phenomenological mixing
(\ref{eq:etamix}-\ref{eq:Metaetap}):
\begin{align}
\tCFF_{\eta}  &
= \cos{\theta_8}\, \tCFF_{\eta_8} - \sin{\theta_1}\, \tCFF_{\eta_1}
\, , \nonumber\\[0.2cm]
\tCFF_{\eta'} &
= \sin{\theta_8}\, \tCFF_{\eta_8} + \cos{\theta_1}\, \tCFF_{\eta_1} 
\, .
\end{align}
Analogous relations hold for the other CMFFs not shown explicitly.

\section{Observables
}
\label{sec:obs}

In the numerical analysis presented in this work, 
we examine the cross sections, 
focusing on their fully differential, 
single-differential, and integrated forms.

\subsection{Fully differential cross-section}
\label{sec:fdcs}

The fully differential cross section
for the
$\gamma \NN \to \gamma \MM \NN'$ process
is given by
\begin{equation}
 \frac{d \sigma}{d\ssM d(-t) \; d\alpha } 
=\frac{|\overline{ \mathcal{M} }|^2 }{32 S^2_{\gamma \NN}  (2\pi)^3}
\, , 
 \label{eq:diff-CS-2}
 \end{equation}
where
\begin{equation}
\label{eq:PPsq+SPsq}
 |\overline{ \mathcal{M} }|^2= 
 |\overline{ \mathcal{M}}^\MA |^2 + |\overline{ \mathcal{M}}^\MV |^2 
\end{equation}
denotes the squared amplitude $\mathcal{M}$ (\ref{eq:MMAMV}),
summed over nucleon helicities and photon polarizations,
and averaged over the polarizations of the incoming particles.
Due to the QED Ward identities,
\begin{equation}
\sum_{\lambda_q,\lambda_k} | \mathcal{M}^\MA(\lambda_q, \lambda_k)  |^2
=
\mathcal{M}^{\MA \: \mu \nu} 
\mathcal{M}^{\MA \, *}_{\mu \nu} \,,
\qquad
\sum_{\lambda_q,\lambda_k} | \mathcal{M}^\MV(\lambda_q, \lambda_k)  |^2
=
\mathcal{M}^{\MV \: \mu \nu} 
\mathcal{M}^{\MV \, *}_{\mu \nu} 
\, .
\label{eq:sumMM}
\end{equation}
Using the standard spinor-sum identities 
with bilinears (\ref{eq:spin_bil_aa}),
and taking into account (\ref{eq:Mmunu}-\ref{eq:TATB5}),
one obtains
\begin{align}
|\overline{\mathcal{M}}^{\MA}|^{2}
&=
2
\left[
(1-\xi^{2})
\left(
2\,|\widetilde{\mathcal{H}}_{A}|^{2}
- 2 \ssM \alpha \balpha\,\mathrm{Re}\!\left(\widetilde{\mathcal{H}}_{A}\widetilde{\mathcal{H}}_{B}^{\ast}\right)
+ \ssM^{2}\alpha^{2}\balpha^2\,|\widetilde{\mathcal{H}}_{B}|^{2}
\right)
\right.
\nn \\[0.4em]
&\quad
-2 \, \xi^{2}
\left(
2\,\mathrm{Re}\!\left(\widetilde{\mathcal{H}}_{A}\widetilde{\mathcal{E}}_{A}^{\ast}\right)
-  \ssM \alpha \balpha\,\mathrm{Re}\!\left(
   \widetilde{\mathcal{H}}_{A}\widetilde{\mathcal{E}}_{B}^{\ast}
   + \widetilde{\mathcal{E}}_{A}\widetilde{\mathcal{H}}_{B}^{\ast}\right)
+  \ssM^{2}\alpha^{2} \balpha^2\,\mathrm{Re}\!\left(\widetilde{\mathcal{H}}_{B}\widetilde{\mathcal{E}}_{B}^{\ast}\right)
\right)
\nn \\[0.4em]
&\quad
\left.
-\frac{\xi^2 \, t}{4 \Mp_{\NN}^{2}}
\left(
2\,|\widetilde{\mathcal{E}}_{A}|^{2}
- 2 \ssM \alpha \balpha \,\mathrm{Re}\!\left(\widetilde{\mathcal{E}}_{A}\widetilde{\mathcal{E}}_{B}^{\ast}\right)
+ \ssM^{2}\alpha^{2} \balpha^2\,|\widetilde{\mathcal{E}}_{B}|^{2}
\right)
\right]
\, ,
\label{eq:Mamp_sq_A}
\end{align}
and
\begin{align}
|\overline{\mathcal{M}}^{\MV}|^{2}
\: = &
\frac{\ssM^{4}\alpha^{2} \balpha^2}{8\xi^{2}}
\left[
(1-\xi^{2})
\left(
|\mathcal{H}_{A5}|^{2}
+ |\mathcal{H}_{B5}|^{2}
\right)
-2\xi^{2}
\left(
\mathrm{Re}\!\left(
\mathcal{H}_{A5}\mathcal{E}_{A5}^{\ast}
+\mathcal{H}_{B5}\mathcal{E}_{B5}^{\ast}\right)
\right)
\right.
\nn \\[0.4em]
&
\left.
-\left(\xi^{2}+\frac{t}{4 \Mp_{\NN}^{2}}\right)
\left(
|\mathcal{E}_{A5}|^{2}
+ |\mathcal{E}_{B5}|^{2}
\right)
\right]
\, .
\label{eq:Mamp_sq_V}
\end{align}
Note that there is no interference between 
parity-odd ($\tCFF$, $\tCFFe$)
and  parity-even ($\CFF$, $\CFFe$)
CMFFs.

\subsection{Single differential and integrated cross-sections}

To determine the single-differential and integrated cross sections,
special care must be taken regarding the cuts
in the phase-space plane $(-t, \alpha)$, or equivalently $(-t, -\usM)$.
This procedure was discussed in detail in
\cite{Boussarie:2016qop,Duplancic:2018bum};
here we summarize the main points.
The common lower limit for $t$ is obtained from \req{eq:genscalars}:
\begin{equation}
(-t)_\smin
 \equiv
(-t_0) 
= 
\frac{4\xi^2 \Mp_{\NN}^2}{1-\xi^2}
\, .
\label{eq:t0}
\end{equation}
According to the factorization conditions, $(-t)$ must remain small,
while, apart from $\ssM$, both $(-\usM)$ and $(-\tsM)$ should be large,
i.e., $-\usM$, $-\tsM \gg \Lambda_{\text{QCD}}$. 
In the language of \cite{Qiu:2022bpq,Qiu:2022pla}, 
this corresponds to large $|\kappa_\perp|$. 
As was demonstrated in \cite{Duplancic:2022ffo}, 
the imposed cuts on the Mandelstam invariants ensure that this is the case.
We therefore adopt the choice
\begin{equation}
\begin{alignedat}{2}
(-t)_\smax&=0.5 \: \mbox{GeV}^2
\, ,
&\qquad 
(-\usM)_\smin
=(-\tsM)_\smin=1 \: \mbox{GeV}^2
\, .
\end{alignedat}
\label{eq:utt-choice}
\end{equation}
These values guarantee $M_{\gamma \MM}^2>1.5$~GeV$^2$
and ensure that the kinematics stay well outside the resonance
region in the collinear limit \cite{Duplancic:2022ffo}.
From \req{eq:stutm} one obtains $(-\usM)$ as a function of $(-t)$:
\begin{equation} 
(-\usM)[-t] =
(-t)+\ssM-(-\tsM)_\smin -m_\MM^2 \, .
\label{eq:ut-fun}
\end{equation}
The corresponding maximum value is
\begin{equation}
(-\usM)_\smax =
\ssM+(-t)_\smax-(-\tsM)_\smin -m_\MM^2 
\, .
\label{eq:umax}
\end{equation}
Consequently, the allowed kinematical domain for $(-\usM)$ is
$(-\usM)_{\smin}<(-\usM)<(-\usM)_{\smax}$,
while the corresponding $\alpha$ limits are derived from \req{eq:alpha}.

The single-differential cross section can be written in the form
\begin{eqnarray}
\frac{d \sigma}{d \xi}
&=& \frac{2(S_{\gamma N}-\Mp_\NN^2)}{(1+\xi)^2}
\left(
\Theta(\ssM_\sMin \leq s' \leq \ssM_\sTrans)
\int_{(-t)[\alpha_\smin]}^{(-t)_\smax}
d(-t)
\int_{\alpha_\smin}^{\alpha[-t]}
d\alpha
\right.
\nonumber \\[0.4cm]
&&
\left.
+
\Theta(\ssM_\sTrans \leq s' \leq \ssM_\sMax)
\int_{(-t)_\smin}^{(-t)_\smax}
d(-t)
\int_{\alpha_\smin}^{\alpha[-t]}
d\alpha
\right)
 \frac{d \sigma}{d\ssM\, d(-t)\, d \alpha } 
\, .
\label{eq:phspaceint}
\end{eqnarray}
Here we introduce $\alpha[-t]$, which follows from \req{eq:ut-fun},
and its inverse, $(-t)[\alpha]$.
This representation is more compact
than the corresponding expression in
Refs.~\cite{Boussarie:2016qop,Duplancic:2018bum},
as in our case both the $\alpha$ and $t$ integrations
are performed numerically.
The integration order was selected
to yield a concise and computationally convenient form
for numerical evaluation.
From \req{eq:stutm} one finds
\begin{equation}
\ssM_\sMin = (-\usM)_\smin + (-\tsM)_\smin + m_\MM^2 - (-t)_\smax
\, ,
\end{equation}
which, combined with \req{eq:xi},
defines the lower kinematic limit
$\xi_\sMin(S_{\gamma \NN}, m_\MM^2)$.
The transition point $\ssM_\sTrans$ follows from the condition
$(-\usM)_\smin = (-\usM)[(-t)_\smin]$,
resulting in a lengthy analytical expression
that depends on $S_{\gamma \NN}$, $m_{\NN}$, $m_{\MM}$,
and on the parameter choices given in \req{eq:utt-choice}.
The upper limit, $\ssM_\sMax$,
is obtained for $(-t)_\smin = (-t)_\smax$:
\begin{equation}
\ssM_\sMax =
\frac{S_{\gamma \NN}-m_\NN^2}{2 m_\NN^2}
\left(
-(-t)_\smax +
\sqrt{(-t)_\smax^2 + 4 m_\NN^2 (-t)_\smax}
\right)
\, ,
\end{equation}
which corresponds to
\begin{equation}
\xi_\sMax = \sqrt{\frac{(-t)_\smax}{4 m_\NN^2 + (-t)_\smax}}
\, ,
\end{equation}
and for the parameter choice of \req{eq:utt-choice}
gives $\xi_\sMax = 0.3526$.
The lower limit $(S_{\gamma \NN})_\smin$
is fixed by the condition $\ssM_\sMin = \ssM_\sMax$,
and depends on the nucleon and meson masses,
and on the same 
parameter set.

Integrating \req{eq:phspaceint} over $\xi$
yields the total cross section as a function of $S_{\gamma \NN}$:
\begin{equation}
\sigma = \int_{\xi_\sMin}^{\xi_\sMax}
d\xi\,
\frac{d\sigma}{d\xi}
\, .
\label{eq:totalsigma}
\end{equation}

\section{Analysis of numerical results}
\label{sec:numres}

This section presents 
the numerical analysis
of photon-meson pair photoproduction,
$\gamma \NN \to \gamma \MM \NN'$,
for selected pseudoscalar mesons
$\MM \in \{\pi^+,\, \pi^-,\, \pi^0,\, \eta,\, \eta'\}$.
The proton is taken as the target nucleon,
except for $\pi^-$ photoproduction, where the target nucleon is a neutron.
The perturbative framework is summarized
in Sec.~\ref{sec:theory},
where only the quark GPD contributions are considered.
In particular, we make use of our analytical results
for the subprocess amplitudes,
Table~\ref{tab:anresPV},
and the integration procedure
described in App.~\ref{app:PV}.
The nonperturbative inputs,
namely the GPDs and DAs,
are discussed in Sec.~\ref{sec:GPDDAs}.
Our numerical analysis focuses on the moderate-$\xi$ region
($\xi > 0.1$).
We use the GK model
\cite{Goloskokov:2007nt,Goloskokov:2009ia,Kroll:2012sm},
and the valence quark approximation%
\footnote{
In this model, only the valence quarks contribute to charged-pion
photoproduction, and in the case of neutral mesons we consistently
omit both sea-quark and gluon contributions due to the relative
importance of valence-quark contributions in this $\xi$-region, as well
as the need for a modified framework when gluons are included.
}.
In addition, the $\pi$-pole contribution
governed by $\tE_{\rm pole}$~\req{eq:pipole}
is included, 
as it is expected to play an important role
in this kinematic domain.
For pions, several DA shapes are tested,
defined either via the truncated Gegenbauer expansion
\req{eq:phiGegen-Q}
or via the beta-function ansatz:
the asymptotic form $\phi_{\pi,{\rm as}}$,
$\phi_{\pi,[0.116]}$~\req{eq:a2-pion},
and the broader parameterizations
$\phi^{(0.5)}_\pi$, $\phi^{(0.4)}_\pi$, and $\phi^{(0.3)}_\pi$
\req{eq:phi_expal}.
The $\eta$ and $\eta'$ mesons are described
in the octet-singlet basis using
$\phi_{\eta_8}$, $\phi_{\eta_1=\eta_{1q}}$,
and the gluon DA $\phi_{\eta_{1g}}$~\req{eq:phiGegen-G}.
Their Gegenbauer coefficients follow
\req{eq:a2-eta},
and the decay-constant mixing \req{eq:etamix} is implemented.
DA evolution is performed according to
Eqs.~(\ref{eq:daGegen-Q}-\ref{eq:andim}),
and the strong coupling constant is evaluated at LO:
\begin{equation}
\alphaS(\mu)
= \frac{4\pi}{\beta_0 \log(\mu^2 / \Lambda_{\rm QCD}^2)},
\qquad
\beta_0 = 11 - \frac{2}{3}\, \tilde{n}_f.
\end{equation}
We take $\tilde{n}_f = 4$
and $\Lambda_{\rm QCD} = 0.22~\text{GeV}$,%
\footnote{Here, $\tilde{n}_f$ differs from
$n_f = 3$, which counts the fixed quark flavours
entering the $\eta$-meson system.}
corresponding to four active flavours at the relevant scales.
The observables analyzed in this section -  
defined previously in Sec.~\ref{sec:obs} - are  
the fully differential cross section \req{eq:diff-CS-2},  
the single differential cross section \req{eq:phspaceint},  
and the total cross section \req{eq:totalsigma}.

There are several key aspects of the analysis that we aim to address
in this section:
\begin{itemize}
\item We assess the relative importance of
$\tGPD$, $\tGPDe$, $\GPD$, and $\GPDe$ GPDs,
i.e., of
$\tCFF$, $\tCFFe$, $\CFF$, and $\CFFe$ Compton-meson form factors,
and in particular the role of $\tGPDe_{\rm pole}$,
i.e., the $\pi$-pole contribution $\tCFFe_{\rm pole}$.
\item We investigate the dependence of our predictions
on the meson structure.
For pions, several DA shapes are tested.
For $\eta$ and $\eta'$ mesons, we put special emphasis
on highlighting the role of two-gluon contributions.
We also examine the sensitivity on the choice of
factorization scale $\muF$.
\item 
We study the dependence of our LO predictions
on the choice of the renormalization scale $\muR$.
\end{itemize}

To illustrate the dependence on the kinematical conditions
relevant for JLab kinematics,
we present a comparison of the fully differential cross-section
evaluated at $t = t_0$~\req{eq:t0}
for selected combinations of
$(\ssM, S_{\gamma \NN})$
and the corresponding $\xi$ values determined from \req{eq:xi}: 
\begin{equation}
\begin{array}{cccccc}
\caseI & (\ssM, S_{\gamma \NN})=(4, 20)\,\text{GeV}^2 & \xi = 0.117 \, ,& 
\quad \caseII & (\ssM, S_{\gamma \NN})=(4, 9)\,\text{GeV}^2 & \xi = 0.327\, ,  \\
\caseIII & (\ssM, S_{\gamma \NN})=(4, 12)\,\text{GeV}^2 & \xi = 0.219\, , & 
\quad \caseIV & (\ssM, S_{\gamma \NN})=(7, 20)\,\text{GeV}^2 & \xi = 0.224\, .  
\end{array}
\label{eq:fourkin}
\end{equation}
Similarly, for the single differential cross section,
we compare predictions at
$S_{\gamma \NN} = 6$, $10$, $20$, and $30~\text{GeV}^2$.
Unless stated otherwise, as a default choice we use
\begin{equation}
\muR^2 = \muF^2 = \ssM \, , 
\qquad \text{and} \qquad 
\phi_\pi = \phi_{\pi,[0.116]} 
\label{eq:default}
\, ,
\end{equation}
the latter corresponding to Eqs. \req{eq:phiGegen-Q} and \req{eq:a2-pion}.
Following common practice in large-angle analyses,
and taking into account the size of the momentum transfer
between the photon and meson \req{eq:pT},
we also consider the alternative scale
\begin{equation}
\mu^2 = \alpha \, \balpha \ssM 
\, ,
\label{eq:muRFchoice}
\end{equation}
for $\muR^2$ and $\muF^2$, taken either separately and combined.

For clarity and compactness of presentation, App.~\ref{app:nres} collects all figures:
Figs.~\ref{fig:fds-pietas-VA}-\ref{fig:fds-etasmu-QG} fully differential,
Figs. \ref{fig:sds-Pip-AV}-\ref{fig:sds-ratio-pole} single differential,
and Figs. \ref{fig:sig-PiEtas-mu}-\ref{fig:sig-EtasMu-QG} integrated cross sections and selected ratios.

%
%

\subsection{GPD contributions and the role of the $\pi$ pole.} 
In this work, we include not only the contributions
of $\tH$ and $H$ GPDs, as in \cite{Duplancic:2018bum,Duplancic:2022ffo},
but also those of the remaining twist-2 chiral-even GPDs, namely $\tE$ and $E$.
Since the value of $\xi$ relevant for our analysis is not small,
there is no a priori justification for neglecting these
terms in (\ref{eq:Mamp_sq_A}-\ref{eq:Mamp_sq_V}).
Moreover, the $t$-dependent factors in front of $\tCFFe$ and $\CFFe$
in Eqs. (\ref{eq:Mamp_sq_A}-\ref{eq:Mamp_sq_V})
enhance the impact of $\tE$ and $E$ in both single-differential
and integrated cross sections, particularly for GPD $E$.
In the fully differential cross section,
the role of $\tE$ is more pronounced than that of $E$.
Still, $\CFF$ and $\tCFF$ remain the dominant contributions,
with the notable exception of $\tCFFe_{\text{pole}}$,
which is included in our analysis
through $\tCFFe \to \tCFFe +\tCFFe_{\text{pole}}$.
Note that the cross section defined in Sec.~\ref{sec:fdcs}
contains interference terms between
$\tCFF$ and $\tCFFe$, as well as between $\CFF$ and $\CFFe$.
However, it is obtained as the sum
of the  partial cross sections,
originating from the axial GPDs $\tH$ and 
$\tE+\tE_{\text{pole}}$,
(i.e., $|\mathcal{M}_{\PP \PP}|^2$, \req{eq:Mamp_sq_A})
and from the vector GPDs $H$ and $E$
(i.e., $|\mathcal{M}_{\SS \PP}|^2$, \req{eq:Mamp_sq_V}).

\vspace*{0.5cm}
\noindent
\textit{Pion-pole in fully differential cross-sections} -
Figure~\ref{fig:fds-pietas-VA} presents
the fully differential cross section as a function of $\alpha$
for the pion, $\eta$, and $\eta'$ mesons,
and with the individual GPD contributions
clearly identified.
The results are evaluated at $t = t_0$~\req{eq:t0}
and $\xi = 0.117$.
The parameter $\alpha$ is related to the photon scattering angle 
through $\alpha=\cos^2(\theta/2)$ \req{eq:theta-alpha}.
Values $\alpha < 0.5$ and $\alpha > 0.5$
correspond to backward and forward kinematics, respectively.
Thin lines represent the axial GPD contributions
and the total results calculated without including the pion-pole term.
In this case, for the $\gamma \pi^+$ channel,
the vector GPDs dominate at backward angles,
whereas the axial ones dominate at forward angles.
For the other channels, the vector contributions,
mainly driven by $H$,
dominate throughout the entire $\alpha$ range.
Thick solid lines represent the full predictions
including $\tE_{\text{pole}}$.
As seen in Fig.~\ref{fig:fds-pietas-VA},
the impact of $\tE_{\text{pole}}$ is significant,
especially for the $\pi^+$ and $\pi^-$ channels.
Unlike DVMP, and analogously to DVCS,
the neutral mesons $\pi^0$, $\eta$, and $\eta'$
also receive a contribution from the pion pole.
The neutral channels correspond to
the $\gamma \pi^0 \to \gamma M^0$ process,
while the charged ones refer to
$\gamma \pi^\pm \to \gamma \pi^\pm$.
In both cases, $\pi^0$ and $\pi^\pm$
are described by $\tE_{\text{pole}} \sim \phi_{\pi}$~\req{eq:pipole},
which contributes only within the ERBL region.

As already noted in \cite{Duplancic:2018bum},
electromagnetic interactions do not preserve
isospin symmetry; consequently,
$\gamma \pi^+$ and $\gamma \pi^-$ photoproduction differ.
At the level of calculation, this difference arises
from the lack of definite $(x \to -x)$ symmetry 
in the contributing GPDs,
i.e., from the $(e_i^2-e_j^2)$ proportional term
in \req{eq:THGPDsym}.
The pion-pole contributions, however,
correspond to charge-conjugated processes
$\gamma \pi^\pm \to \gamma \pi^\pm$
and involve the symmetric $\tE_{\text{pole}}$ GPD.
They are therefore equal in magnitude,
except for interference with the non-symmetric $\tH$.

The $\pi$-pole contributions, and consequently the total
differential cross section, are enhanced in the $\gamma \pi^\pm$ channel at backward angles
($\alpha < 0.5$ in Fig.~\ref{fig:fds-pietas-VA}, i.e., $\cos(\theta)<0$).
A similar backward enhancement is observed in wide-angle Compton scattering on the pion
\cite{Maina:1993uq,Tamazouzt:1988sn,Coriano:1994nh,Zeng:2002mw},
and is attributed to angular momentum conservation%
\footnote{For a spin-0 meson with helicity-conserving quarks,
photon-helicity-flip amplitudes require two units of
orbital angular momentum, which are kinematically enhanced near
$\theta \to \pi$.}~\cite{Maina:1993uq}.
The origin of backward dominance in our results
can be sketched at a qualitative level.
Re-examining the subprocess amplitude decomposition
\req{eq:THGPDsym} together with the results of
Table~\ref{tab:anresPV}, one observes%
\footnote{One should also keep in mind
Eqs.~(\ref{eq:Mamp_sq_A}-\ref{eq:Mamp_sq_V}) with the
$\alpha\balpha$ factors appearing in the relevant terms.}
that the terms proportional to $(e_i - e_j)^2$
and $(e_i^2 - e_j^2)$ are predominantly controlled by
$1/\alpha = 1/\cos^2(\theta/2)$ and
$1/\balpha = 1/\sin^2(\theta/2)$, respectively.
For the pion-pole contribution in the $\gamma \pi^\pm$ channel,
the $(e_i^2 - e_j^2)$ proportional term in \req{eq:THGPDsym} vanishes,
since $\tE_{\text{pole}}$, expressed through the pion DA,
is symmetric under $x \to -x$.
Note that the $\delta(y)$ terms from Table~\ref{tab:anresPV}
vanish for $\tE_{\text{pole}}$ because the pion DA is zero
at the end points.
As a result, the $(e_i - e_j)^2$ proportional term
in \req{eq:THGPDsym} enhances the backward-angle region
($\theta \to \pi$) in the pion-pole contribution to
$\gamma \pi^\pm$ production.
When the terms in Table~\ref{tab:anresPV} are convoluted
with other GPDs, the $(e_i^2 - e_j^2)$ proportional term
in \req{eq:THGPDsym} also contributes, enhancing the
forward-angle region ($\theta \to 0$) and making the
total result more uniform across the $\alpha$
(i.e., $\theta$) range.

Only the third term in \req{eq:THGPDsym},
proportional to $e_i e_j$, contributes
to the photoproduction of neutral mesons.
Consequently, the pion-pole contribution in neutral-meson production
is less pronounced than for charged pions.
This observation agrees qualitatively with theoretical and experimental
results for the crossed process $\gamma \gamma \to \pi \MM$,
with the $\sigma(\pi^0 \pi^0)/\sigma(\pi^+ \pi^-)$
cross-section ratio being $\approx 0.1$ theoretically
and $\approx 0.3$-$0.5$ experimentally \cite{Belle:2009ylx,Belle:2009xpa}.

Next, we examine the $\xi$ dependence of the pion-pole contributions.
Figure~\ref{fig:fds-pietas-VA-smallxi}
illustrates that the pion-pole contribution to $\gamma \pi^+$ 
photoproduction is very small (and, interestingly, destructive) 
for $\xi = 0.01$ at $(\ssM, S_{\gamma \NN}) = (4, 200)$~GeV$^2$, 
and becomes entirely negligible for $\xi = 0.002$ 
at $(\ssM, S_{\gamma \NN}) = (7, 2000)$~GeV$^2$.
This shows that the pion-pole contribution is indeed small at low $\xi$, 
and that the small-$\xi$ predictions given in
\cite{Duplancic:2018bum, Duplancic:2022ffo}
would not be affected by it.

Figure~\ref{fig:fds-pis-pole} compares the fully differential cross sections
for $\gamma \pi^+$, $\gamma \pi^-$, and $\gamma \pi^0$ final states
at four kinematical settings
\req{eq:fourkin} with $\xi \in \{0.117, 0.219, 0.224, 0.327\}$.
Thin lines represent the results obtained without the
$\pi$-pole contribution, while thick lines show the full results,
allowing us to assess the relative importance of the $\pi$-pole term
under these conditions.
The overall behaviour is similar across all kinematical cases.
The $\gamma \pi^-$ cross sections are slightly higher than those for $\gamma \pi^+$,
whereas $\gamma \pi^0$ dominates at forward angles.
In all cases, the $\pi$-pole contribution is very significant for charged
pions and remains relevant for $\pi^0$, increasing with $\xi$.
Although $\xi$ is comparable,
the predictions for case~$\caseIV$ are smaller than those for case~$\caseIII$
because of the suppression with $S_{\gamma N}$
in \req{eq:diff-CS-2}.
Analogously, 
Figure~\ref{fig:fds-etas-pole} and
Figure~\ref{fig:fds-etasmu-pole} 
compare the fully differential cross sections
for $\gamma \eta$ and $\gamma \eta'$ final states
at four kinematical settings and for two different choices of
$\mu_R$ and $\mu_F$.
Here too, the pion-pole contributions are sizable and increase with $\xi$.

Figure~\ref{fig:fds-ratio-pole} illustrates
the impact of the pion-pole contributions via
the ratio of the fully differential cross section
\req{eq:diff-CS-2}
calculated with and without $\tE_{\text{pole}}$
for four kinematical settings
\req{eq:fourkin} with $\xi \in \{0.117, 0.219, 0.224, 0.327\}$,
and for all pseudoscalar mesons considered:
$\pi^+$, $\pi^-$, $\pi^0$, $\eta$, and $\eta'$.
Since the dependence on $S_{\gamma \NN}$ and $\ssM$ cancels,
the ratio depends only on $\xi$ and $\alpha$.
The ratio increases with $\xi$
and is large for $\pi^+$ and $\pi^-$,
reaching values between $10$ and $60$ in the backward region.
It is smaller and more symmetric in $\alpha$
for neutral mesons.
The ratios for $\pi^0$ and $\eta$ exhibit similar behaviour,
with a maximum of around $3$ and $1.9$, respectively,
while the ratio for $\eta'$ is slightly modified by $gg$ contributions,
reaching values up to $2.4$.
The results for $\xi = 0.219$ $\caseIII$
and $\xi = 0.224$ $\caseIV$ are nearly identical,
except for $\eta'$, where the $gg$ contribution leads to a small deviation.

\vspace*{0.5cm}
\noindent
\textit{Pion-pole in single differential and integrated cross-sections} -
We now turn to the predictions for the single differential cross section.
Figures~\ref{fig:sds-Pip-AV} and \ref{fig:sds-Pi0-AV}
show the single differential cross section
for $\gamma \pi^+$ and $\gamma \pi^0$
photoproduction evaluated at
$S_{\gamma \NN} = 30$, $20$, $10$, and $6~\text{GeV}^2$,
with the vector and axial-vector GPD contributions
indicated as in Fig.~\ref{fig:fds-pietas-VA}.
A comparison between our results with the pion-pole included
(thick lines) and those obtained without the pion-pole term
(thin lines for axial and total contributions) illustrates the
impact of the pion-pole on this observable.
The $\gamma \pi^-$ channel exhibits a similar pattern to $\gamma \pi^+$,
but with systematically higher values, and is therefore not shown.
Similarly, Figures~\ref{fig:sds-Eta-AV} and \ref{fig:sds-Etap-AV}
present the results for $\gamma \eta$ and $\gamma \eta'$
at $S_{\gamma \NN} = 10$ and $6~\text{GeV}^2$,
for two choices of renormalization and factorization scales,
\req{eq:default} and \req{eq:muRFchoice}.
As discussed for the fully differential cross section,
the pion-pole contributions increase with $\xi$.
This is evident from the results at lower $S_{\gamma \NN}$,
where $\xi_{\text{min}}$ is already large.
Our analysis is not complete at small $\xi$,
and predictions for $\xi \lesssim 0.1$ should therefore be viewed only as indicative.
One observes a pronounced decrease in the total cross section
with increasing $\xi$.
Still, the single-differential results confirm that 
the pion-pole contributions are more important for charged pions, 
less pronounced but still present for $\pi^0$, $\eta$, and $\eta'$.
To illustrate this effect more clearly,
Figure~\ref{fig:sds-ratio-pole} presents the ratio
of the single differential cross section
obtained with and without the $\pi$-pole contribution,
for $\gamma \pi^+$ and $\gamma \eta$ photoproduction.
The ratio increases with $\xi$;
for $\xi \in [0.2, 0.3]$ it grows from
$3$ ($1.2$) to $8$ ($1.5$) for $\pi^+$ ($\eta$),
at $S_{\gamma \NN} = 20$~GeV$^2$.

Finally, we discuss the integrated cross-section.
Figure \ref{fig:sig-PiEtas-mu} presents the cross-sections
for the photoproduction of
$\gamma \pi^+$, $\gamma \pi^-$, $\gamma \pi^0$,
$\gamma \eta$, and $\gamma \eta'$ pairs
as functions of the center of mass energy of the incoming
photon-nucleon system, $S_{\gamma \NN}$,
for two choices of the renormalization and factorization scales.
Thin lines denote the results obtained
without the pion-pole contribution.
The ratio of the total cross-section
with and without the pion-pole contribution,
as a function of $S_{\gamma \NN}$,
is displayed in Figure \ref{fig:sig-ratio-pole}.
While the pion-pole effects enhance the integrated cross-section
less strongly than the differential ones,
their impact remains significant for charged pions
and still noticeable for light neutral pseudoscalars.
For $S_{\gamma \NN} \in [6,30]$ GeV$^2$,
the ratio decreases from about $2$-$2.5$ to $1.4$ for charged pions,
and from $1.3$-$1.6$ to $1.06$ for $\pi^0$, $\eta$, and $\eta'$.

\subsection{Meson structure} 

Photon-meson photoproduction is sensitive to both
the nucleon and meson internal structure.

\vspace*{0.5cm}
\noindent
\textit{Pion distribution amplitudes} -
For $\gamma \pi$ photoproduction, we present predictions
based on five pion DAs shown in Fig.~\ref{fig:DAs}.
Their impact on the fully differential cross-section
is illustrated on the left panel of Figure~\ref{fig:fds-ratio-DAs},
where we show the ratio of the fully differential
cross-section for the $\gamma \pi^+$ channel,
calculated with a given pion DA and normalized
to the result obtained with the asymptotic DA.
Thin lines indicate the results obtained
without the pion-pole contribution.
Corresponding ratios remain nearly constant across
the entire $\alpha$ range, taking values of about $1.25$
for the default DA $\phi_{[0.116]}$, and up to a factor of $3$
for the broadest investigated DA $\phi^{(0.3)}$.
When the pion-pole contribution is included,
the ratios increase markedly for broader DAs.
For $\phi^{(0.3)}$, the ratio ranges from $7.1$ to $3.85$
across the $\alpha$ interval under consideration.
Since the form of $\tE_{\text{pole}}$ \req{eq:pipole} depends
on the pion DA, the resulting backward enhancement is fully
consistent with the behaviour described in the preceding subsection.
The $\gamma \pi^-$ channel shows the same sensitivity to the DA shape
(not shown for brevity), with the broadest DA yielding ratios 
between $6$ and $4$, whereas in the $\gamma \pi^0$ channel the ratios are
flatter and somewhat lower, around $3$ -- $3.5$.

The right panel of Figure~\ref{fig:fds-ratio-DAs}
shows the ratio of the full differential cross-sections
calculated with and without the pion pole for each DA.
This confirms the backward-angle dominance
of the charged-pion predictions, which becomes
even more pronounced for broader DAs.
The $\gamma \pi^-$ channel follows a similar pattern,
while for $\gamma \pi^0$ the ratios are somewhat smaller
and symmetric in $\alpha$, as are the corresponding
cross-sections themselves. 

In the remainder of this work, we use $\phi_{[0.116]}$ as the default
DA, whose shape is supported by moment-based lattice results.
The numerical predictions obtained with this DA are relatively close
to those for the asymptotic DA 
(the ratio for all $\pi$ channels lies in the range $1.2$ -- $1.4$).
We note that, if a broader pion DA were to be confirmed and quantified,
the predicted $\gamma \pi$ photoproduction cross sections would increase
accordingly, as discussed above.

\vspace*{0.5cm}
\noindent
\textit{Quark and gluon components of $\eta$ and $\eta'$} -
In describing the $\gamma \eta$ and $\gamma \eta'$
channels, two types of mixing must be taken into account.
The phenomenological mixing between
the flavour-octet $\eta_8$ and flavour-singlet $\eta_1$ states
is taken into account through
(\ref{eq:etamix}-\ref{eq:Metaetap}).
The flavour-singlet state $\eta_1$ itself
contains quark ($\eta_{1q}$) and gluon ($\eta_{1g}$)
components that mix under evolution.
Including the gluonic contributions
and exploring their impact on the structure of
the $\eta$ and $\eta'$ mesons
was one of the main motivations for this study.
Figure \ref{fig:fds-etasmu-QG} displays the fully differential
cross-sections for $\gamma \eta$ (thin lines) and $\gamma \eta'$ (thick lines)
photoproduction as functions of $\alpha$,
for two choices of renormalization and factorization scales.
Predictions that include only quark contributions are
shown as dot-dashed lines, while solid lines represent
the complete results with gluons included.
For the quark-only case, the $\gamma \eta$
cross-sections are higher than those for $\gamma \eta'$,
reflecting the relative size of the decay constants.
The inclusion of gluons modifies the $\gamma \eta$ results
destructively at backward ($\alpha<0.5$) and constructively 
at forward ($\alpha>0.5$) angles.
The effect is even stronger for $\gamma \eta'$,
where constructive interference lifts the
$\gamma \eta'$ predictions above the $\gamma \eta$ ones
in the forward region for the default
scale choice \req{eq:default}, and entirely above
for \req{eq:muRFchoice}.

An analogous behaviour is shown in Figures
\ref{fig:sds-EtaMu-QG} and \ref{fig:sds-EtapMu-QG},
where the single differential cross-sections are presented
for selected values of $S_{\gamma \NN}=20$ and $10$ GeV$^2$
and for two choices of the renormalization and factorization scales,
$\muR$ and $\muF$.
In addition to the quark-only predictions (dot-dashed lines)
and the full results including gluons (solid lines),
we also display the gluon-only contributions (dotted lines).
A comparison of the quark-only and total results shows that
for $\gamma \eta$, the gluon effect is relatively small,
reducing the prediction at low $\xi$ and increasing it at large $\xi$.
For $\gamma \eta'$, in contrast, the gluon contribution
is sizable and increases the result across the entire $\xi$ range.
The integrated cross-section shown in Figure
\ref{fig:sig-EtasMu-QG}
confirms the same trend:
the gluon effect on $\gamma \eta$ is moderate and reduces the
quark-only prediction, while for $\gamma \eta'$ it is large and
enhances it.
These findings demonstrate that $\gamma \eta'$ photoproduction
is highly sensitive to the gluonic component of the $\eta'$ meson.

\vspace*{0.5cm}
\noindent
\textit{Evolution and dependence on factorization scale} -
In this work, we have examined the impact of the
evolution of meson distribution amplitudes
and the sensitivity of the predictions to the choice of factorization scale.
Two representative choices of $\muF$ are singled out:
\req{eq:default} and
\req{eq:muRFchoice}.
The effect of DA evolution
\req{eq:evLO8} 
-and consequently the dependence of
the leading-order (LO) photoproduction prediction on the
$\muF$ scale-%
\footnote{The hard-scattering subamplitude $\Mamp$
develops an explicit dependence on $\muF$ only at NLO.}
is found to be small for the default
pion DA choice (second equation in \req{eq:default}),
even when using
\req{eq:muRFchoice}.
The influence of evolution for the broader DAs
is expected to be more pronounced,
but including the full evolution requires either
an approximation of the infinite Gegenbauer expansion
or a numerical solution of the evolution equation.
We leave its implementation for future work.

The DA evolution plays a crucial role in the description
of $\eta$ and $\eta'$ mesons, where it mixes
the quark and gluon components of $\eta_1$
\req{eq:evolution}, while $\eta_8$
obeys the same evolution as pions.
Figures
\ref{fig:fds-etas-pole} and
\ref{fig:fds-etasmu-pole},
together with the left and right panels of Figure \ref{fig:fds-etasmu-QG},
compare the fully differential cross-sections
for $\gamma \eta$ and $\gamma \eta'$ photoproduction
for renormalization and factorization scales
chosen as
$\muF^2= \ssM$
\req{eq:default} and
$\muF^2= \ssM \alpha \balpha$
\req{eq:muRFchoice}.
The renormalization scale, discussed in 
Section \ref{subsec:normalisation-scale-dependence},
mainly affects the overall normalization,
whereas the change in the shape of the predictions,
i.e., their dependence on $\alpha$,
originates from the DA evolution to different factorization scales.
For $\gamma \eta$,
which is mainly governed by the flavour-octet component $\eta_8$,
the shape remains almost unchanged,
whereas for $\gamma \eta'$ photoproduction,
driven by the flavour-singlet component $\eta_1$,
the difference becomes clearly visible.
In particular, one observes a pronounced backward dominance of
the $\gamma \eta'$ predictions at the lower renormalization scale
\req{eq:muRFchoice}.
This feature is only partly an artefact of the dependence of
$\alpha_{\max}$ on the meson mass \req{eq:umax},
which restricts the available $\alpha$ range in the forward region.
A more detailed inspection of individual contributions
shows that the gluon terms themselves are backward-dominated,
due to the dominant $\Mamp_{\Delta B5}$ contribution
listed in Tables \ref{tab:anres} and \ref{tab:anresPV}
and convoluted by vector GPDs.
Consequently, the lower $\muF$ scale exposes the dominant gluon
contributions both in shape and magnitude
(as seen from the comparison of the quark-only and total predictions).
As expected from the evolution equation,
the gluon contributions decrease with increasing scale.

Figures
\ref{fig:sds-Eta-AV} and \ref{fig:sds-Etap-AV}
show that the individual GPD contributions
to the single differential cross-sections
retain the same shape
for $\gamma \eta$ photoproduction for the two scale choices considered,
whereas they are significantly modified
in the case of $\gamma \eta'$ photoproduction.
Figure
\ref{fig:sds-EtaMu-QG}
clearly illustrates
that the corresponding curves for $\gamma \eta$ photoproduction
(except at very low $S_{\gamma \NN}$)
remain nearly identical
when evaluated at different factorization scales.
Figure \ref{fig:sds-EtapMu-QG}, by contrast,
shows a distinct change in both the shape and the
magnitude of the gluon contributions,
which become larger at smaller factorization scales
and thereby modify the interplay
between the quark and gluon components.
Finally, the integrated cross-section
shown in Figure
\ref{fig:sig-EtasMu-QG}
exhibits essentially the same pattern:
a change in the factorization scale does not affect
the shape of the $\gamma \eta$ predictions,
whereas for $\gamma \eta'$, the relation between the quark-only and
total results differs markedly between scales,
once again emphasizing the enhanced
role of the gluon contributions at lower scales.

\subsection{Normalization and scale dependence}
\label{subsec:normalisation-scale-dependence}

In this work, the amplitudes are calculated at LO
in the strong coupling constant.
Consequently, the resulting predictions are proportional to
$\alphaS(\muR)$ and exhibit a strong dependence
on the choice of the renormalization scale.
To stabilize this dependence, one must include the
NLO corrections,%
\footnote{
NLO corrections contain both constant and
$\log(\muR)$-dependent terms, which
reduce the sensitivity to $\muR$.
},
a technically demanding task.
Similarly, the dependence on the factorization scale $\muF$
enters the subprocess amplitudes only at NLO.
We have already discussed the dependence of the meson DAs
on $\muF$ and its implications,
while the evolution of the GPDs with $\muF$
is left for a future extension of this work.
To illustrate the sensitivity of our predictions
to the choice of renormalization scale,
we show the results obtained with the default setting
$\muR^2=\ssM$ \req{eq:default}
and with $\muR^2=\ssM \alpha (1-\alpha)$ \req{eq:muRFchoice}.

The ratio of the fully differential cross-sections obtained with
\req{eq:muRFchoice} and \req{eq:default} follows the expected
$\alphaS^2(\ssM \alpha \bar{\alpha}) / \alphaS^2(\ssM)$ scaling,
which in itself explains most of the observed enhancement by
a factor of $2.2$--$4.5$ for pions at $\ssM = 4$~GeV$^2$ across the
allowed $\alpha$ range.
Assuming $\muR=\muF$,
the remaining discrepancy arises from the
mild factorization-scale dependence discussed above.
By comparing Figures \ref{fig:fds-etas-pole}
and \ref{fig:fds-etasmu-pole}, as well as
the panels in Figure \ref{fig:fds-etasmu-QG},
one finds that the corresponding ratio
is about $2$ -- $3$ for $\eta$,
and increases to $3.6$ -- $10$ for $\eta'$.
Similarly, the ratios of the single differential cross-sections
evaluated at \req{eq:muRFchoice} and \req{eq:default}
are $2.5$ -- $3$ for $\pi^\pm$ and
$2.5$ -- $3.5$ for $\pi^0$.
The single differential cross-sections shown
in Figs. \ref{fig:sds-Eta-AV}-\ref{fig:sds-EtapMu-QG}
indicate ratios of $2.2$ -- $2.7$ for $\eta$ and $4$ -- $6.5$ for $\eta'$.
Finally, the integrated cross-sections
presented in Figures \ref{fig:sig-PiEtas-mu}
and \ref{fig:sig-EtasMu-QG}
yield approximately constant ratios across the relevant
$S_{\gamma \NN}$ range:
$2.6$ -- $2.7$ for pions,
$2.1$ -- $2.5$ for $\eta$,
and
$5.1$ -- $5.2$ for $\eta'$.
The systematically larger enhancement observed for $\eta'$ across all
considered observables arises from the factorization-scale dependence
of the gluon contributions and their interference with the quark
contributions.

Figures \ref{fig:sig-PiEtas-mu} and \ref{fig:sig-EtasMu-QG}
show that the integrated cross-sections, obtained for our chosen
set of parameters, DAs, and GPDs, reach values of a few tens of pb.
These absolute values should not be interpreted too strictly,
as our approach is applicable only at moderate $\xi$
and does not include the contributions relevant at small $\xi$,
such as those from sea quarks and gluons in the nucleon.
Since the subprocess amplitudes, GPDs, and DAs are identical,
the observed ordering of the pion cross-sections,
$\sigma_{\pi^-}>\sigma_{\pi^+}>\sigma_{\pi^0}$,
is entirely determined by the meson flavour structure,
i.e. the charge factors.
The cross-sections for $\eta$ and $\eta'$ are somewhat smaller,
with
$\sigma_{\eta}>\sigma_{\eta'}$ for the default scale choice \req{eq:default},
whereas for \req{eq:muRFchoice} the $\eta'$ cross-section
exceeds that of $\eta$
at higher values of $S_{\gamma \NN}$.
Similarly, App. \ref{app:sds} shows that the single differential cross-sections
$d\sigma/d\xi$ for $\xi>0.1$ reach values of a few hundred pb,
while the fully differential cross-sections presented in App.
\ref{app:fds} are of the order of a few tens of pb~GeV$^{-4}$.
Both vary strongly depending on the specific kinematical conditions.

\subsection{Numerical summary}

We have carried out a detailed analysis of
the fully differential, single differential 
and integrated cross-sections
for
$\gamma \MM$ photoproduction,
valid in the moderate-$\xi$ region ($\xi>0.1$).
We want to stress the following findings
regarding the key aspects of this analysis
stated at the beginning of this section:
\begin{itemize}
\item
Our results show that in the moderate-$\xi$ region the pion-pole term
$\tCFFe_{\text{pole}}$ plays a particularly significant role,
most notably in $\gamma \pi^\pm$ photoproduction, where its
contribution becomes especially pronounced at backward
scattering angles. Pion-pole effects are also present in the
photoproduction of neutral mesons.
Apart from the pion-pole term, the vector GPD contributions,
in particular $\CFF$, dominate the observable.
See, e.g., Fig.\ref{fig:fds-pietas-VA}.
\item
Photoproduction of $\gamma \pi$ pairs is particularly sensitive
to the shape of the pion DAs, and in the moderate-$\xi$ region
this sensitivity
is further enhanced by the pion-pole contributions \req{eq:pipole}.
The results obtained with the default moment-based lattice DA
$\phi_{[0.116]}$ \req{eq:a2-pion} are close to those obtained with
the commonly used asymptotic DA, whereas the predictions obtained with
the broader DAs \req{eq:phi_expal} can be significantly larger.
For the fully differential cross section, the average ratio of the
predictions obtained with the broad DA $\phi^{(0.3)}_\pi$ and the
asymptotic DA is about $5.5$ for $\pi^+$, $4.5$ for $\pi^-$, and $3$
for $\pi^0$. 
See, e.g., Fig.\ref{fig:fds-ratio-DAs}.
\item
The study of $\gamma \eta$ and
$\gamma \eta'$ channels provides valuable insight into
the gluon content of these mesons.
In particular, the $\gamma \eta'$ channel
shows sizeable gluon contributions in $\eta'$ meson 
and a strong sensitivity to the factorization-scale choice.
See, e.g., Fig.\ref{fig:fds-etasmu-QG}.
\item
The variation of the LO predictions with the renormalization scale
introduces potentially 
large theoretical uncertainties, highlighting the
importance of 
NLO corrections. 
We have compared the results
at two physically motivated large-scale choices for the renormalization
scale: the $\gamma \MM$ invariant mass \req{eq:default} and the
square of relative momentum transfer between the photon and meson \req{eq:muRFchoice}.
For our kinematic settings, the corresponding predictions exhibit a
sizeable $\muR$ dependence, with the latter choice producing an
enhancement of about $2$--$4$ times. In the $\eta'$ channel, the strong
dependence on the factorization scale 
amplifies this effect.
See, e.g., Figs.~\ref{fig:sds-EtapMu-QG} and \ref{fig:sig-PiEtas-mu}.
\end{itemize}

\section{Conclusions}
\label{sec:sum}

In this work, we revisit the hard exclusive
photoproduction of photon--meson pairs,
$\gamma \NN \to \gamma \MM\, \NN'$,
and evaluate the corresponding cross sections
at leading twist and leading order in perturbative QCD.
The core of the analytical calculation, summarized in
Section~\ref{sec:ares}, consists of compact expressions for the
subprocess amplitudes in the pseudoscalar channels
$\MM \in \{\pi^\pm, \pi^0, \eta, \eta'\}$, including only the quark
GPD contributions while also taking into account the two--gluon
components of the $\eta$ and $\eta'$ distribution amplitudes.
These results provide a transparent structure of the amplitudes
and a practical basis for efficient numerical implementation.

The phenomenological analysis is performed in the moderate--$\xi$
region ($\xi > 0.1$), with the adopted valence–quark approximation,
in which the valence contributions are numerically sizable, while
the sea–quark and gluon contributions are consistently omitted. 
The latter would require a modified factorisation framework for their reliable
inclusion.
The pion–pole contribution plays a central role in this kinematic
domain. It provides the main contribution in $\gamma\pi^\pm$
production and has a visible impact in the neutral channels as well.
Including the pion–pole term is therefore essential in the
moderate--$\xi$ region relevant for JLab and COMPASS/AMBER measurements
\cite{Duplancic:2022ffo, Arrigo2025_TowardsImprovedHadronFemto}, 
whereas for small $\xi$, its effect becomes negligible.
All leading--twist chiral--even GPDs are included in our analysis,
and the vector contributions, in particular $\CFF$, generally
dominate once the pion–pole term is absent.

Our analysis shows a strong sensitivity of
$\gamma\pi$ production to the shape of the pion distribution amplitude,
with broad DAs possibly enhancing the predictions several times
in comparison with the asymptotic DA.
The $\gamma\eta$ and $\gamma\eta'$ channels probe the gluon content of
the mesons, with gluons contributing already at LO in this process.
This is in contrast to the
$\gamma^* \gamma \to \eta(\eta')$ transition form factors,
which are used to constrain the meson DAs from experiment,
where gluon contributions appear only at NLO.
Interestingly, in deeply virtual production of $\eta$ and $\eta'$ mesons,
the two--gluon contributions vanish at LO for kinematic reasons
\cite{Duplancic:2016bge}.
The presence of gluon contributions already at LO
therefore makes photon-meson photoproduction
particularly suited for studying the gluonic components
of the $\eta$ and $\eta'$ mesons.
In this case, the $\gamma\eta'$ channel shows sizeable
two--gluon contributions and a strong dependence
on the factorization scale.

Being a LO QCD prediction,
the numerical results are strongly dependent on the choice of
the renormalization scale,
indicating that the inclusion of NLO
corrections is essential for stabilizing this dependence
and achieving a quantitatively reliable description.
At the subprocess level, photon-meson photoproduction
is closely related to wide-angle meson-pair production,
$\gamma \gamma \to M M'$,
and valuable insights can be gained by comparing the two processes.
In particular, the pseudoscalar-pseudoscalar ($\PP \PP$) channel
has been analyzed at NLO \cite{Duplancic:2006nv},
and extending this analysis,
involving a few hundred diagrams,
to photoproduction kinematics as well as to the
scalar-pseudoscalar ($\SS \PP$) channel
remains an ambitious task.
Nevertheless, taking lessons from DVCS and DVMP,
for which a consistent description based on universal GPDs
emerges only at NLO \cite{Cuic:2023mki},
we view this as a well-motivated and natural extension.
In the absence of NLO corrections,
we therefore compare our predictions
at two physically motivated scale choices and find
markedly different results, differing by a factor of a few.

The present study establishes a consistent and transparent
leading-order framework for photon–meson photoproduction,
with the quark GPD contributions included.
This process is complementary to DVCS and DVMP in providing
information on GPDs, but is more sensitive to their
$x$-dependence due to the additional kinematic variables
and the mixing of $x$ with external scales in the propagators
\cite{Qiu:2023mrm}.
The analytical and numerical results presented here provide
a reference baseline for forthcoming experimental studies
at JLab, 
COMPASS/AMBER, and future facilities,
which will ultimately enable a more precise determination
of GPDs and improve our understanding of nucleon structure.
%



\subsection*{Acknowledgments}
We thank P. Kroll, Pawel Nadel-Turonski and Justin Stevens for discussions.
This publication was partially supported by 
the Croatian Science Foundation project IP-2019-04-9709 
and N. Crnković was supported by the Croatian
Science Foundation project DOK-2020-01-9883.
L. Szymanowski was supported by the Grant No. 2024/53/B/ST2/00968 of
the National Science Centre in Poland.
He thanks IJCLab and the "P2I - Graduate School of Physics", in the framework ``Investissements d’Avenir'' (ANR-11-IDEX-0003-01) 
managed by the Agence Nationale de la Recherche (ANR), France, for support.
The work of S. N. was supported by the Science
and Technology Facilities Council (STFC) under Grant No. ST/X00077X/1, and by the Royal Society
through Grant No. URF/R1/201500.
\clearpage

\appendix

\makeatletter
\renewcommand{\theequation}{\Alph{section}.\arabic{equation}}
\renewcommand{\theHequation}{\theHsection.\arabic{equation}} 
\makeatother

\section{Kinematics}
\label{app:kin}

We choose the light-cone vectors $p$ and $n$ as
\begin{equation}
p^{\mu}= \frac{\sqrt{\ss}}{2}(1,0,0,1), 
\quad 
n^{\mu}= \frac{\sqrt{\ss}}{2}(1,0,0,-1)
\, ,
\label{eq:fvpn}
\end{equation}
which satisfy $p^2 = n^2 = 0$
and $p \cdot n = \ss/2$,
while $\ss$ is the auxiliary scale.
A generic four-momentum $v^{\mu}$ can then be decomposed
in the Sudakov basis as
\begin{equation}
v^{\mu} = a\,n^{\mu} + b\,p^{\mu} + v_{\perp}^{\mu}
\, ,
\quad
\text{with}
\quad
v_{\perp}^{\mu} = (0, v^x, v^y, 0)
\, ,
\end{equation}
so that $v^2 = a b s_0 + v_{\perp}^2$.

With this convention, and adopting the usual definition
$\Delta = p_2 - p_1$ given in Eq.~\eqref{eq:Delta},
the particle momenta for the process \eqref{eq:proc}
can be expressed as
\begin{equation}
\begin{alignedat}{2}
p_1^{\mu} &= (1+\xi)\,p^{\mu} 
+ \frac{\Mp_{\NN}^2}{\ss(1+\xi)}\,n^{\mu} \, , 
&\qquad
p_2^{\mu} &= (1-\xi)\,p^{\mu}
+ \frac{\Mp_{\NN'}^2-\Delta_{\perp}^2}{\ss(1-\xi)}\,n^{\mu} 
+ \Delta_{\perp}^{\mu} \, , 
\\[0.3em]
q^{\mu}   &= n^{\mu} \, , 
&\qquad
k^{\mu}   &= 
\alpha\,n^{\mu}
- \frac{(\pT_{\perp}-\Delta_{\perp}/2)^2}{\ss\,\alpha}\,p^{\mu} 
+ \pT_{\perp}^{\mu}
- \frac{\Delta_{\perp}^{\mu}}{2} \, , \\[0.3em]
&&
p_{\MM}^{\mu} &= 
\alpha_{\MM}\,n^{\mu}
+ \frac{-(\pT_{\perp}+\Delta_{\perp}/2)^2+m_{\MM}^2}{\ss\,\alpha_{\MM}}\,p^{\mu} 
- \pT_{\perp}^{\mu}
- \frac{\Delta_{\perp}^{\mu}}{2} \, .
\end{alignedat}
\label{eq:fvectors}
\end{equation}
Here, $\Mp_{\NN} \approx \Mp_{\NN'}$ and $m_{\MM}$ denote the nucleon and meson masses, respectively,
while $\pT_\perp$ represents the relative transverse momentum between the outgoing photon
and the meson.
We introduce the dimensionless variables $\alpha$, $\alpha_{\MM}$, and $\xi$.
The latter, as usual, denotes the skewness 
\req{eq:xidef}.

From \eqref{eq:fvectors} follows that scalars
\eqref{eq:scalars} amount to
\begin{equation}
\begin{alignedat}{2}
S_{\gamma \NN} &= (1+\xi)\,\ss + \Mp_{\NN}^2 
\, , 
&\qquad
t &=\Delta^2= -\frac{4\xi^2 \Mp_{\NN}^2}{1-\xi^2}
+ \frac{1+\xi}{1-\xi}\,\Delta_\perp^2
\, ,
\\[0.4em]
s' = M_{\gamma \MM}^2
&=2 \xi \ss + t
\, ,
&\qquad
t' &= \frac{(\pT_\perp-\Delta_\perp/2)^2}{\alpha} 
\, ,
\qquad
u' = \frac{(\pT_\perp+\Delta_\perp/2)^2 -(1-\alpha_{\MM})m_{\MM}^2}{\alpha_{\MM}} 
\, ,
\label{eq:genscalars}
\end{alignedat}
\end{equation}
while a similar lengthy expression for $M_{\MM \NN'}^2$ can be found in 
\cite{Boussarie:2016qop,Duplancic:2018bum}. 
Skewness then takes the general form
\begin{equation}
\xi = \frac{\tau}{2-\tau} 
\qquad
\mbox{with}
\qquad
\tau=\frac{\ssM-t}{S_{\gamma \NN}-\Mp_{\NN}^2}
\, .
\label{eq:xigen}
\end{equation}
According to factorization, 
illustrated in Fig. \ref{fig:fact-gMgM},
$\pT_{\perp}$ needs to be large,
implying large $\ssM=M^2_{\gamma \MM}$ and $\tsM$
and
one neglects $\Delta_\perp$ in front of $\pT_{\perp}$.
In the collinear approximation, 
or so-called generalized Bjorken limit, 
needed for the determination of subprocess amplitudes 
(\ref{eq:sub-qqqq}, \ref{eq:sub-qqgg}),
one also neglects the hadronic masses.
Thus
\begin{align}
\ssM=M^2_{\gamma \MM} &\approx -\frac{\pT_{\perp}^2}{\alpha \bar{\alpha}}
= 2 \xi s_0 
\, , 
&\: 
-t' &\approx 
-\frac{\pT_{\perp}^2}{\alpha}
=\bar{\alpha} \ssM 
\, , 
&\: 
-u' &\approx 
-\frac{\pT_{\perp}^2}{\balpha}
=\alpha \ssM
\, , 
\label{eq:stucol}
\end{align}
with $\alpha_{\MM} \approx 1-\alpha \equiv \balpha$,
while $t=0$.

\section{Distribution amplitude formalism: convention and evolution}
\label{app:DAev}


The definition \req{eq:defphiqFT} 
of meson DAs describes the projection of a collinear
$q \bar{q}$ state onto a pseudoscalar meson $\MM$.
It is applied to evaluation
of Feynman diagrams, as in Fig.~\ref{fig:diagQ},
by replacing the quark and antiquark spinors%
\footnote{Spinors are normalized as
$u^\dagger (p,\lambda) u(p,\lambda') = 
\sqrt{2}\, p^+ \, \delta_{\lambda \lambda'}$.}
with the projector
\begin{equation}
{\cal P}_{\MM,q}^{ij,kl}  
= {\cal C}_\MM^{ij} \, \frac{\delta_{kl}}{4 N_c} \,
\gamma_5 \not\!p_\MM\,,
\label{eq:qqproj}
\end{equation}
where $(i,k)$ and $(j,l)$ denote the (flavour, colour) indices
of the quark and antiquark, respectively.
For the diagrams in Fig.~\ref{fig:diagG},
a collinear $gg$ state of two outgoing gluons%
\footnote{The complex-conjugated expression applies
for an incoming pseudoscalar meson.
}
with colour indices $a$ and $b$,
Lorentz indices $\mu$ and $\nu$,
and momentum fractions $z$ and $\bz$,
is projected onto the flavour-singlet pseudoscalar state $\eta_1$ as
\begin{equation}
{\cal P}_{\eta_1,g}^{\mu \nu, ab}  
=
\frac{1}{2 \sqrt{n_f}} \, \frac{\delta_{ab}}{4 N_c} \,
\, i \,
\veps^{\mu \nu}_{\perp}
\frac{1}{z \bz} 
\, ,
\qquad
\veps^{\mu \nu}_{\perp}
=
\veps^{\mu \nu}_{\gamma \delta}
\frac{k^{\gamma} p_{\MM}^{\delta}}{k \cdot p_\MM}
\, ,
\label{eq:ggproj}
\end{equation}
with $k$ here satisfying the standard light–cone requirements, being
collinear to $p_\MM$ in the light–cone sense (non–zero minus and
vanishing transverse components), so that it serves as the reference
vector in the projection.
The factor $1/(z \bz)$
arises from converting the field-strength tensor matrix element
\req{eq:defphigFT}
to the gauge-field representation
$\langle \eta_1(p_\MM)|
A^{\nu}(-\zcor)
A^{\mu}(\zcor)
|0\rangle$,
which appears in the amplitude evaluation.
Throughout this work 
for the Levi-Civita tensor
we adopt the convention
\begin{equation}
\label{eq:LC-convention}
\epsilon^{0123} = +1
\, .
\end{equation}
The projectors
\req{eq:qqproj} and \req{eq:ggproj}
are then used in the calculation of the hard-scattering subprocess
amplitudes, which are subsequently convoluted with
$f_\MM \, \phi_\MM$ and $f_{\eta_1} \, \phi_{\eta_1,g}$, respectively.

For completeness, we note that in the case of scalar mesons,
the definitions are modified as follows:
$\gamma_5$ is omitted in
\req{eq:defphiqFT} and \req{eq:qqproj};
the dual field-strength tensor is replaced by
$\widetilde{G}^{\mu \nu} \to G^{\mu \nu}$,
and
$
i\, \veps^{\mu \nu}_{\perp}
\;\to\;
g^{\mu \nu}_\perp =
g^{\mu \nu}
- \frac{k^{\mu} p^{\nu} + k^{\nu} p^{\mu}}{k \cdot p_\MM}
$
in
\req{eq:defphigFT} and \req{eq:ggproj}, respectively.
The symmetry properties given in \req{eq:DAsym} change sign for
scalar mesons.


The definition of the gluon DA and the corresponding
flavor-singlet evolution is convention dependent%
\footnote{%
A detailed discussion of the various conventions
found in the literature is provided in
Ref.~\cite{Kroll:2002nt}.
The evolution kernels corresponding to the present convention
were summarized in Ref.~\cite{Kroll:2012gsh},
where $\eta$ and $\eta'$ transition form factors
were also employed to update the numerical values
of the DA coefficients.
The same formalism and DAs were later applied
to other processes in Refs.~\cite{Kroll:2019nby,Kroll:2021zss},
where a slightly modified normalization-affecting the DA coefficients
-was introduced for convenience.
}.
In this work we adopt the convention of
Ref.~\cite{Kroll:2019nby},
developed and employed in
Refs.~\cite{Kroll:2002nt,Kroll:2012gsh,Kroll:2021zss}.
For completeness, we summarize the LO
evolution equations for quark and gluon DAs
of pseudoscalar mesons represented as Gegenbauer expansions:
\begin{eqnarray}
\phi_{\MM}(z;\muF) &=& 6 z (1-z)
\left(
1 + \sum_{n=2, 4, \ldots}\!\, a_n^{\MM}(\muF)
                           \; C_n^{3/2}(2 z-1)
\right),
\label{eq:daGegen-Q}
\\[0.3cm]
\phi_{\eta_{1g}}(z;\muF) &=& 30 z^2 (1-z)^2
 \sum_{n=2, 4, \ldots}\!\, a_n^{g}(\muF)
                           \; C_{n-1}^{5/2}(2 z-1),
\label{eq:daGegen-G}
\end{eqnarray}
where the flavour-octet pseudoscalar mesons
$\MM=\{\pi^+, \pi^-, \pi^0, \eta_8\}$
evolve according to
\begin{equation}
a_{n}^\MM(\mu_F)
= a_{n}^\MM(\mu_0)\, L^{\gamma_n^{qq}/\beta_0},
\qquad
L = \frac{\alphaS(\mu_0)}{\alphaS(\mu_F)}.
\label{eq:evLO8}
\end{equation}
For the flavour-singlet $\eta_1$,
the evolution mixes the quark 
$\phi_{\eta_{1q}}\equiv \phi_{\eta_{1}}$
and gluon DAs
$\phi_{\eta_{1g}}$
:
\begin{align}
a_{n}^1(\mu_F)
&= \frac{1}{1-\rho_n^{(+)}\rho_n^{(-)}}
\Big[
  (L^{\gamma_n^{(+)}/\beta_0}
   - \rho_n^{(+)}\rho_n^{(-)} L^{\gamma_n^{(-)}/\beta_0})
   a_{n}^1(\mu_0^2)
 + (L^{\gamma_n^{(-)}/\beta_0}
   - L^{\gamma_n^{(+)}/\beta_0})
   \rho_n^{(-)}\, a_{n}^g(\mu_0^2)
\Big],
\nonumber \\[0.4em]
a_{n}^g(\mu_F)
&= \frac{1}{1-\rho_n^{(+)}\rho_n^{(-)}}
\Big[
  (L^{\gamma_n^{(-)}/\beta_0}
   - \rho_n^{(+)}\rho_n^{(-)} L^{\gamma_n^{(+)}/\beta_0})
   a_{n}^{g}(\mu_0^2)
 + (L^{\gamma_n^{(+)}/\beta_0}
   - L^{\gamma_n^{(-)}/\beta_0})
   \rho_n^{(+)}\, a_{n}^{1}(\mu_0^2)
\Big],
\label{eq:evolution}
\end{align}
with
\begin{equation}
\rho_n^{(+)} = \frac{1}{5}
\frac{\gamma_n^{gq}}{\gamma_n^{(+)} - \gamma_n^{gg}},
\qquad
\rho_n^{(-)} = 5
\frac{\gamma_n^{qg}}{\gamma_n^{(-)} - \gamma_n^{qq}},
\label{eq:rho}
\end{equation}
and eigenvalues of the anomalous-dimension matrix
\begin{equation}
\gamma^{(\pm)}_n =
\frac{1}{2}
\left[
  \gamma^{qq}_n + \gamma^{gg}_n
  \pm \sqrt{
    (\gamma^{qq}_n - \gamma^{gg}_n)^2
    + 4\, \gamma^{qg}_n \gamma^{gq}_n
  }
\right].
\label{eq:gamma+-}
\end{equation}
The LO anomalous dimensions used in this convention read
\begin{equation}
\begin{alignedat}{2}
\gamma^{qq}_n &= C_F \!\left[
   3 + \frac{2}{(n+1)(n+2)}
   - 4 \sum_{i=1}^{n+1} \frac{1}{i}
\right],
&\qquad
\gamma^{qg}_n &= C_F\,
  \frac{n (n+3)}{3 (n+1)(n+2)},
  \quad n \ge 2,
\\[0.3em]
\gamma^{gq}_n &= 
n_f \,
  \frac{12}{(n+1)(n+2)},
  \quad n \ge 2,
&\qquad
\gamma^{gg}_n &= \beta_0
  + N_c \!\left[
      \frac{8}{(n+1)(n+2)}
      - 4 \sum_{i=1}^{n+1} \frac{1}{i}
    \right],
  \quad n \ge 2.
\end{alignedat}
\label{eq:andim}
\end{equation}
where $C_F = (N_c^2 - 1)/(2 N_c)$ and $\beta_0=(11/3)N_c-(2/3) n_f$.
One should distinguish between $n_f = 3$, used in \req{eq:andim} and
in the DA definitions \req{eq:defphiqFT} and \req{eq:defphigFT},
which counts the fixed valence-quark content in the meson’s
flavor-singlet combination, and $\tilde{n}_f$, the number of active
flavors at a given scale, which enters the $\beta$ functions and
governs the running of $\alphaS$.

\clearpage

\section{Hard-scattering amplitudes for mesons with non-symmetric DAs}
\label{app:THq}

For completeness, this section provides  
the hard-scattering amplitudes for the  
$\gamma q_\a \to (q_\a \bar{q}_\b) q_\b$  
(or equivalently, $\gamma (q_\a \bar{q}_\b) \to (q_\a \bar{q}_\b) $)  
subprocess, without taking into account the symmetry properties  
of the meson DA.  
Unlike the symmetrized expressions in  
Sec.~\ref{sec:ares}, these results apply to  
pseudoscalar mesons with non-symmetric DAs,  
such as kaons.  
The full hard-scattering amplitude for the  
$\gamma (P) \to \gamma P$ subprocess,  
which couples to the parity-odd (axial) GPD, reads
\begin{eqnarray}
\MampPP^{(\a \b)} &=&
\left( \ei \, \MAi +
\ej \, \MAj +
\eij \, \MAij \right) T_A 
+
\left( \ei \, \MBi +
\ej \, \MBj +
\eij \, \MBij \right) T_B 
\, ,
\end{eqnarray}
where
\begin{eqnarray}
\MAi(y,z) &=&
\frac{4}{\ssM \, \balpha}
\left(
\frac{\alpha-z}{z \bz \, (y+\Ieta)}
+\frac{1 - \alpha z}{\alpha \, z \bz \, (y-\Ieta)}
+\frac{\balpha^2-2 \alpha z}{\alpha \, z \bz \, (\by-\Ieta)}
\right)
\, ,
\nonumber \\[0.3cm] 
\MAij(y,z) &=&
\frac{4}{\ssM \, \alpha}
\left(
\frac{\alpha-z}{z \bz \, (y+\Ieta)}
-\frac{1- \alpha \bz}{z \bz \, (\by-\Ieta)}
- \frac{(1-\balpha z)^2 +\alpha}{z \bz (\mixAzy  +\Ieta)}
\right)
+ \Big\{(y, z)\to (\by, \bz) \Big\}
\, ,
\qquad
\end{eqnarray}
\begin{eqnarray}
\MBi(y,z) &=&
\frac{8}{\ssM^2 \, \alpha}
\left(
-\frac{1}{\balpha\, \bz \, (y+\Ieta)}
+\frac{1}{\alpha \balpha \, \bz \, (y-\Ieta)}
+\frac{1}{\alpha \, z \, (\by-\Ieta)}
\right)
\, ,
\nonumber\\[0.3cm] 
\MBij(y,z) &=&
\frac{8}{\ssM^2 \alpha^2}
\left(
-\frac{1}{\bz \, (y+\Ieta)}
-\frac{1}{z (\by-\Ieta)}
\right)
+ \Big\{(y, z)\to (\by, \bz) \Big\}
\, ,
\end{eqnarray}
while
\begin{equation}
\MAj(y,z) = \MAi(\by,\bz) \, ,
\qquad
\MBj(y,z) = \MBi(\by,\bz) \, .
\end{equation}
The 
amplitude for the  
$\gamma (S) \to \gamma P$ subprocess  
couples to the parity-even (vector) GPD and is given by
\begin{eqnarray}
\MampSP^{(\a \b)} &=&
\left( \ei \, \MAiFive +
\ej \,\MAjFive  +
\eij \, \MAijFive \right) T_{A5} 
+
\left( \ei \, \MBiFive +
\ej \,\MBjFive  +
\eij \, \MBijFive \right) T_{B5} 
\, ,
\end{eqnarray}
with
\begin{eqnarray}
\MAiFive(y,z) &=&
\frac{-16 \xi i}{\ssM^3 \, \alpha \, \balpha}
\left(
\frac{\alpha (z-\bz) - z}{ \balpha \, z \bz  \, (y+\Ieta)}
+\frac{1-\alpha z}{ \alpha \balpha \, z \bz \, (y-\Ieta)}
-\frac{1+\alpha(z-\bz)}{\alpha \, z \bz \, (\by-\Ieta)}
\right)
\, ,
 \\[0.3cm] 
\MAijFive(y,z) &=&
\frac{-16 \xi i}{\ssM^3 \, \alpha^2}
\left(
\frac{\alpha (z-\bz) -z}{\balpha \, z \bz \, (y+\Ieta)}
+\frac{1-\alpha \bz}{\balpha \, z \bz \, (\by-\Ieta)}
+
\frac{\alpha z^2-\bz^2}{ z \bz\, (\mixAzy +\Ieta)}
\right)
\nn
- \Big\{(y, z)\to (\by, \bz) \Big\}
\, ,
\end{eqnarray}
\begin{eqnarray}
\MBiFive(y,z) &=&
\frac{-16 \xi i}{\ssM^3 \, \alpha \, \balpha}
\left(
\frac{\alpha -z}{ \balpha \, z \bz  \, (y+\Ieta)}
+\frac{z-\bz -\alpha z}{ \alpha \balpha \, z \bz \, (y-\Ieta)}
-\frac{\balpha-2 z}{\alpha \, z \bz \, (\by-\Ieta)}
\right)
\, ,
\nonumber \\[0.3cm] 
\MBijFive(y,z) &=&
- \MAijFive(y,z) 
+
\frac{32 \xi i}{\ssM^3 \, \alpha^2}
\left(
\frac{1}{ \, \bz \, (y+\Ieta)}
-\frac{1}{ \, z \, (\by-\Ieta)}
- \Big\{(y, z)\to (\by, \bz) \Big\}
\right)
\nonumber \\[0.2cm] &=& 
- \MAijFive(y,z) 
+ \MBijFiveDelta(y,z) 
\, ,
\end{eqnarray}
while
\begin{equation}
\MAjFive(y,z) = - \MAiFive(\by,\bz) \, ,
\qquad
\MBjFive(y,z) = - \MBiFive(\by,\bz) \, .
\end{equation}

\newpage

\section{Calculation of PV integrals}
\label{app:PV}

Here, as defined in \req{eq:yxix}, we adopt the compact notation
$
y = \tfrac{\xi + x}{2\xi}
$, 
$\by = \tfrac{\xi - x}{2\xi}$,
and use
$\otimes \equiv \stackrel{x}{\otimes} = \stackrel{y}{\otimes}$
and $\stackrel{z}{\otimes}$
to denote the convolution integrals defined in \req{eq:otimes}.
We explain here the PV treatment of the
elementary contributions \req{eq:bb}.

Using the standard distribution identity
\begin{equation}
\frac{1}{y \pm \Ieta}
= \PV \frac{1}{y} 
\mp i \pi \, \delta(y) 
\, ,
\label{eq:1y-PV}
\end{equation}
the convolution with a GPD yields
\begin{eqnarray}
\PV \frac{1}{y} \otimes \GPD(x,\xi)
&=&
\int_{-1}^{1} dx \frac{1}{\xi+x}
\left[
\GPD(x,\xi)-\GPD(-\xi,\xi)
\right]
+
\ln \frac{1+\xi}{1-\xi}
\GPD(-\xi,\xi)
\, ,
\label{eq:PVy}
\\[0.3cm]
\delta(y) \otimes \GPD(x,\xi)
&=&
\GPD(-\xi,\xi)
\, .
\label{eq:deltay}
\end{eqnarray}
For mixed denominators of the form $\mixAzy$, with
$\alpha,z\in[0,1]$, we rewrite the propagator in a form suitable for
PV decomposition and using \req{eq:1y-PV} get
\begin{equation}
\frac{1}{\mixAzy  +\Ieta}
\to 
\frac{-1}{y + \alpha \by}
\left(
 \PV \frac{1}{z-\frac{y}{y+\alpha \by}} 
+ i \pi
\, \delta\left( z-\frac{y}{y+\alpha \by} \right) 
\right)
\, .
\label{eq:mix-PV}
\end{equation}
The PV contribution in this case leads to a combined convolution
over $x$ and $z$, naturally split into the DGLAP and ERBL regions:
\begin{eqnarray}
\lefteqn{\phi(z)\,  \stackrel{z}{\otimes} \, f(y,z)\, \PV \frac{1}{z-\frac{y}{y+\alpha \by}} 
\, \otimes \, \GPD(x,\xi)}
\nonumber \\[0.2cm]
&=&
\left[ \int_{-1}^{-\xi} + \int_{\xi}^{1} \right] 
\frac{dx}{2 \xi}
 \;
\GPD(x,\xi)
\int_0^1 dz \, \phi(z)\, f(y,z)\, \frac{1}{z-\frac{y}{y+\alpha \by}} 
\nonumber \\[0.2cm] &&
+ \int_{-\xi}^{\xi} 
\frac{dx}{2 \xi}
\;
\GPD(x,\xi) \,
\int_0^1 dz\,  
\left[ \phi(z)\, f(y,z)
- \phi\left( \frac{y}{y+\alpha \by} \right) \, 
f\left(y,\frac{y}{y+\alpha \by}\right) \right]
\, \frac{1}{z-\frac{y}{y+\alpha \by}} 
\nonumber \\[0.2cm] &&
- \int_{-\xi}^{\xi} 
\frac{dx}{2 \xi}
\;
\GPD(x,\xi) \; \ln \frac{y}{\alpha \by} \;
 \phi\left( \frac{y}{y+\alpha \by} \right) \, 
f\left(y,\frac{y}{y+\alpha \by}\right) 
\, ,
\end{eqnarray}
while the imaginary part resides in ERBL region
\begin{equation}
\phi(z) \, \stackrel{z}{\otimes} \, f(y,z)\, \delta\left( z-\frac{y}{y+\alpha \by} \right) 
\, \otimes \, \GPD(x,\xi)
=
 \int_{-\xi}^{\xi} 
\frac{dx}{2 \xi}
\;
\GPD(x,\xi) \;
 \phi\left( \frac{y}{y+\alpha \by} \right) \, 
f\left(y,\frac{y}{y+\alpha \by}\right) 
\, .
\end{equation}
In order to keep the expressions above as compact as possible,
we use both $y$ and $x$ notation whenever convenient%
\footnote{
Note that from $x=\xi(2y-1)$ follows
$
 \int_{-\xi}^{\xi} 
\frac{dx}{2\xi} =
 \int_{0}^{1} 
dy
\, ,
$
and it is easy to see from
\req{eq:mix-PV}
 that the LO contribution to
the s-channel process
$\gamma \MM_1 \to \gamma \MM_2$ ($0<y<1$)
develops an imaginary part.
In contrast, for the corresponding
t-channel process
$\gamma \gamma \to M_1 M_2$,
the LO contribution is real within the
physical region, while imaginary parts
arise only for values of $y$ outside $[0,1]$
(i.e., $y<0$ or $y>1$),
as in the subprocess of
$\pi N \to \gamma \gamma \NN'$.
This parallels the observations given in
\cite{Qiu:2023mrm,Qiu:2024mny},
namely the enhanced GPD sensitivity in the ERBL region for
$\gamma \NN \to \gamma \MM \NN'$,
and in the DGLAP region for
$\pi N \to \gamma \gamma \NN'$.
} 
interchangably.

\section{Compendium of numerical results}
\label{app:nres}

In this section, we collect the figures
showing the numerical results discussed in Section~\ref{sec:numres}.

\subsection{Fully differential cross-section}
\label{app:fds}

%
\begin{figure}[!h]
\centering
\begin{tabular}{cc}
\includegraphics[width=0.45\textwidth]{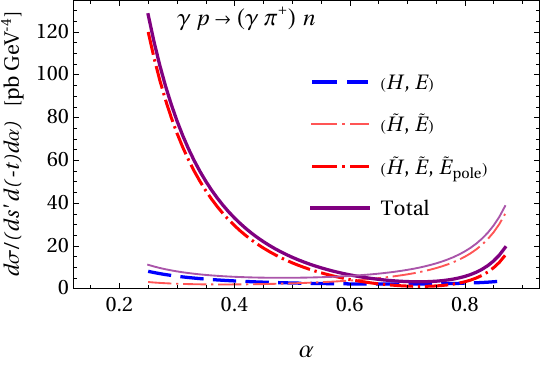}
&
\hspace*{-0.75cm}
\includegraphics[width=0.45\textwidth]{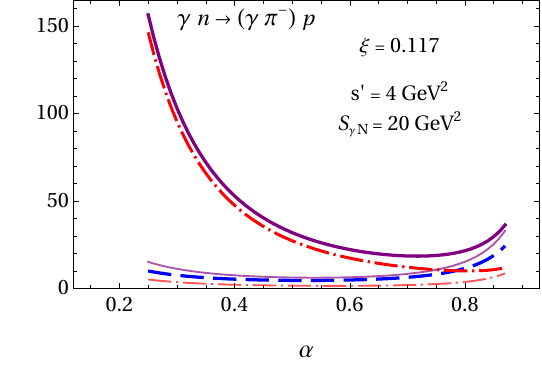}
\\
\includegraphics[width=0.45\textwidth]{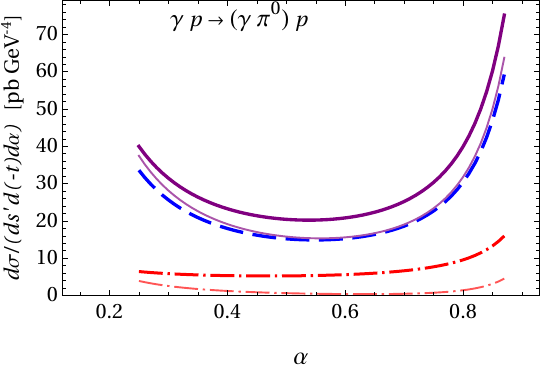}
&
\hspace*{-0.75cm}
\includegraphics[width=0.45\textwidth]{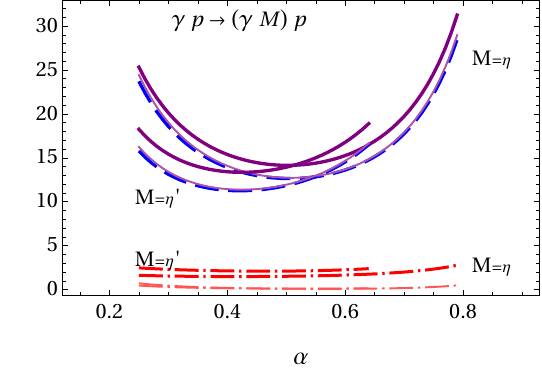}
\end{tabular}
\caption{Fully differential cross-section for $\gamma \NN \to \gamma \MM \, \NN'$,
with $M=\pi^+, \pi^-, \pi^0, \eta, \eta'$.
Dashed and dash-dotted lines correspond to the vector and axial-vector GPD contributions,
respectively, while solid lines show their sum, i.e., the total result.
Thin lines represent axial and total results without the $\pi$-pole contribution.
The kinematics are specified by $S_{\gamma \NN}=20$ GeV$^2$, $s'=4$ GeV$^2$,
yielding $\xi=0.117$, while $(-t)=(-t_0)$ \req{eq:t0}.}
\label{fig:fds-pietas-VA}
\end{figure}
%

\vspace*{0.7cm}
%
\begin{figure}[!h]
\centering
\begin{tabular}{cc}
\includegraphics[width=0.41\textwidth]{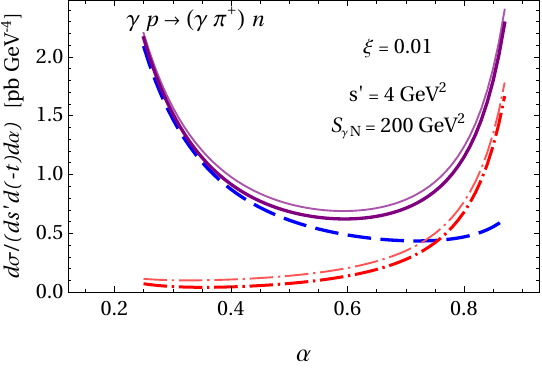}
&
\hspace*{-0.75cm}
\includegraphics[width=0.41\textwidth]{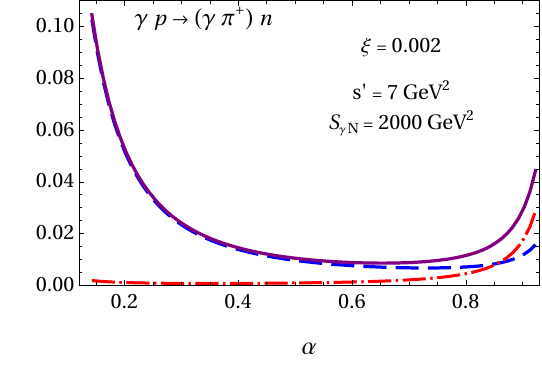}
\end{tabular}
\caption{Illustration of the size of the pion-pole contribution to
fully differential cross-section for $\gamma p \to \gamma \pi^+ n $
at small $\xi$ values. Line designations as 
in Fig. \ref{fig:fds-pietas-VA}, and $(-t)=(-t_0)$.}
\label{fig:fds-pietas-VA-smallxi}
\end{figure}
%
%
\begin{figure}[h]
\centering
\begin{tabular}{cc}
\includegraphics[width=0.45\textwidth]{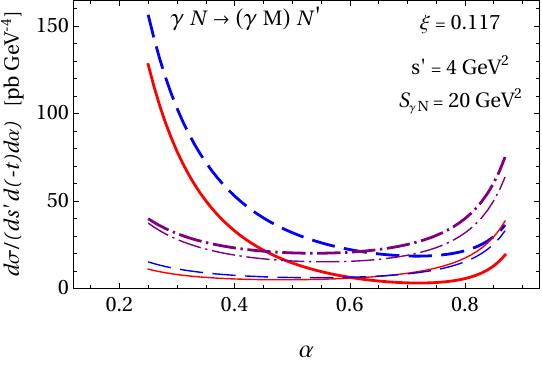}
&
\hspace*{-0.75cm}
\includegraphics[width=0.45\textwidth]{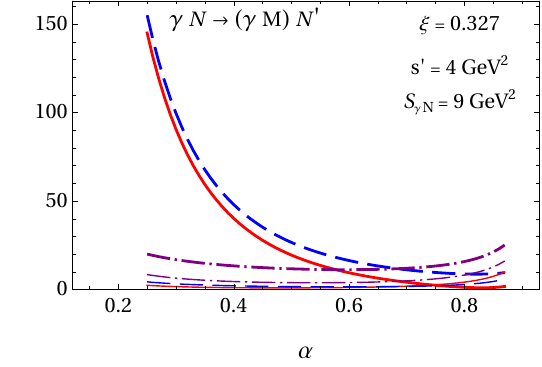}
\\
\includegraphics[width=0.45\textwidth]{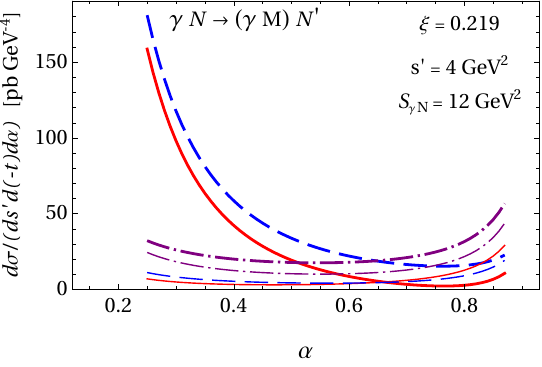}
&
\hspace*{-0.75cm}
\includegraphics[width=0.45\textwidth]{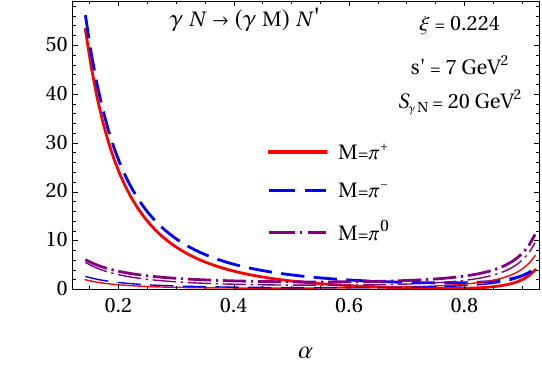}
\end{tabular}
\caption{Fully differential cross-section for $\gamma \NN \to \gamma \pi N'$
at $(-t)=(-t_0)$ and four different kinematics:
($s'=4$, $S_{\gamma \NN}=20$) GeV$^2$ with $\xi=0.117$;
($s'=4$, $S_{\gamma \NN}=12$) GeV$^2$ with $\xi=0.219$;
($s'=7$, $S_{\gamma \NN}=20$) GeV$^2$ with $\xi=0.224$;
and ($s'=4$, $S_{\gamma \NN}=9$) GeV$^2$ with $\xi=0.327$.
Here $\pi^+, \pi^-$, and $\pi^0$ are represented by solid, dashed,
and dot-dashed lines, respectively. Thin lines denote results obtained
without the $\pi$-pole contribution.}
\label{fig:fds-pis-pole}
\end{figure}
%
%
\begin{figure}[!h]
\centering
\begin{tabular}{cc}
\includegraphics[width=0.41\textwidth]{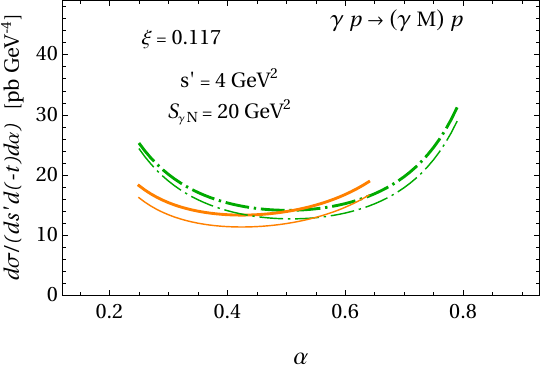}
&
\hspace*{-0.75cm}
\includegraphics[width=0.41\textwidth]{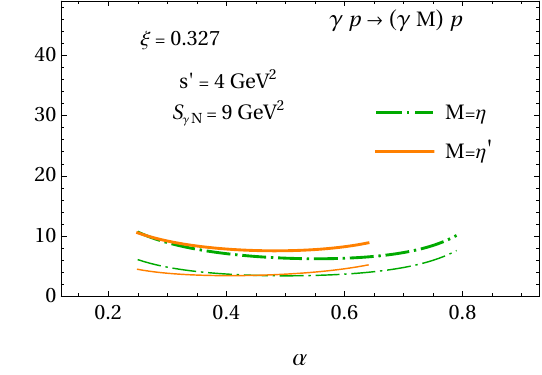}
\\
\includegraphics[width=0.41\textwidth]{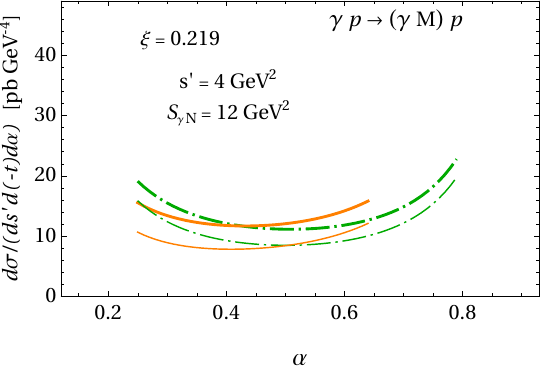}
&
\hspace*{-0.75cm}
\includegraphics[width=0.41\textwidth]{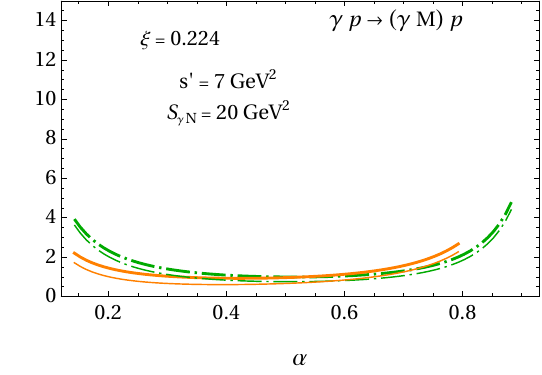}
\end{tabular}
\caption{Fully differential cross-section for 
$\gamma p \to \gamma \eta (\eta') p$
at  $(-t)=(-t_0)$ and four different kinematics:
($s'=4$, $S_{\gamma \NN}=20$) GeV$^2$ with $\xi=0.117$;
($s'=4$, $S_{\gamma \NN}=12$) GeV$^2$ with $\xi=0.219$;
($s'=7$, $S_{\gamma \NN}=20$) GeV$^2$ with $\xi=0.224$;
and ($s'=4$, $S_{\gamma \NN}=9$) GeV$^2$ with $\xi=0.327$.
Here $\eta$, and $\eta'$ are represented by dot-dashed,
and solid lines, respectively. Thin lines denote results obtained
without the $\pi$-pole contribution.
The scales are set to $\mu_R^2=\mu_F^2=s'$.}
\label{fig:fds-etas-pole}
\end{figure}
%
%
\begin{figure}[!h]
\centering
\begin{tabular}{cc}
\includegraphics[width=0.41\textwidth]{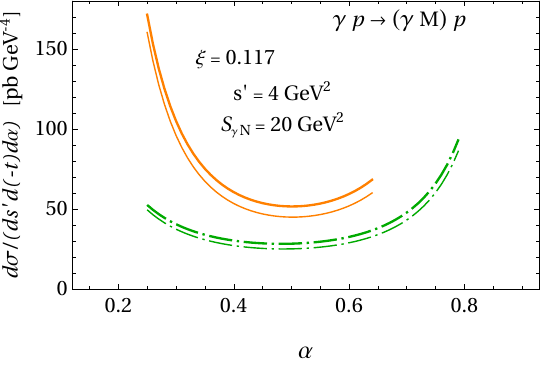}
&
\hspace*{-0.75cm}
\includegraphics[width=0.41\textwidth]{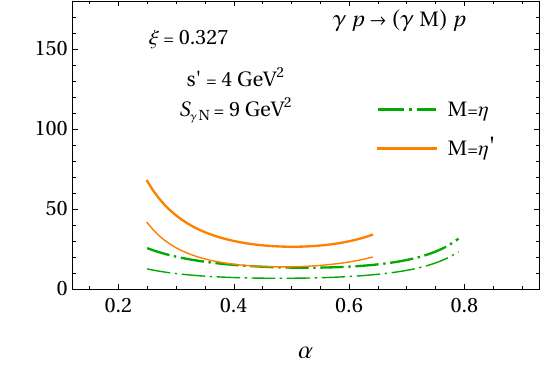}
\\
\includegraphics[width=0.41\textwidth]{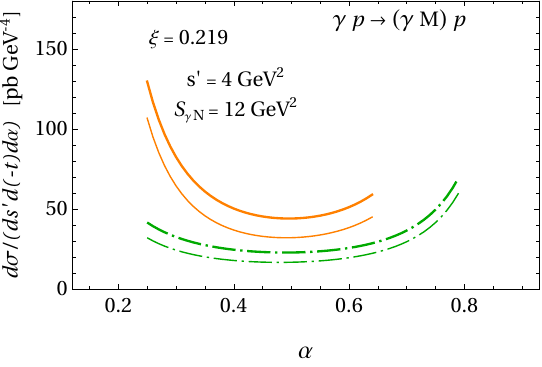}
&
\hspace*{-0.75cm}
\includegraphics[width=0.41\textwidth]{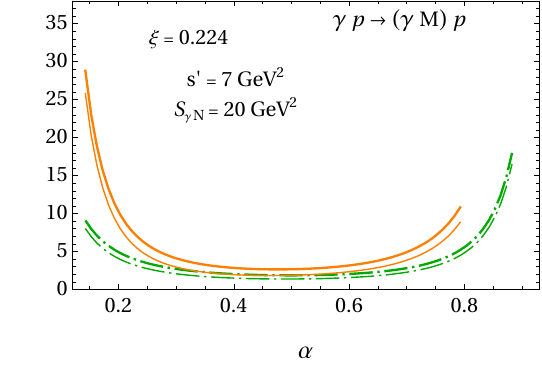}
\end{tabular}
\caption{Fully differential cross-section 
$\gamma p \to \gamma \eta (\eta') p$
at  $(-t)=(-t_0)$ and four different kinematics.
Line designations as in Fig. \ref{fig:fds-etas-pole}.
Here the scales are set to
$\mu_R^2=\mu_F^2=s'\alpha(1-\alpha)$.}
\label{fig:fds-etasmu-pole}
\end{figure}
%
%
\begin{figure}
\centering
\begin{tabular}{cc}
\includegraphics[width=0.45\textwidth]{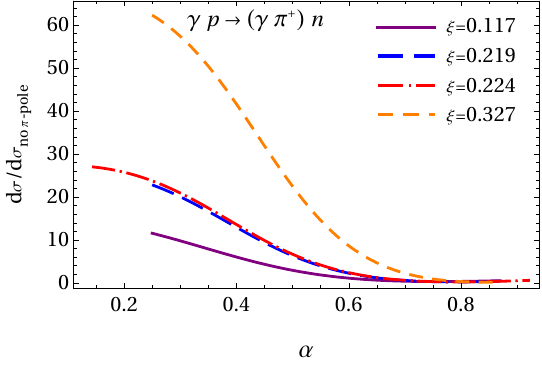}
&
\hspace*{-0.75cm}
\includegraphics[width=0.45\textwidth]{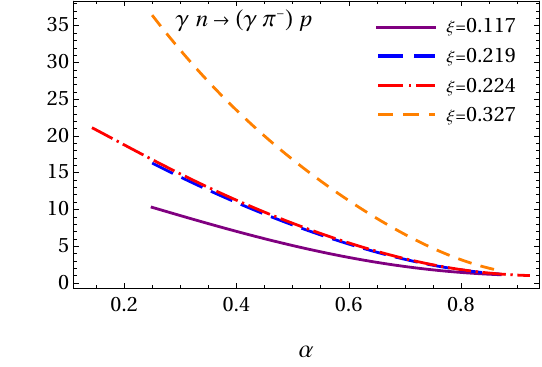}
\\
\includegraphics[width=0.45\textwidth]{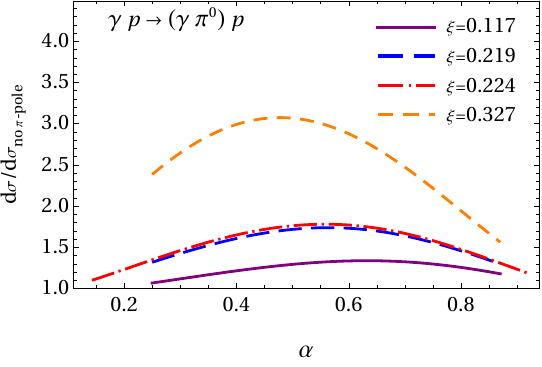}
&
\hspace*{-0.75cm}
\includegraphics[width=0.45\textwidth]{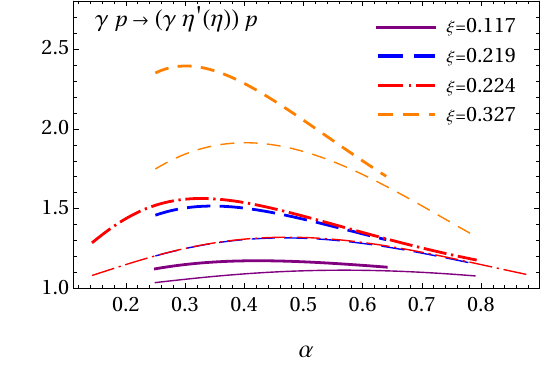}
\end{tabular}
\caption{Ratio of the fully differential cross-section
 \req{eq:diff-CS-2}
for the process $\gamma \NN \to \gamma \MM \, \NN'$
($M=\pi^+, \pi^-, \pi^0, \eta, \eta'$)
obtained with and without the $\pi$-pole contribution.
Different kinematics are compared, specified by
$\xi=0.117$, $0.219$, $0.224$, and $0.327$,
while $(-t)=(-t_0)$.
Thin lines in the bottom-right panel correspond to $\eta$,
while thick lines represent the $\eta'$ results.}
\label{fig:fds-ratio-pole}
\end{figure}
%
%
\begin{figure}
\centering
\begin{tabular}{cc}
\hspace*{-0.3cm}\includegraphics[width=0.45\textwidth]{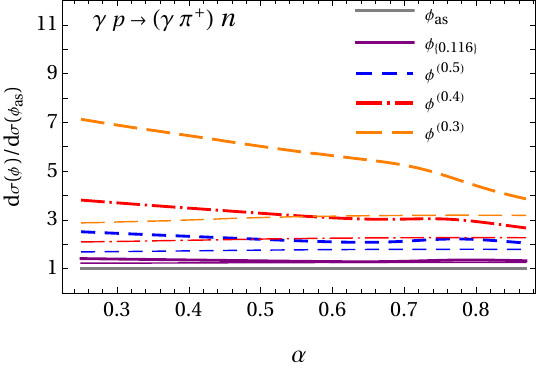}
&
\includegraphics[width=0.45\textwidth]{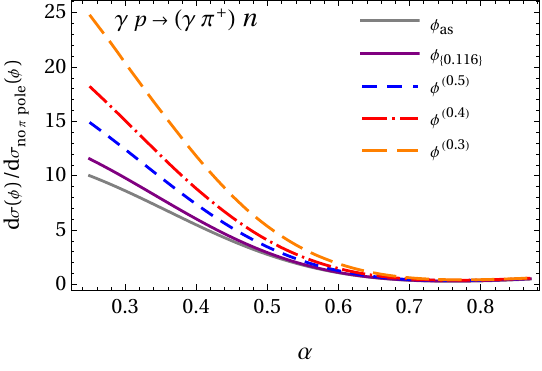}
\end{tabular}
\caption{Left: Ratio of the fully differential cross-section
 \req{eq:diff-CS-2}
for $\gamma p \to \gamma \pi^+ n$, obtained with a given
pion DA and normalized to the result
with the asymptotic DA. Thick and thin lines denote results
with and without the $\pi$-pole contribution, respectively.
Right: Ratio of the full differential cross-section for the
same process, comparing different pion DAs, obtained with
and without the $\pi$-pole contribution. Both figures are
evaluated at $s'=4$ GeV$^2$ and $S_{\gamma \NN}=20$ GeV$^2$,
yielding $\xi=0.117$,
and at $(-t)=(-t_0)$.}
\label{fig:fds-ratio-DAs}
\end{figure}
%
%
\begin{figure}
\centering
\begin{tabular}{cc}
\includegraphics[width=0.45\textwidth]{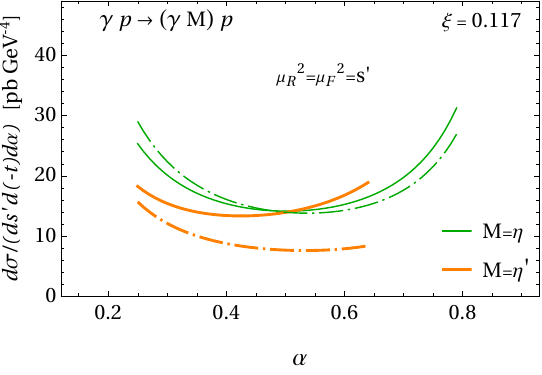}
&
\hspace*{-0.75cm}
\includegraphics[width=0.45\textwidth]{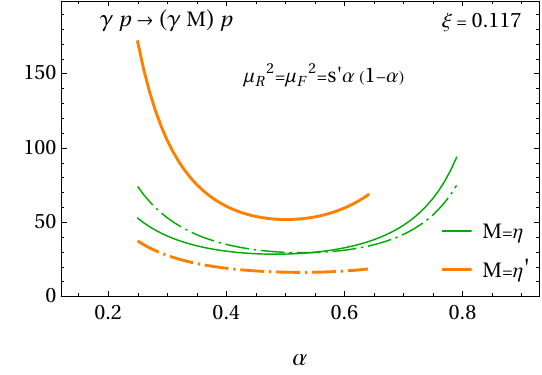}
\end{tabular}
\caption{Fully differential cross-section for $\gamma p \to \gamma \eta (\eta') p'$,
with $\eta$ and $\eta'$ shown by thin and thick lines, respectively.
Dot-dashed lines represent quark-only contributions, while the solid lines 
also include the two-gluon contributions to $\eta$ and $\eta'$. 
In the left figure the renormalization and factorization
scales are set to $\mu_R^2=\mu_F^2=s'$, whereas in the right figure
$\mu_R^2=\mu_F^2=s'\alpha(1-\alpha)$. Both figures are evaluated at
$s'=4$ GeV$^2$ and $S_{\gamma \NN}=20$ GeV$^2$, yielding $\xi=0.117$,
and at $(-t)=(-t_0)$.
}.
\label{fig:fds-etasmu-QG}
\end{figure}
%

\clearpage

\subsection{Single differential cross-section}
\label{app:sds}

%
\begin{figure}[h]
\centering
\begin{tabular}{cc}
\includegraphics[width=0.41\textwidth]{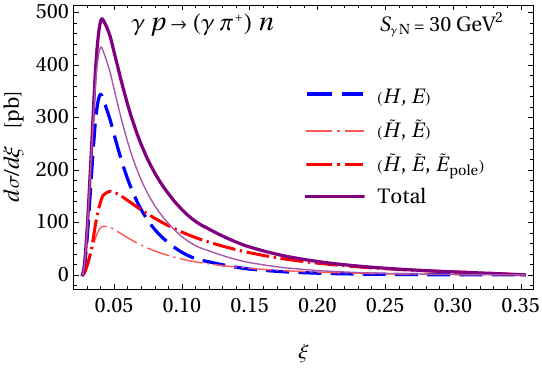}
&
\hspace*{-0.75cm}
\includegraphics[width=0.41\textwidth]{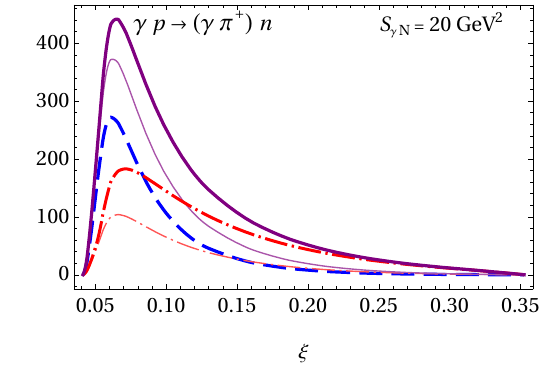}
\\
\includegraphics[width=0.41\textwidth]{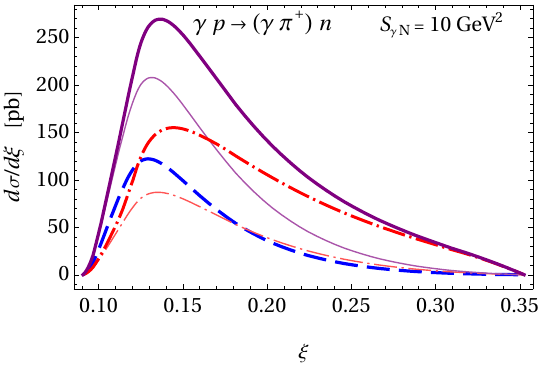}
&
\hspace*{-0.75cm}
\includegraphics[width=0.41\textwidth]{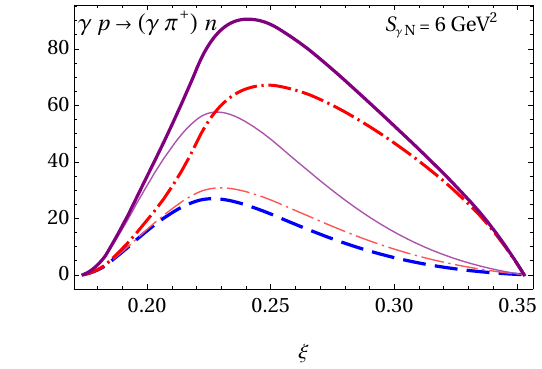}
\end{tabular}
\caption{Single differential cross-section for
$\gamma p \to \gamma \pi^+ n$ at
$S_{\gamma \NN}=30$, $20$, $10$, and $6$ GeV$^2$.
Dashed and dash-dotted lines correspond to the vector
and axial-vector GPD contributions, respectively,
while solid lines show their sum, i.e., the total result.
Thin lines denote axial and total results obtained without the $\pi$-pole contribution.}
\label{fig:sds-Pip-AV}
\end{figure}
%
%
%
%
\begin{figure}[!h]
\centering
\begin{tabular}{cc}
\includegraphics[width=0.41\textwidth]{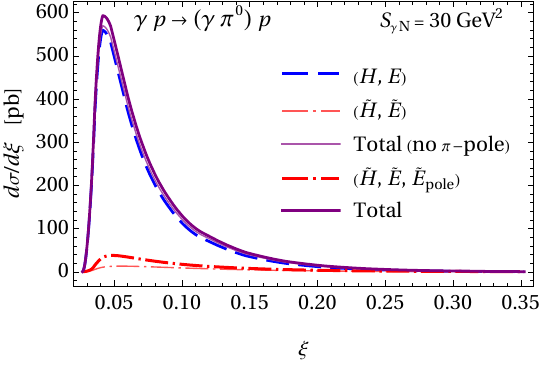}
&
\hspace*{-0.75cm}
\includegraphics[width=0.41\textwidth]{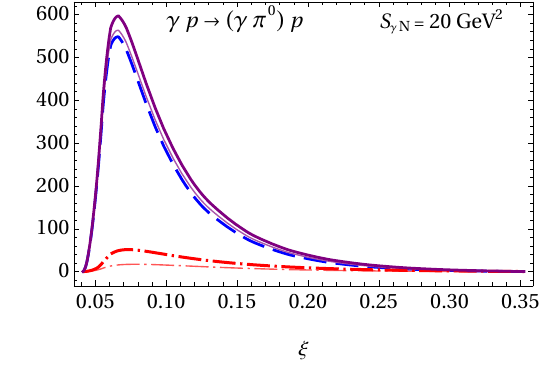}
\\
\includegraphics[width=0.41\textwidth]{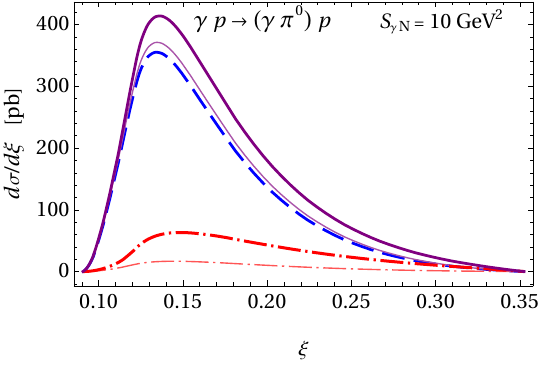}
&
\hspace*{-0.75cm}
\includegraphics[width=0.41\textwidth]{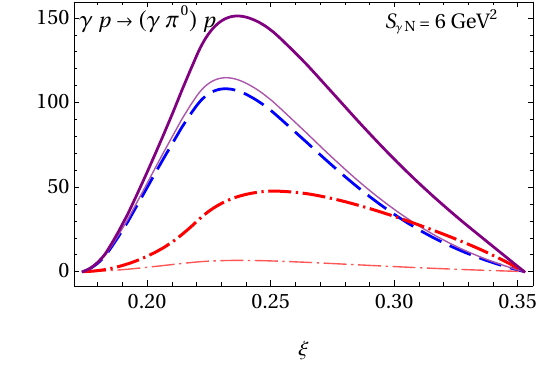}
\end{tabular}
\caption{Single differential cross-section for
$\gamma p \to \gamma \pi^0 p$
at $S_{\gamma \NN}=30$, $20$, $10$, and $6$ GeV$^2$.
Line designations as in Fig.~\ref{fig:sds-Pip-AV}.}
\label{fig:sds-Pi0-AV}
\end{figure}
%

%
\begin{figure}
\centering
\begin{tabular}{cc}
\includegraphics[width=0.41\textwidth]{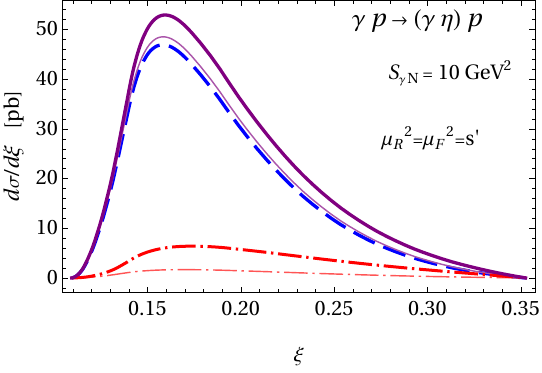}
&
\hspace*{-0.75cm}
\includegraphics[width=0.41\textwidth]{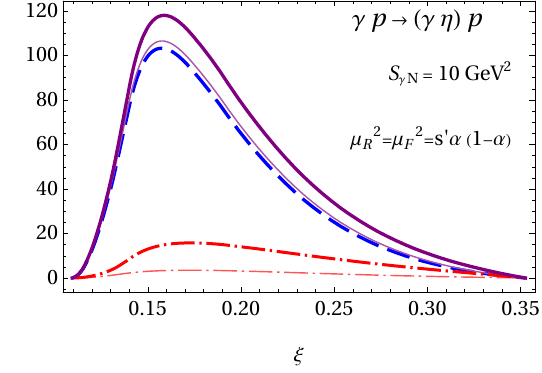}
\\
\includegraphics[width=0.41\textwidth]{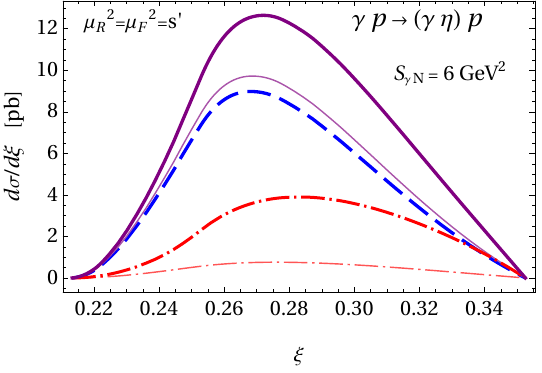}
&
\hspace*{-0.75cm}
\includegraphics[width=0.41\textwidth]{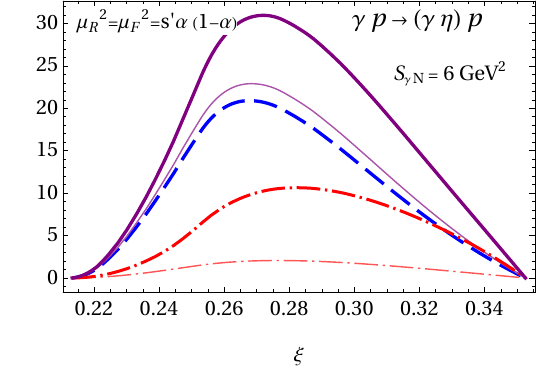}
\end{tabular}
\caption{Single differential cross-section for
$\gamma p \to (\gamma \eta) p$ at
$S_{\gamma \NN}=10$ and $6$ GeV$^2$,
compared for the scale choices
$\mu_R^2=\mu_F^2=s'$ and
$\mu_R^2=\mu_F^2=s'\alpha(1-\alpha)$.
Line designations as in Fig.~\ref{fig:sds-Pip-AV}.
}
\label{fig:sds-Eta-AV}
\end{figure}
%
%
\begin{figure}[!h]
\centering
\begin{tabular}{cc}
\includegraphics[width=0.41\textwidth]{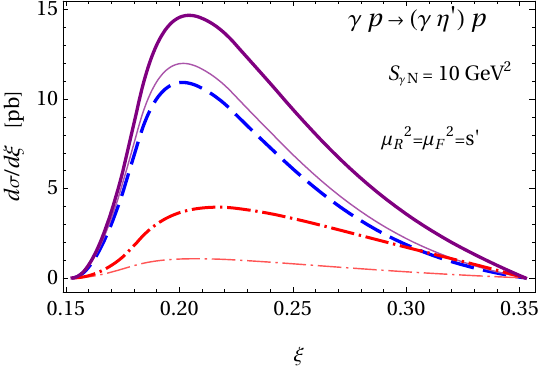}
&
\hspace*{-0.75cm}
\includegraphics[width=0.41\textwidth]{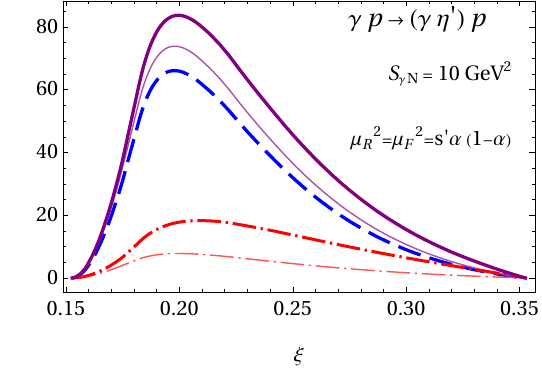}
\\
\includegraphics[width=0.41\textwidth]{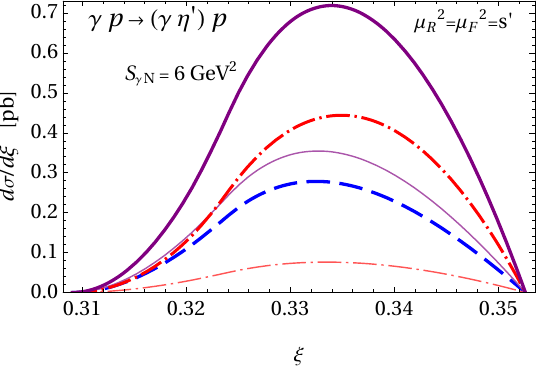}
&
\hspace*{-0.75cm}
\includegraphics[width=0.41\textwidth]{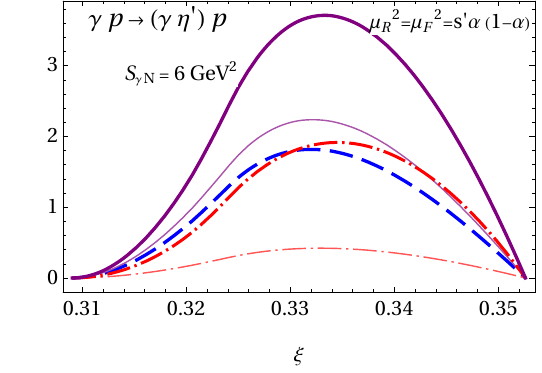}
\end{tabular}
\caption{Single differential cross-section for
$\gamma p \to (\gamma \eta') p$ at
$S_{\gamma \NN}=10$ and $6$ GeV$^2$,
compared for the scale choices
$\mu_R^2=\mu_F^2=s'$ and 
$\mu_R^2=\mu_F^2=s'\alpha(1-\alpha)$.
Line designations as in Fig.~\ref{fig:sds-Pip-AV}.
}
\label{fig:sds-Etap-AV}
\end{figure}
%
%
%
\begin{figure}
\centering
\begin{tabular}{cc}
\includegraphics[width=0.41\textwidth]{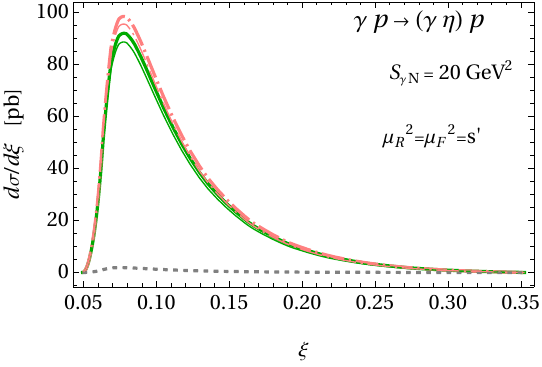}
&
\hspace*{-0.75cm}
\includegraphics[width=0.41\textwidth]{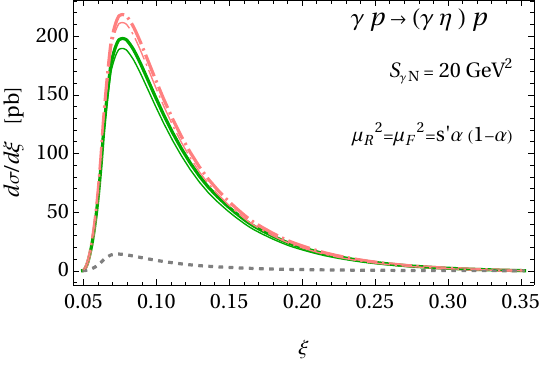}
\\
\includegraphics[width=0.41\textwidth]{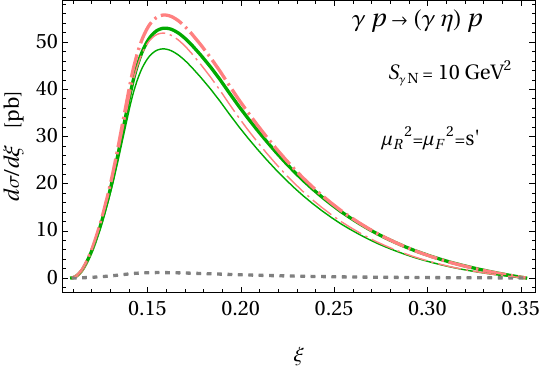}
&
\includegraphics[width=0.41\textwidth]{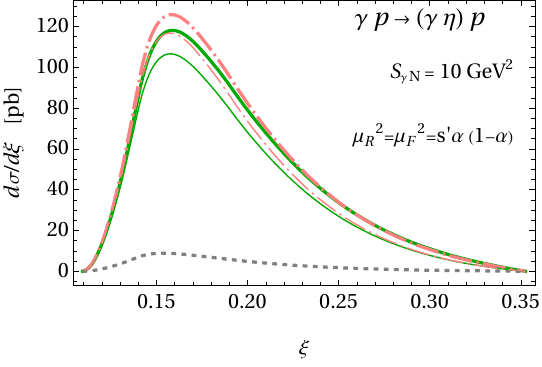}
\\
\end{tabular}
\caption{Single differential cross-section for
$\gamma p \to (\gamma \eta) p$ at
$S_{\gamma \NN}=20$, and $10$ GeV$^2$,
compared for the scale choices
$\mu_R^2=\mu_F^2=s'$ and
$\mu_R^2=\mu_F^2=s'\alpha(1-\alpha)$.
Dot-dashed lines represent quark-only contributions, 
while the total contributions, denoted by solid lines also include 
the two-gluon contributions to $\eta$. 
For comparison, dotted lines correspond to gluon-only contributions.
Thin lines represent results obtained without the $\pi$-pole contribution.
}
\label{fig:sds-EtaMu-QG}
\end{figure}
%
%
\begin{figure}[p]
\centering
\begin{tabular}{cc}
\includegraphics[width=0.41\textwidth]{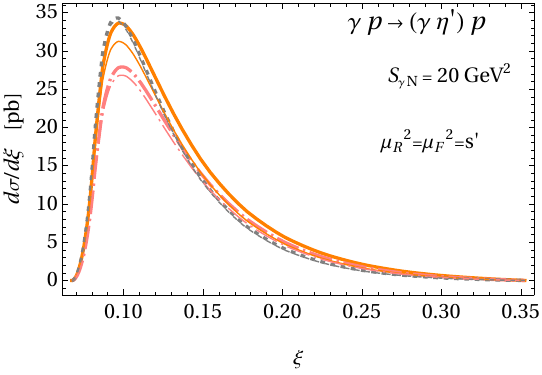}
&
\includegraphics[width=0.41\textwidth]{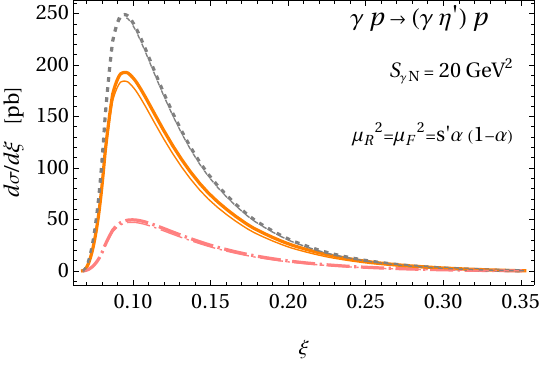}
\\
\includegraphics[width=0.41\textwidth]{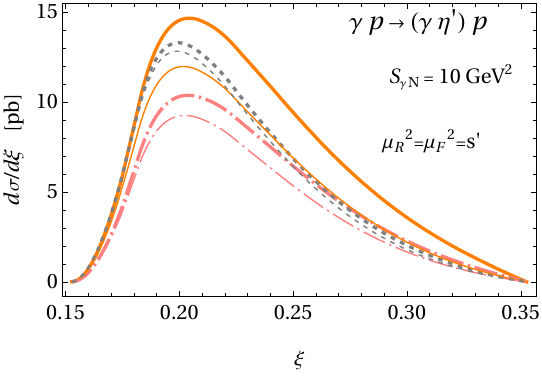}
&
\includegraphics[width=0.41\textwidth]{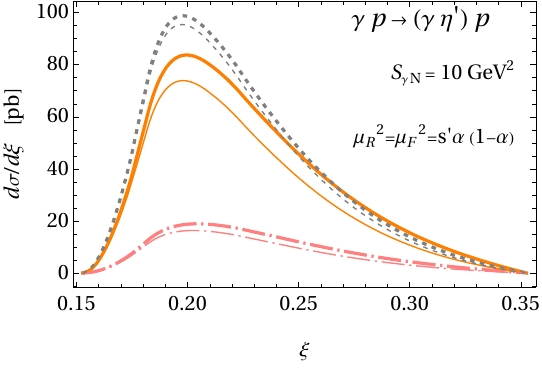}
\end{tabular}
\caption{Single differential cross-section for
$\gamma p \to (\gamma \eta') p$ at
$S_{\gamma \NN}= 20$ and $10$ GeV$^2$,
compared for the scale choices
$\mu_R^2=\mu_F^2=s'$ and
$\mu_R^2=\mu_F^2=s'\alpha(1-\alpha)$.
Line designations as in Fig.~\ref{fig:sds-EtaMu-QG}.
}
\label{fig:sds-EtapMu-QG}
\end{figure}
%
\clearpage
%
\begin{figure}[h!]
\centering
\begin{tabular}{cc}
\includegraphics[width=0.45\textwidth]{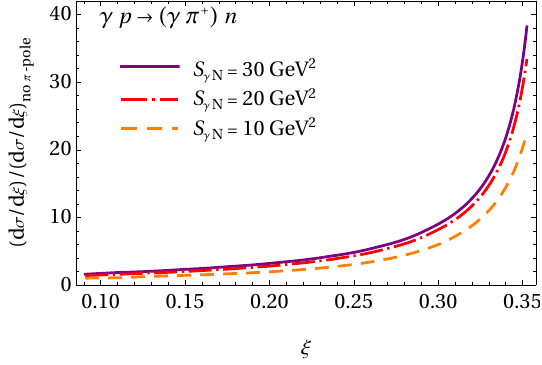}
&
\hspace*{-0.75cm}
\includegraphics[width=0.45\textwidth]{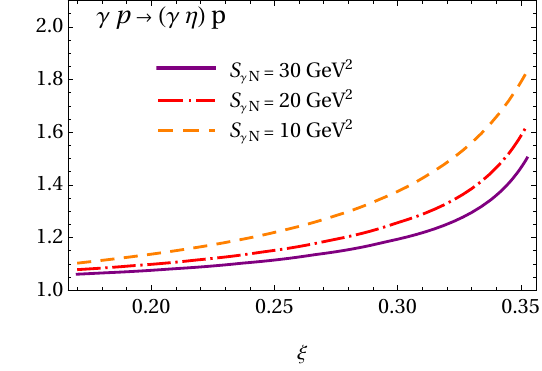}
\end{tabular}
\caption{Ratio of the single differential cross-section
\req{eq:phspaceint}
calculated with and without the $\pi$-pole contribution.
Results at different kinematical settings are compared,
specified by
$S_{\gamma\NN} = 30$, $20$, and $10$~GeV$^2$.
Left: $\gamma p \to \gamma \pi^+ n$.
Right: $\gamma p \to \gamma \eta p$.
}
\label{fig:sds-ratio-pole}
\end{figure}
%
%

\subsection{Integrated cross-section}
\label{app:sig}

%
\begin{figure}[h]
\centering
\begin{tabular}{cc}
\includegraphics[width=0.48\textwidth]{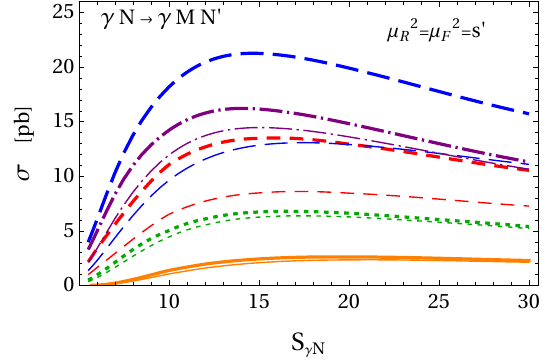}
&
\hspace*{-0.75cm}
\includegraphics[width=0.48\textwidth]{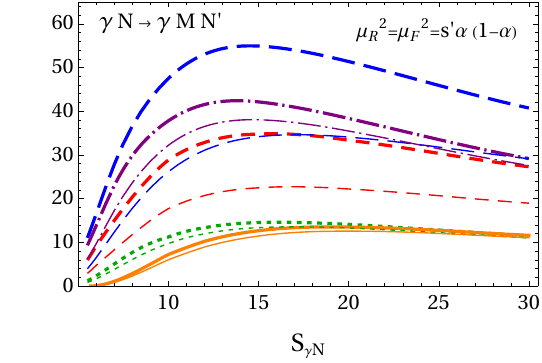}
\end{tabular}
\caption{Cross- section for $\gamma \NN \to \gamma \MM \, \NN'$,
with $M=\pi^+, \pi^-, \pi^0, \eta$, and $\eta'$.
Thick and thin lines denote results with and without
the $\pi$-pole contribution, respectively.
In the left figure the renormalization and factorization
scales are set to $\mu_R^2=\mu_F^2=s'$, while in the right
figure $\mu_R^2=\mu_F^2=s'\alpha(1-\alpha)$.
Line designations as in Fig. \ref{fig:sig-ratio-pole}.}
\label{fig:sig-PiEtas-mu}
\end{figure}
%
%
\begin{figure}[!h]
\centering
\begin{tabular}{cc}
\includegraphics[width=0.5\textwidth]{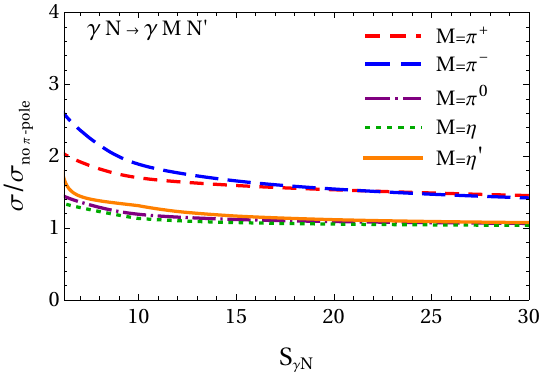}
\end{tabular}
\caption{Ratio of the cross-sections
\req{eq:totalsigma}
for $\gamma \NN \to \gamma \MM \NN'$,
with $\MM = \pi^+, \pi^-, \pi^0, \eta$, and $\eta'$,
calculated with and without the $\pi$-pole contribution.}
\label{fig:sig-ratio-pole}
\end{figure}
%
%
\begin{figure}
\centering
\begin{tabular}{cc}
\includegraphics[width=0.45\textwidth]{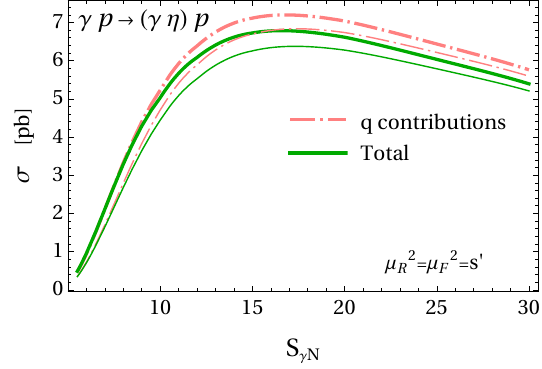}
&
\hspace*{-0.75cm}
\includegraphics[width=0.45\textwidth]{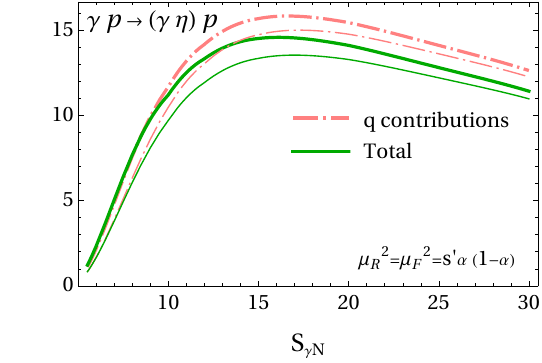}
\\
\includegraphics[width=0.45\textwidth]{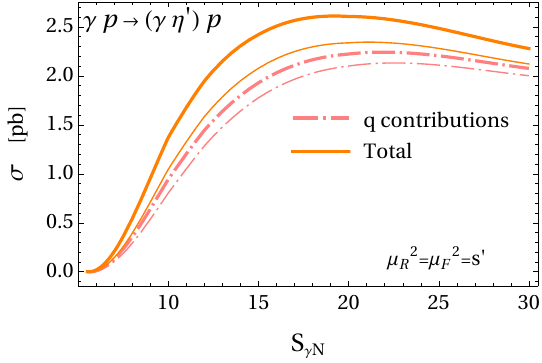}
&
\hspace*{-0.75cm}
\includegraphics[width=0.45\textwidth]{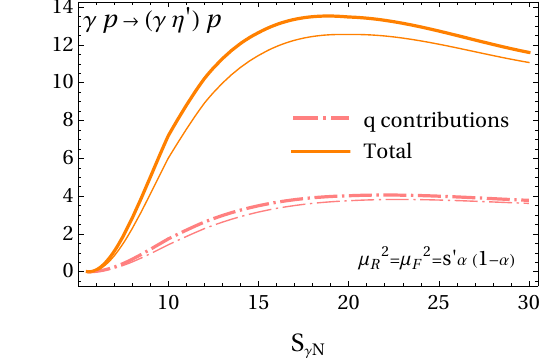}
\end{tabular}
\caption{Cross-section for $\gamma p \to \gamma \eta (\eta') p$.
Thick and thin lines denote results with and without
the $\pi$-pole contribution, respectively.
Dot-dashed lines represent quark-only contributions, 
while the total contributions, denoted by solid lines also include 
the two-gluon contributions to $\eta$ and $\eta'$. 
In the left column
the renormalization and factorization scales are set to
$\mu_R^2=\mu_F^2=s'$, whereas in the right column
$\mu_R^2=\mu_F^2=s'\alpha(1-\alpha)$.}
\label{fig:sig-EtasMu-QG}
\end{figure}
%

\clearpage

\bibliographystyle{JHEP-2}                                                                                
\bibliography{ref-gM1gM2}

@article{Ji:1998xh,
    author = "Ji, Xiang-Dong and Osborne, Jonathan",
    title = "{One loop corrections and all order factorization in deeply virtual Compton scattering}",
    eprint = "hep-ph/9801260",
    archivePrefix = "arXiv",
    reportNumber = "UMD-PP-98-074, DOE-ER-40762-139",
    doi = "10.1103/PhysRevD.58.094018",
    journal = "Phys. Rev. D",
    volume = "58",
    pages = "094018",
    year = "1998"
}

@article{Collins:1998be,
    author = "Collins, John C. and Freund, Andreas",
    title = "{Proof of factorization for deeply virtual Compton scattering in QCD}",
    eprint = "hep-ph/9801262",
    archivePrefix = "arXiv",
    reportNumber = "PSU-TH-192",
    doi = "10.1103/PhysRevD.59.074009",
    journal = "Phys. Rev. D",
    volume = "59",
    pages = "074009",
    year = "1999"
}

@article{Collins:1996fb,
    author = "Collins, John C. and Frankfurt, Leonid and Strikman, Mark",
    title = "{Factorization for hard exclusive electroproduction of mesons in QCD}",
    eprint = "hep-ph/9611433",
    archivePrefix = "arXiv",
    reportNumber = "CERN-TH-96-314, PSU-TH-168",
    doi = "10.1103/PhysRevD.56.2982",
    journal = "Phys. Rev. D",
    volume = "56",
    pages = "2982--3006",
    year = "1997"
}

@article{Diehl:2003ny,
    author = "Diehl, M.",
    title = "{Generalized parton distributions}",
    eprint = "hep-ph/0307382",
    archivePrefix = "arXiv",
    reportNumber = "DESY-THESIS-2003-018",
    doi = "10.1016/j.physrep.2003.08.002",
    journal = "Phys. Rept.",
    volume = "388",
    pages = "41--277",
    year = "2003"
}

@article{dHose:2016mda,
    author = "d'Hose, Nicole and Niccolai, Silvia and Rostomyan, Armine",
    title = "{Experimental overview of Deeply Virtual Compton Scattering}",
    doi = "10.1140/epja/i2016-16151-9",
    journal = "Eur. Phys. J.",
    volume = "A52",
    number = "6",
    pages = "151",
    year = "2016",
    group  = "GPD",
    comments  = "Experimental review - partner of our DVCS review in EPJA."
}

@article{Kumericki:2016ehc,                                                                                        author         = "Kumerički, K. and Liuti, S. and Moutarde, H.",                                             title          = "{GPD phenomenology and DVCS fitting}",                                                     journal        = "Eur. Phys. J.",                                                                            volume         = "A52",                                                                                      year           = "2016",                                                                                     number         = "6",                                                                                        pages          = "157",                                                                                
      doi            = "10.1140/epja/i2016-16157-3",
      eprint         = "1602.02763",
      archivePrefix  = "arXiv",
      primaryClass   = "hep-ph",
      SLACcitation   = "%%CITATION = ARXIV:1602.02763;%%",
      recid  = "1420559",
      group  = "GPD",
      comments  = ""
}

@article{Favart:2015umi,
    author = "Favart, L. and Guidal, M. and Horn, T. and Kroll, P.",
    title = "{Deeply Virtual Meson Production on the nucleon}",
    eprint = "1511.04535",
    archivePrefix = "arXiv",
    primaryClass = "hep-ph",
    doi = "10.1140/epja/i2016-16158-2",
    journal = "Eur. Phys. J. A",
    volume = "52",
    number = "6",
    pages = "158",
    year = "2016",
      group  = "GPD",
      group  = "DVMP",
      comments  = "EPJA review."
}

@article{Baier:1982vlv,
    author = "Baier, V. N. and Grozin, A. G.",
    title = "{Certain exclusive processes in QCD taking into account two-gluon states}",
    journal = "Sov. J. Nucl. Phys.",
    volume = "35",
    pages = "596--605",
    year = "1982"
}

@article{Baier:1985pd,
    author = "Baier, V. N. and Grozin, A. G.",
    title = "{Hard hadron processes in QCD}",
    journal = "Fiz. Elem. Chast. Atom. Yadra",
    volume = "16",
    pages = "5--58",
    year = "1985"
}

@article{Baier:1985wv,
    author = "Baier, V. N. and Grozin, A. G.",
    title = "{Gluonic contributions to the exclusive amplitudes}",
    doi = "10.1007/BF01571399",
    journal = "Z. Phys. C",
    volume = "29",
    pages = "161--165",
    year = "1985"
}

@article{Brodsky:1981rp,
    author = "Brodsky, Stanley J. and Lepage, G. Peter",
    title = "{Large Angle Two Photon Exclusive Channels in Quantum Chromodynamics}",
    reportNumber = "SLAC-PUB-2733",
    doi = "10.1103/PhysRevD.24.1808",
    journal = "Phys. Rev. D",
    volume = "24",
    pages = "1808",
    year = "1981"
}

@article{Nizic:1987sw,
    author = "Nižić, B.",
    title = "{Beyond leading order perturbative QCD corrections to gamma gamma ---{\ensuremath{>}} M+ M- (M = pi, K)}",
    doi = "10.1103/PhysRevD.35.80",
    journal = "Phys. Rev. D",
    volume = "35",
    pages = "80--101",
    year = "1987"
}

@article{Duplancic:2006nv,
    author = "Duplančić, Goran and Nižić, Bene",
    title = "{NLO perturbative QCD predictions for gamma gamma ---{\ensuremath{>}} M+ M- (M = pi, K)}",
    eprint = "hep-ph/0607069",
    archivePrefix = "arXiv",
    reportNumber = "IRB-TH-2-06",
    doi = "10.1103/PhysRevLett.97.142003",
    journal = "Phys. Rev. Lett.",
    volume = "97",
    pages = "142003",
    year = "2006"
}

@article{Benayoun:1989ng,
    author = "Benayoun, Maurice and Chernyak, V. L.",
    title = "{SU(3) Symmetry Breaking Effects in $\gamma \gamma \to$ Two Mesons Processes}",
    reportNumber = "LPC-89-01",
    doi = "10.1016/0550-3213(90)90143-2",
    journal = "Nucl. Phys. B",
    volume = "329",
    pages = "285--309",
    year = "1990"
}

@article{Boussarie:2016qop,
    author = "Boussarie, R. and Pire, B. and Szymanowski, L. and Wallon, S.",
    title = "{Exclusive photoproduction of a $\gamma\,\rho$ pair with a large invariant mass}",
    eprint = "1609.03830",
    archivePrefix = "arXiv",
    primaryClass = "hep-ph",
    reportNumber = "LPT-ORSAY-16-58, CPHT-RR038.072016, LPT-Orsay-16-58",
    doi = "10.1007/JHEP02(2017)054",
    journal = "JHEP",
    volume = "02",
    pages = "054",
    year = "2017",
    note = "[Erratum: JHEP 10, 029 (2018)]"
}

@article{Duplancic:2018bum,
    author = "Duplan{\v{c}}i{\'c}, G. and Passek-Kumeri{\v{c}}ki, K. and Pire, B. and Szymanowski, L. and Wallon, S.",
    title = "{Probing axial quark generalized parton distributions through exclusive photoproduction of a $\gamma\,\pi^\pm$ pair with a large invariant mass}",
    eprint = "1809.08104",
    archivePrefix = "arXiv",
    primaryClass = "hep-ph",
    reportNumber = "LPT-Orsay-18-84, CPHT-RR096.092018",
    doi = "10.1007/JHEP11(2018)179",
    journal = "JHEP",
    volume = "11",
    pages = "179",
    year = "2018"
}

@article{Duplancic:2022ffo,
    author = "Duplan{\v{c}}i{\'c}, Goran and Nabeebaccus, Saad and Passek-Kumeri{\v{c}}ki, Kornelija and Pire, Bernard and Szymanowski, Lech and Wallon, Samuel",
    title = "{Accessing chiral-even quark generalised parton distributions in the exclusive photoproduction of a $ \gamma \pi ^{\pm} $ pair with large invariant mass in both fixed-target and collider experiments}",
    eprint = "2212.00655",
    archivePrefix = "arXiv",
    primaryClass = "hep-ph",
    doi = "10.1007/JHEP03(2023)241",
    journal = "JHEP",
    volume = "03",
    pages = "241",
    year = "2023"
}

@article{Duplancic:2023kwe,
    author = "Duplan{\v{c}}i{\'c}, Goran and Nabeebaccus, Saad and Passek-Kumeri{\v{c}}ki, Kornelija and Pire, Bernard and Szymanowski, Lech and Wallon, Samuel",
    title = "{Probing chiral-even and chiral-odd leading twist quark generalized parton distributions through the exclusive photoproduction of a {\ensuremath{\gamma}}{\ensuremath{\rho}} pair}",
    eprint = "2302.12026",
    archivePrefix = "arXiv",
    primaryClass = "hep-ph",
    doi = "10.1103/PhysRevD.107.094023",
    journal = "Phys. Rev. D",
    volume = "107",
    number = "9",
    pages = "094023",
    year = "2023"
}

@article{Nabeebaccus:2023rzr,
    author = "Nabeebaccus, Saad and Schoenleber, Jakob and Szymanowski, Lech and Wallon, Samuel",
    title = "{Breakdown of collinear factorization in the exclusive photoproduction of a {\ensuremath{\pi}}0{\ensuremath{\gamma}} pair with large invariant mass}",
    eprint = "2311.09146",
    archivePrefix = "arXiv",
    primaryClass = "hep-ph",
    doi = "10.1103/PhysRevD.111.034040",
    journal = "Phys. Rev. D",
    volume = "111",
    number = "3",
    pages = "034040",
    year = "2025"
}

@article{Nabeebaccus:2024mia,
    author = "Nabeebaccus, Saad and Schoenleber, Jakob and Szymanowski, Lech and Wallon, Samuel",
    title = "{Demonstration of collinear factorization breaking due to collinear-to-soft Glauber exchanges for a 2{\textrightarrow}3 exclusive process at leading twist}",
    eprint = "2409.16067",
    archivePrefix = "arXiv",
    primaryClass = "hep-ph",
    doi = "10.1103/PhysRevD.111.L091502",
    journal = "Phys. Rev. D",
    volume = "111",
    number = "9",
    pages = "L091502",
    year = "2025"
}

@article{Qiu:2022bpq,
    author = "Qiu, Jian-Wei and Yu, Zhite",
    title = "{Exclusive production of a pair of high transverse momentum photons in pion-nucleon collisions for extracting generalized parton distributions}",
    eprint = "2205.07846",
    archivePrefix = "arXiv",
    primaryClass = "hep-ph",
    reportNumber = "JLAB-THY-22-3617, MSUHEP-22-018",
    doi = "10.1007/JHEP08(2022)103",
    journal = "JHEP",
    volume = "08",
    pages = "103",
    year = "2022"
}

@article{Qiu:2022pla,
    author = "Qiu, Jian-Wei and Yu, Zhite",
    title = "{Single diffractive hard exclusive processes for the study of generalized parton distributions}",
    eprint = "2210.07995",
    archivePrefix = "arXiv",
    primaryClass = "hep-ph",
    reportNumber = "MSUHEP-22-032, JLAB-THY-22-3742, JLAB-THY-22-3742, MSUHEP-22-032",
    doi = "10.1103/PhysRevD.107.014007",
    journal = "Phys. Rev. D",
    volume = "107",
    number = "1",
    pages = "014007",
    year = "2023"
}

@article{Qiu:2023mrm,
    author = "Qiu, Jian-Wei and Yu, Zhite",
    title = "{Extraction of the Parton Momentum-Fraction Dependence of Generalized Parton Distributions from Exclusive Photoproduction}",
    eprint = "2305.15397",
    archivePrefix = "arXiv",
    primaryClass = "hep-ph",
    reportNumber = "JLAB-THY-23-3828, JLAB-THY-23-3828, MSUHEP-23-015",
    doi = "10.1103/PhysRevLett.131.161902",
    journal = "Phys. Rev. Lett.",
    volume = "131",
    number = "16",
    pages = "161902",
    year = "2023"
}

@article{Qiu:2024mny,
    author = "Qiu, Jian-Wei and Yu, Zhite",
    title = "{Extracting transition generalized parton distributions from hard exclusive pion-nucleon scattering}",
    eprint = "2401.13207",
    archivePrefix = "arXiv",
    primaryClass = "hep-ph",
    reportNumber = "JLAB-THY-24-3985, MSUHEP-24-001",
    doi = "10.1103/PhysRevD.109.074023",
    journal = "Phys. Rev. D",
    volume = "109",
    number = "7",
    pages = "074023",
    year = "2024"
}

@article{Siddikov:2024blb,
    author = "Siddikov, Marat",
    title = "{Exclusive photoproduction of {\ensuremath{\eta}}c{\ensuremath{\gamma}} pairs with large invariant mass}",
    eprint = "2408.01822",
    archivePrefix = "arXiv",
    primaryClass = "hep-ph",
    doi = "10.1103/PhysRevD.110.056043",
    journal = "Phys. Rev. D",
    volume = "110",
    number = "5",
    pages = "056043",
    year = "2024"
}

@article{Ivanov:2002jj,
    author = "Ivanov, D. Yu. and Pire, B. and Szymanowski, L. and Teryaev, O. V.",
    title = "{Probing chiral odd GPD's in diffractive electroproduction of two vector mesons}",
    eprint = "hep-ph/0209300",
    archivePrefix = "arXiv",
    reportNumber = "CPHT-RR-059-0602",
    doi = "10.1016/S0370-2693(02)02856-3",
    journal = "Phys. Lett. B",
    volume = "550",
    pages = "65--76",
    year = "2002"
}

@article{Enberg:2006he,
    author = "Enberg, R. and Pire, B. and Szymanowski, L.",
    title = "{Transversity GPD in photo- and electroproduction of two vector mesons}",
    eprint = "hep-ph/0601138",
    archivePrefix = "arXiv",
    reportNumber = "CPHT-RR005-0106, LBNL-59084",
    doi = "10.1140/epjc/s2006-02545-3",
    journal = "Eur. Phys. J. C",
    volume = "47",
    pages = "87--94",
    year = "2006"
}

@article{ElBeiyad:2010pji,
    author = "El Beiyad, M. and Pire, B. and Segond, M. and Szymanowski, L. and Wallon, S.",
    title = "{Photoproduction of a pi rhoT pair with a large invariant mass and transversity generalized parton distribution}",
    eprint = "1001.4491",
    archivePrefix = "arXiv",
    primaryClass = "hep-ph",
    reportNumber = "CPHT-RR001.0110, LPT-10-07",
    doi = "10.1016/j.physletb.2010.02.086",
    journal = "Phys. Lett. B",
    volume = "688",
    pages = "154--167",
    year = "2010"
}

@article{Siddikov:2022bku,
    author = "Siddikov, Marat and Schmidt, Ivan",
    title = "{Exclusive production of quarkonia pairs in collinear factorization framework}",
    eprint = "2212.14019",
    archivePrefix = "arXiv",
    primaryClass = "hep-ph",
    doi = "10.1103/PhysRevD.107.034037",
    journal = "Phys. Rev. D",
    volume = "107",
    number = "3",
    pages = "034037",
    year = "2023"
}

@article{Pedrak:2017cpp,
    author = "Pedrak, A. and Pire, B. and Szymanowski, L. and Wagner, J.",
    title = "{Hard photoproduction of a diphoton with a large invariant mass}",
    eprint = "1708.01043",
    archivePrefix = "arXiv",
    primaryClass = "hep-ph",
    reportNumber = "CPHT-RR-047-082017",
    doi = "10.1103/PhysRevD.96.074008",
    journal = "Phys. Rev. D",
    volume = "96",
    number = "7",
    pages = "074008",
    year = "2017",
    note = "[Erratum: Phys.Rev.D 100, 039901 (2019)]"
}

@article{Pedrak:2020mfm,
    author = "Pedrak, A. and Pire, B. and Szymanowski, L. and Wagner, J.",
    title = "{Electroproduction of a large invariant mass photon pair}",
    eprint = "2003.03263",
    archivePrefix = "arXiv",
    primaryClass = "hep-ph",
    reportNumber = "CPHT-RR010.022020",
    doi = "10.1103/PhysRevD.101.114027",
    journal = "Phys. Rev. D",
    volume = "101",
    number = "11",
    pages = "114027",
    year = "2020"
}

@article{Grocholski:2021man,
    author = "Grocholski, Oskar and Pire, Bernard and Sznajder, Pawe{\l} and Szymanowski, Lech and Wagner, Jakub",
    title = "{Collinear factorization of diphoton photoproduction at next to leading order}",
    eprint = "2110.00048",
    archivePrefix = "arXiv",
    primaryClass = "hep-ph",
    reportNumber = "CPHT-RR077.092021",
    doi = "10.1103/PhysRevD.104.114006",
    journal = "Phys. Rev. D",
    volume = "104",
    number = "11",
    pages = "114006",
    year = "2021"
}

@article{Grocholski:2022rqj,
    author = "Grocholski, Oskar and Pire, Bernard and Sznajder, Pawe{\l} and Szymanowski, Lech and Wagner, Jakub",
    title = "{Phenomenology of diphoton photoproduction at next-to-leading order}",
    eprint = "2204.00396",
    archivePrefix = "arXiv",
    primaryClass = "hep-ph",
    reportNumber = "CPHT-RR020.032022, DESY-22-061",
    doi = "10.1103/PhysRevD.105.094025",
    journal = "Phys. Rev. D",
    volume = "105",
    number = "9",
    pages = "094025",
    year = "2022"
}

@article{Kroll:2021zss,
    author = "Kroll, Peter and Passek-Kumeri{\v{c}}ki, Kornelija",
    title = "{Wide-angle photoproduction of the {\ensuremath{\eta}}'-meson and its gluon content}",
    eprint = "2111.08965",
    archivePrefix = "arXiv",
    primaryClass = "hep-ph",
    reportNumber = "RBI-ThPhys-2021-39, WUB/21-2",
    doi = "10.1103/PhysRevD.105.034005",
    journal = "Phys. Rev. D",
    volume = "105",
    number = "3",
    pages = "034005",
    year = "2022"
}

@article{Kroll:2019nby,
    author = "Kroll, Peter and Passek-Kumeri{\v{c}}ki, Kornelija",
    title = "{On some implications of the BaBar data on the $\gamma^*\eta'$ transition form factor}",
    eprint = "1903.06650",
    archivePrefix = "arXiv",
    primaryClass = "hep-ph",
    reportNumber = "WU-B-19-2",
    doi = "10.1016/j.physletb.2019.04.054",
    journal = "Phys. Lett. B",
    volume = "793",
    pages = "195--199",
    year = "2019"
}

@article{Kroll:2012gsh,
    author = "Kroll, Peter and Passek-Kumerički, Kornelija",
    title = "{The {\ensuremath{\eta}} ({\ensuremath{\eta}}') gamma transition form factor and the gluon{\textendash}gluon distribution amplitude}",
    eprint = "1206.4870",
    archivePrefix = "arXiv",
    primaryClass = "hep-ph",
    reportNumber = "WU-B-12-01, WU B 12-01",
    doi = "10.1088/0954-3899/40/7/075005",
    journal = "J. Phys. G",
    volume = "40",
    pages = "075005",
    year = "2013"
}

@article{Kroll:2002nt,
    author = "Kroll, Peter and Passek-Kumerički, Kornelija",
    title = "{The Two gluon components of the eta and eta-prime mesons to leading twist accuracy}",
    eprint = "hep-ph/0210045",
    archivePrefix = "arXiv",
    reportNumber = "WU-B-02-02",
    doi = "10.1103/PhysRevD.67.054017",
    journal = "Phys. Rev. D",
    volume = "67",
    pages = "054017",
    year = "2003"
}

@article{Feldmann:1998vh,
    author = "Feldmann, T. and Kroll, P. and Stech, B.",
    title = "{Mixing and decay constants of pseudoscalar mesons}",
    eprint = "hep-ph/9802409",
    archivePrefix = "arXiv",
    reportNumber = "WU-B-98-2, HD-THEP-98-5",
    doi = "10.1103/PhysRevD.58.114006",
    journal = "Phys. Rev. D",
    volume = "58",
    pages = "114006",
    year = "1998"
}

@article{RQCD:2019osh,
    author = {Bali, Gunnar S. and Braun, Vladimir M. and B{\"u}rger, Simon and G{\"o}ckeler, Meinulf and Gruber, Michael and Hutzler, Fabian and Korcyl, Piotr and Sch{\"a}fer, Andreas and Sternbeck, Andr{\'e} and Wein, Philipp},
    collaboration = "RQCD",
    title = "{Light-cone distribution amplitudes of pseudoscalar mesons from lattice QCD}",
    eprint = "1903.08038",
    archivePrefix = "arXiv",
    primaryClass = "hep-lat",
    doi = "10.1007/JHEP08(2019)065",
    journal = "JHEP",
    volume = "08",
    pages = "065",
    year = "2019",
    note = "[Addendum: JHEP 11, 037 (2020)]"
}

@article{LatticeParton:2022zqc,
    author = "Hua, Jun and others",
    collaboration = "Lattice Parton",
    title = "{Pion and Kaon Distribution Amplitudes from Lattice QCD}",
    eprint = "2201.09173",
    archivePrefix = "arXiv",
    primaryClass = "hep-lat",
    doi = "10.1103/PhysRevLett.129.132001",
    journal = "Phys. Rev. Lett.",
    volume = "129",
    number = "13",
    pages = "132001",
    year = "2022"
}

@article{Gao:2022vyh,
    author = "Gao, Xiang and Hanlon, Andrew D. and Karthik, Nikhil and Mukherjee, Swagato and Petreczky, Peter and Scior, Philipp and Syritsyn, Sergey and Zhao, Yong",
    title = "{Pion distribution amplitude at the physical point using the leading-twist expansion of the quasi-distribution-amplitude matrix element}",
    eprint = "2206.04084",
    archivePrefix = "arXiv",
    primaryClass = "hep-lat",
    reportNumber = "JLAB-THY-22-3626",
    doi = "10.1103/PhysRevD.106.074505",
    journal = "Phys. Rev. D",
    volume = "106",
    number = "7",
    pages = "074505",
    year = "2022"
}

@article{Li:2022qul,
    author = "Li, Hsiang-nan",
    title = "{Dispersive derivation of the pion distribution amplitude}",
    eprint = "2205.06746",
    archivePrefix = "arXiv",
    primaryClass = "hep-ph",
    doi = "10.1103/PhysRevD.106.034015",
    journal = "Phys. Rev. D",
    volume = "106",
    number = "3",
    pages = "034015",
    year = "2022"
}

@article{Brodsky:2006uqa,
    author = "Brodsky, Stanley J. and de Teramond, Guy F.",
    title = "{Hadronic spectra and light-front wavefunctions in holographic QCD}",
    eprint = "hep-ph/0602252",
    archivePrefix = "arXiv",
    reportNumber = "SLAC-PUB-11716",
    doi = "10.1103/PhysRevLett.96.201601",
    journal = "Phys. Rev. Lett.",
    volume = "96",
    pages = "201601",
    year = "2006"
}

@article{Maina:1993uq,
    author = "Maina, Ezio and Torasso, Roberto",
    title = "{Pion Compton scattering in perturbative QCD}",
    eprint = "hep-ph/9309314",
    archivePrefix = "arXiv",
    reportNumber = "DFTT-35-93",
    doi = "10.1016/0370-2693(94)90666-1",
    journal = "Phys. Lett. B",
    volume = "320",
    pages = "337--345",
    year = "1994"
}

@article{Tamazouzt:1988sn,
    author = "Tamazouzt, Mourad",
    title = "{Primakoff effect at high-energy}",
    reportNumber = "LPC-88-03",
    doi = "10.1016/0370-2693(88)91896-5",
    journal = "Phys. Lett. B",
    volume = "211",
    pages = "477--480",
    year = "1988"
}

@article{Coriano:1994nh,
    author = "Coriano, Claudio and Li, Hsiang-Nan",
    title = "{The Transition to perturbative QCD in Compton scattering}",
    eprint = "hep-ph/9405295",
    archivePrefix = "arXiv",
    reportNumber = "ANL-HEP-PR-94-19, CCUTH-94-03",
    doi = "10.1016/0550-3213(94)00504-8",
    journal = "Nucl. Phys. B",
    volume = "434",
    pages = "535--564",
    year = "1995"
}

@article{Zeng:2002mw,
    author = "Zeng, Ding-Fang and Ma, Bo-Qiang",
    title = "{Recalculation of pion Compton scattering in perturbative QCD}",
    eprint = "hep-ph/0202220",
    archivePrefix = "arXiv",
    doi = "10.1016/S0370-2693(02)02278-5",
    journal = "Phys. Lett. B",
    volume = "542",
    pages = "55--64",
    year = "2002"
}

@article{Belle:2009ylx,
    author = "Uehara, S. and others",
    collaboration = "Belle",
    title = "{High-statistics study of neutral-pion pair production in two-photon collisions}",
    eprint = "0903.3697",
    archivePrefix = "arXiv",
    primaryClass = "hep-ex",
    reportNumber = "KEK-PREPRINT-2008-50, BELLE-PREPRINT-2009-4",
    doi = "10.1103/PhysRevD.79.052009",
    journal = "Phys. Rev. D",
    volume = "79",
    pages = "052009",
    year = "2009"
}

@article{Belle:2009xpa,
    author = "Uehara, S. and others",
    collaboration = "Belle",
    title = "{High-statistics study of eta pi0 production in two-photon collisions}",
    eprint = "0906.1464",
    archivePrefix = "arXiv",
    primaryClass = "hep-ex",
    doi = "10.1103/PhysRevD.80.032001",
    journal = "Phys. Rev. D",
    volume = "80",
    pages = "032001",
    year = "2009"
}

@article{Kroll:2012sm,
    author = "Kroll, Peter and Moutarde, Herve and Sabatie, Franck",
    title = "{From hard exclusive meson electroproduction to deeply virtual Compton scattering}",
    eprint = "1210.6975",
    archivePrefix = "arXiv",
    primaryClass = "hep-ph",
    reportNumber = "IRFU-12-174, WUB-12-22, IRFU-12-174; WUB/12-22",
    doi = "10.1140/epjc/s10052-013-2278-0",
    journal = "Eur. Phys. J. C",
    volume = "73",
    number = "1",
    pages = "2278",
    year = "2013"
}

@article{Goloskokov:2005sd,
    author = "Goloskokov, S. V. and Kroll, P.",
    title = "{Vector meson electroproduction at small Bjorken-x and generalized parton distributions}",
    eprint = "hep-ph/0501242",
    archivePrefix = "arXiv",
    reportNumber = "WU-B-05-01, WU B 05-01",
    doi = "10.1140/epjc/s2005-02298-5",
    journal = "Eur. Phys. J. C",
    volume = "42",
    pages = "281--301",
    year = "2005"
}

@article{Goloskokov:2007nt,
    author = "Goloskokov, S. V. and Kroll, P.",
    title = "{The Role of the quark and gluon GPDs in hard vector-meson electroproduction}",
    eprint = "0708.3569",
    archivePrefix = "arXiv",
    primaryClass = "hep-ph",
    reportNumber = "WU-B-07-07, WU B 07-07",
    doi = "10.1140/epjc/s10052-007-0466-5",
    journal = "Eur. Phys. J. C",
    volume = "53",
    pages = "367--384",
    year = "2008"
}

@article{Goloskokov:2009ia,
    author = "Goloskokov, S. V. and Kroll, P.",
    title = "{An Attempt to understand exclusive pi+ electroproduction}",
    eprint = "0906.0460",
    archivePrefix = "arXiv",
    primaryClass = "hep-ph",
    reportNumber = "WU-B-09-06, WU B 09-06",
    doi = "10.1140/epjc/s10052-009-1178-9",
    journal = "Eur. Phys. J. C",
    volume = "65",
    pages = "137--151",
    year = "2010"
}

@article{Musatov:1999xp,
    author = "Musatov, I. V. and Radyushkin, A. V.",
    title = "{Evolution and models for skewed parton distributions}",
    eprint = "hep-ph/9905376",
    archivePrefix = "arXiv",
    reportNumber = "JLAB-THY-99-12",
    doi = "10.1103/PhysRevD.61.074027",
    journal = "Phys. Rev. D",
    volume = "61",
    pages = "074027",
    year = "2000"
}

@article{Belitsky:2001ns,
    author = "Belitsky, Andrei V. and Mueller, Dieter and Kirchner, A.",
    title = "{Theory of deeply virtual Compton scattering on the nucleon}",
    eprint = "hep-ph/0112108",
    archivePrefix = "arXiv",
    reportNumber = "DOE-ER-40762-009, UMD-PP-02-011, YITP-SB-01-51",
    doi = "10.1016/S0550-3213(02)00144-X",
    journal = "Nucl. Phys. B",
    volume = "629",
    pages = "323--392",
    year = "2002"
}

@article{Deja:2023ahc,
    author = "Deja, K. and Martinez-Fernandez, V. and Pire, B. and Sznajder, P. and Wagner, J.",
    title = "{Phenomenology of double deeply virtual Compton scattering in the era of new experiments}",
    eprint = "2303.13668",
    archivePrefix = "arXiv",
    primaryClass = "hep-ph",
    reportNumber = "CPHT-RR012.032022",
    doi = "10.1103/PhysRevD.107.094035",
    journal = "Phys. Rev. D",
    volume = "107",
    number = "9",
    pages = "094035",
    year = "2023"
}

@article{Cuic:2023mki,
    author = "{\v{C}}ui{\'c}, Marija and Duplan{\v{c}}i{\'c}, Goran and Kumeri{\v{c}}ki, Kre{\v{s}}imir and Passek-K., Kornelija",
    title = "{NLO corrections to the deeply virtual meson production revisited: impact on the extraction of generalized parton distributions}",
    eprint = "2310.13837",
    archivePrefix = "arXiv",
    primaryClass = "hep-ph",
    doi = "10.1007/JHEP12(2023)192",
    journal = "JHEP",
    volume = "12",
    pages = "192",
    year = "2023",
    note = "[Erratum: JHEP 02, 225 (2024)]"
}

@article{Duplancic:2016bge,
    author = {Duplan{\v{c}}i{\'c}, G. and M{\"u}ller, D. and Passek-Kumeri{\v{c}}ki, K.},
    title = "{Next-to-leading order corrections to deeply virtual production of pseudoscalar mesons}",
    eprint = "1612.01937",
    archivePrefix = "arXiv",
    primaryClass = "hep-ph",
    doi = "10.1016/j.physletb.2017.05.097",
    journal = "Phys. Lett. B",
    volume = "771",
    pages = "603--610",
    year = "2017"
}

@article{Mankiewicz:1998kg,
    author = "Mankiewicz, L. and Piller, G. and Radyushkin, A.",
    title = "{Hard exclusive electroproduction of pions}",
    eprint = "hep-ph/9812467",
    archivePrefix = "arXiv",
    reportNumber = "TUM-T39-98-33, JLAB-THY-99-02",
    doi = "10.1007/s100529900045",
    journal = "Eur. Phys. J. C",
    volume = "10",
    pages = "307--312",
    year = "1999"
}

@article{Vanderhaeghen:1999xj,
    author = "Vanderhaeghen, M. and Guichon, Pierre A. M. and Guidal, M.",
    title = "{Deeply virtual electroproduction of photons and mesons on the nucleon: Leading order amplitudes and power corrections}",
    eprint = "hep-ph/9905372",
    archivePrefix = "arXiv",
    doi = "10.1103/PhysRevD.60.094017",
    journal = "Phys. Rev. D",
    volume = "60",
    pages = "094017",
    year = "1999"
}

@article{Botts:1989kf,
    author = "Botts, James and Sterman, George F.",
    title = "{Hard Elastic Scattering in QCD: Leading Behavior}",
    reportNumber = "ITP-SB-89-7",
    doi = "10.1016/0550-3213(89)90372-6",
    journal = "Nucl. Phys. B",
    volume = "325",
    pages = "62--100",
    year = "1989"
}

@inproceedings{Fucilla:2025wow,
    author = "Fucilla, Michael and Nabeebaccus, Saad and Szymanowski, Lech and Wallon, Samuel and Yarwick, Joseph",
    title = "{Exclusive photoproduction of a $\pi^0\gamma$ pair in the saturation framework}",
    eprint = "2511.11516",
    archivePrefix = "arXiv",
    primaryClass = "hep-ph",
    month = "11",
    year = "2025"
}

@article{Berger:2001xd,
    author = "Berger, Edgar R. and Diehl, M. and Pire, B.",
    title = "{Time - like Compton scattering: Exclusive photoproduction of lepton pairs}",
    eprint = "hep-ph/0110062",
    archivePrefix = "arXiv",
    reportNumber = "CPHT-S010-0201, DESY-01-119",
    doi = "10.1007/s100520200917",
    journal = "Eur. Phys. J. C",
    volume = "23",
    pages = "675--689",
    year = "2002"
}

@inproceedings{Arrigo2025_TowardsImprovedHadronFemto,
  author      = {Shelby Arrigo},
  title       = {Gamma+pi0 production in GlueX},
  booktitle   = {Towards Improved Hadron Femtography with Hard Exclusive Reactions, Edition IV},
  year        = {2025},
  address     = {Virginia Tech / Jefferson Lab, USA},
  url         = {https://indico.phys.vt.edu/event/66/contributions/1603/},
  note        = {Conference talk},
  urldate     = {2025-11-19}
}


\end{document}